\documentclass[11pt]{article}

 \usepackage{graphics}
 \usepackage{lscape}
 \usepackage{listings}
 \usepackage[hidelinks]{hyperref}

\usepackage{enumitem}
\newlength\titlebox 
\newcommand{\reg}[1]{\textrm{#1} }

\newcounter{cdef}
\newenvironment{definition}[1]{\refstepcounter{cdef}
\ \\
\noindent\textbf{Definition $\arabic{cdef}$. #1}}
{\nopagebreak\hfill (End of definition)\\}

\newcounter{cex}
\newcommand{\funion}{\textit{union}}
\newcommand{\fconcat}{\textit{concat}}
\newcommand{\fstar}{\textit{iter}}
\newcommand{\rep}{\textit{rep}}
\newcommand{\repit}{\textit{rep}}

\newcommand{\nl}{\textit{nl}}

\newcommand{\size}{\ensuremath{\textit{size}}}
\newcommand{\sizerm}{\ensuremath{\textrm{size}}}
\newcommand{\merge}{\ensuremath{\textit{reduce}}}
\newcommand{\reduce}{\ensuremath{\textit{reduce}}}

\newcommand{\concat}{\ensuremath{\odot}}
\newcommand{\union}{\ensuremath{\oplus}}

\newcommand{\iter}{\ensuremath{{\star}}}

\newcommand{\card}{\ensuremath{\sharp}}

\newcommand{\E}{\ensuremath{E}}

\newcommand{\Lang}{\ensuremath{\mathcal{L}}}

\newtheorem{theorem}{Theorem}

\newcommand{\MultiList}{\texttt{MultiList}}
\newcommand{\OneList}{\texttt{OneList}}
\newcommand{\TwoLists}{\texttt{TwoLists}}
\newcommand{\first}{\texttt{first}}
\newcommand{\next}{\texttt{succ}}
\newcommand{\pred}{\texttt{pred}}
\newcommand{\liste}{\textit{list}}
\newcommand{\unify}{\textit{unify}}

\newcommand{\tabNexpr}{\texttt{tabNexpr}}
\newcommand{\type}{\texttt{type}}
\newcommand{\tabCode}{\texttt{tabCode}}
\newcommand{\iExprList}{\texttt{iExprList}}
\newcommand{\hashTable}{\texttt{hashTable}}
\newcommand{\iB}{\textit{iB}}
\newcommand{\iC}{\textit{iC}}
\newcommand{\iE}{\textit{iE}}
\newcommand{\iF}{\textit{iF}}
\newcommand{\iEq}{\textit{iEq}}
\newcommand{\iR}{\textit{iR}}
\newcommand{\id}{\textit{id}}

\newcommand{\nEq}{\textit{nEq}}

\newcommand{\Li}{\textit{L}}
\newcommand{\PU}{\textit{PU}}
\newcommand{\PUI}{\textit{PUI}}
\newcommand{\RI}{\textit{RI}}

\newcommand{\tree}{\texttt{tree}}
\newcommand{\sizett}{\texttt{size}}

\newcommand{\ZERO}{\texttt{ZERO}}
\newcommand{\ONE}{\texttt{ONE}}
\newcommand{\LETTER}{\texttt{LETTER}}
\newcommand{\UNION}{\texttt{UNION}}
\newcommand{\CONCAT}{\texttt{CONCAT}}
\newcommand{\STAR}{\texttt{STAR}}

\newcommand{\mod}{\textrm{mod}}

\newcommand{\Ccal}{\mathcal{C}}
\newcommand{\Ecal}{\mathcal{E}}

\newcommand{\Eq}{\textit{Eq}}

\newcommand{\rpart}{\textit{rpart}}

\newcommand{\tIE}{\texttt{tIE}}
\newcommand{\tabTIE}{\texttt{tabTIE}}

\newcommand{\tabIE}{\texttt{tabIE}}
\newcommand{\tabIR}{\texttt{tabIR}}
\newcommand{\hashEq}{\texttt{hashEq}}
\newcommand{\hashTIE}{\texttt{hashTIE}}
\newcommand{\nextIR}{\texttt{nextIR}}
\newcommand{\nextIEq}{\texttt{nextIEq}}
\newcommand{\listIEQIE}{\texttt{list\_IEQ\_IE}}
\newcommand{\listIEQIEx}{\texttt{list\_IEQ\_IE\_x}}
\newcommand{\listIEQIR}{\texttt{list\_IEQ\_IR}}
\newcommand{\listIRtwoIEq}{\texttt{list\_IR\_2IEq}}
\newcommand{\listIEtwoIEq}{\texttt{list\_IE\_2IEq}}

\newcommand{\findIEq}{\textit{findIEq}}
\newcommand{\findIR}{\textit{findIR}}
\newcommand{\addEq}{\textit{addEq}}
\newcommand{\removeEq}{\textit{removeEq}}
\newcommand{\substitute}{\textit{substitute}}

\title{
\textbf{A Layered Implementation Framework\\ for  Regular Languages}}
\author{Baudouin Le~Charlier\\
Universit\'e catholique de Louvain, Belgium\\
baudouin.lecharlier@uclouvain.be}

\begin{document}

\maketitle

\begin{abstract}
I present the most fundamental features of an implemented system designed to manipulate representations of regular languages. Some functionalities of the system have been presented in previous papers without describing the low level data structures and algorithms ensuring its efficiency. This latter point is the main subject of the present paper. 
The system is structured into two layers, allowing regular languages to be represented in an increasingly compact, efficient, and integrated way.
Both layers are first presented at a high level, adequate to design and prove the correctness of abstract algorithms. Then, their low-level implementations are described meticulously. Algorithms using the system must be written at this level.

At the high level, the first layer offers a notion of normalized regular expressions ensuring that the set of all syntactic derivatives of an expression is finite. This is convenient to design high-level algorithms to compute automata from expressions, compare expressions for inclusion or equivalence, or even simplify expressions. At the low level, normalized expressions are uniquely represented by identifiers, i.e.\ by standard integers so that checking syntactic equality boils down to checking equality of integers. High-level operations on normalized expressions are implemented by algorithms working on integers. High-level algorithms working on normalized expressions can therefore be straightforwardly translated into programs manipulating integers.
The contribution of this layer is the simplification of existing algorithms, at the high level, and a significant increase in efficiency, at the low level.

The second layer, called the background, introduces additional notions to record, integrate, and simplify things computed within the first layer, or even outside the system. Therefore, results of previous computations can be reused for solving new tasks more simply and quickly. At the high level, normalized expressions denoting the same regular language can be unified by grouping them into equivalence classes. One shortest expression is chosen in each class as its representative, which can be used to form equations relating expressions to their derivatives. Sets of equations can be used to represent deterministic finite automata (DFAs). Equations are also used as equality constraints over the regular languages denoted by expressions. Solving such constraints may reduce the number of equivalence classes, and, therefore, the sizes of DFAs. At the low level, equations are uniquely represented by integers, similarly to normalized expressions. 

In this paper, I focus on describing the low-level data structures and the main algorithms of the two layers without describing applicative algorithms. Some of the latter have been described in previous papers; others are topics for future work.
Nevertheless, this paper also presents extensive experimental results to demonstrate the usefulness of the proposed framework and, in particular, the fact that it makes it possible to represent large sets of regular languages in a unified way where distinct identifiers designate different languages, represented by both a small expression and a minimal deteministic automaton (MDFA). It is also shown that such large sets of regular languages can provide interesting statistical information on the way large expressions can be minimized or simplified.

\end{abstract}

\tableofcontents
\addcontentsline{toc}{section}{List of Tables}
\listoftables
\addcontentsline{toc}{section}{List of Figures}
\listoffigures

\normalsize
\newpage
\section{Introduction}
\label{introduction}

According to Set theory \cite{Halmos}, all mathematical objects are sets. So, for mathematicians, sets are a ubiquitous notion, sufficient to define and reason about anything we can imagine. Nevertheless, to write down definitions and reasonings, we also need symbols and formulas. But these are not always distinguished from the sets themselves, in common practice and elementary textbooks \cite{Calculus}. Some mathematicians even consider that the sets \emph{are} the formulas, actually (see, for instance, \cite{Godement}, Section 0.2.\ \emph{Le langage r\'eel des Math\'ematiques}).

In this paper, we consider regular languages. They are possibly infinite sets of strings of letters, which cannot be actually written on paper by enumeration. Instead, we can use finite symbol groupings called regular expressions. Also in this area, some authors confuse the expressions and the sets. This was especially true in the early days of computer science (for example, in \cite{Brzozowski,Conway}), 
when relatively simple regular expressions were used to help designing electronic circuits by hand \cite{Brzozowski}. In \cite{Conway}, Conway freely uses regular languages as states of an automaton recognizing a regular language. Although defined as sets of strings, the states can be represented on paper by simple bullets (\cite{Conway}, page $42$), confirming that large sets can be viewed as atomic objects. To compute the sets represented by the bullets, it is not practical to reason on their set-theoretical definition but rules are set out, making it possible to compute expressions representing them. Sets are represented by symbols in the rules. This method is completely similar to the derivation rules given in \cite{Calculus} for mathematical functions. Conway writes that \emph{the implied algorithm is extremely efficient and, in many cases, gives the minimal machine directly to anyone skilled in input differentiation} (\cite{Conway}, page $43$). The method presents a number of difficulties however: Although it is proven in \cite{Conway} that the number of derivatives of a regular language is finite, as sets, a mechanical application of the rules may generate infinitely many expressions; many of them represent the same regular language but it can be difficult to check which ones. As for mathematical functions, it is at least necessary to apply simplification rules, otherwise bigger and bigger expressions are generated, making the method impractical.

The methods proposed in \cite{Brzozowski,Conway} may be considered applicable only to relatively small examples and by people ``skilled in input differentiation'', i.e., primarily, skilled in simplifying expressions. A major improvement to these approaches was proposed later in \cite{Antimirov}, which lays the foundations for an efficient automation of the method. Here, we want to go further by introducing an implemented framework in which regular languages really are ``as simple as bullets'', which can be easily and efficiently manipulated for solving problems such as simplifying regular expressions \cite{BLC4,Kahrs,Stoughton}, computing deterministic finite automata \cite{Parsing,Brzozowski,BLC5}, checking inclusion or equivalence of regular languages \cite{Antimirov2,Pous}, or studying statistical properties of regular languages \cite{Rotondo}.

In more technical terms, the key idea behind the work presented in this paper is to use \emph{unique identifiers} to stand for any kind of objects we want to represent in our system. Identifiers actually are \emph{integers}, which makes it possible to design efficient data structures to relate the objects to their identifiers. Describing these data structures and demonstrating their usefulness is one of the main goals of this work. The framework that we introduce is made of two layers of increasing power and complexity:

\begin{enumerate}
\item The first layer is concerned with the notion of \emph{normalized regular expressions}. A high-level definition of these expressions is given as well as operations to work on them.
Plain regular expressions can be efficiently translated to normalized expressions and, above all, many equivalent expressions are translated to the same syntactically unique normalized expression (see Theorem \ref{theorem2}).  At the implementation level, unique identifiers are assigned to normalized expressions, using hashing techniques, so that high-level operations are implemented by algorithms working on integers. This allows an efficient implementation of high-level algorithms working on normalized expressions. As an example, the algorithm of \cite{Brzozowski} can be implemented, without checking sufficient conditions for the equivalence of expressions as in \cite{Brzozowski}, by testing the equality of their identifiers.

\item The second layer, called \emph{the background}, maintains an ideally large set of normalized expressions partitioned into classes of equivalent ones, i.e.\ expressions denoting the same regular language. In every equivalence class, a best \emph{representative} is chosen. Additionally, the background may contain \emph{equations} relating some representatives to the representatives of their direct derivatives. These equations are used as constraints on the equivalence classes and they can be grouped to form deterministic automata (DFA) recognizing the regular languages represented in the background.
More information can be added to the background by merging equivalence classes and reducing the set of equations when equivalent expressions  are discovered by any algorithm working with the objects in the background. On the other hand, the information contained in the background can be used to simplify the task of algorithms applied to objects newly added to it. Equivalence classes are implemented using the Union-Find method \cite{GallerF64} applied to identifiers while equations are given identifiers that are useful to use them efficiently as constraints on the equivalence classes.
\end{enumerate}

\noindent The rest of this paper is organized as follows. Sections \ref{first.layer} and \ref{second.layer} respectively deal with normalized expressions and the background, presented at both a high level and an implementation level, where the usefulness of identifiers and low-level data structures is stressed.
Section \ref{experiments} describes experiments. Since the goal of this paper is to describe the foundations of the approach independently of any specific application, the experiments concentrate on the background itself, i.e.\ on what it provides ``for free''. For instance, it is shown that the equivalence classes of the background can be refined, i.e.\ merged, in such a way that expressions in two different classes denote different regular languages. In addition, the representatives of classes are small, sometimes minimal, expressions, obtained using the sole fact that two expressions denoting the same regular language belong to the same equivalence class. This method, applied to large sets of randomly generated expressions, gives us precise statistical information on the distribution of regular expressions with respect to their minimal size.

\section{First layer: normalized expressions}
\label{first.layer}

We first give a high-level definition of normalized expressions and their operations in Section \ref{normalized.expressions.normalized}. Section \ref{normalized.expressions.implementation} describes their low-level implementation with a complexity analysis and a short experimental evaluation.

\subsection{High-level description of normalized expressions and their operations}
\label{normalized.expressions.normalized}

\subsubsection{Regular expressions and normalized expressions}
\label{normalized.expressions}

\paragraph{Basic concepts} We
 assume a finite set of \emph{letters}. In practice, in this paper, only lowercase letters, i.e.\ $a$, $b$, \dots, $z$ are used. (For the examples, mainly $a$ and $b$.)
A string of letters, denoted by $t$, $u$, or $w$, is a finite sequence $x_1 x_2 \dots x_n$ where $x_1, x_2, \dots, x_n$ are letters ($n$ is the length of the string). The empty string ($n=0$) is denoted by $1$. We identify every letter $x$ with the string of length $1$ containing $x$ only. The concatenation of two strings $u$ and $w$, denoted by $u \cdot w$ or, more simply, by $u\,w$, is the string $y_1 y_2 \dots y_m x_1 x_2 \dots x_n$ where $u = y_1 y_2 \dots y_m$ and $w = x_1 x_2 \dots x_n$. 
Let $S$, $S_1$, and $S_2$ be sets of strings. The concatenation of $S_1$ and $S_2$, denoted by $S_1 \cdot S_2$, is the set of strings $\{ w_1 \cdot w_2\:|\: w_1\in S_1\: \& \: w_2\in S_2\}$. The iteration of $S$, denoted by $S^*$, is the set of strings $\{ w_1\, \cdot \,w_2\, \cdot \,\dots\,  \cdot w_n\:|\:  w_1\,,\,w_2\,,\,\dots\, ,w_n\in S\: (n \ge 0)\}$. %
The derivative of $S$ with respect to the string $w$ is the set $S_w = \{ u \:|\: w\cdot u \in S\}$. If $w$ is a letter $x$, we say that $S_x$ is the direct derivative of $S$ with respect to $x$.

\begin{definition}{Regular expressions, normalized expressions}
\label{definition.expressions}
\begin{enumerate}
\item A (plain) regular expression $E$ is either the symbol $0$, or the symbol $1$, or a letter, or a \emph{union}, of the form $E_1 + E_2$, or a \emph{concatenation}, of the form $E_1 \cdot E_2$, or an \emph{iteration}, of the form $E^*$, where $E$,  $E_1$ and $E_2$ are (simpler) regular expressions. A regular expression denotes a set of strings  in the ``obvious way'' (see, e.g., \cite{Parsing,Conway,Stoughton}). This set is denoted by $\Lang(\E)$ and is called a regular set.

\item A \emph{normalized} regular expression is either the symbol $0$, or the symbol $1$, or a letter, or a \emph{union}, of the form $E_1\, + E_2\, + \, \dots \, + \, E_n$, where $n\ge 2$ and $E_1,\, E_2,\, \dots,\, E_n$ are syntactically different normalized expressions that are neither unions nor equal to $0$, or a \emph{concatenation}, of the form $E_1 \cdot E_2$, where $E_1$ and $E_2$ are not equal to $0$ or $1$ and $E_1$ is not a concatenation, or an \emph{iteration}, of the form $E^*$, where $E$ is not equal to $0$ or $1$ and is not an iteration. 

We say that $E_1,\, E_2,\, \dots,\, E_n$,  $E_1$ and $E_2$, and $E$ respectively are \emph{the direct subexpressions} of the normalized expressions $E_1\, + E_2\, + \, \dots \, + \, E_n$, $E_1 \cdot E_2$, and $E^*$. The symbols $0$, $1$, and the letters have an empty set of subexpressions. If $E'$ is a direct subexpression of $E$, we also say that $E$ is a \emph{superexpression} of $E'$.

We sometimes write meta-expressions such as ``$E_1\, + E_2\, + \, \dots \, + \, E_n$, where $n\ge 0$''. This means that the intended expression either is $0$ (if $n = 0$), or is neither a union nor $0$ (if $n = 1$), or is a union (if $n\ge 2$).

We assume a total ordering on the set of normalized expressions.\footnote{In practice, this ordering is fixed by the implementation (see Section \ref{normalized.expressions.implementation}): At the low level, 
representations of normalized expressions are created one by one in some order, which progressively defines the ordering on expressions.
 It will become clear later that this choice is the most efficient for implementing the basic operations on normalized expressions.}
Using this ordering, we impose the additional constraint that, in a union $E_1\, + E_2\, + \, \dots \, + \, E_n$, the sequence $E_1,\, E_2,\, \dots,\, E_n$ is strictly sorted, in ascending order, with respect to this ordering. 
\end{enumerate}
\end{definition}
\paragraph{Note} My implemented system\cite{BLC4} also considers \emph{extended} regular expressions and \emph{normalized extended} regular expressions, which are of the form $E_1 \,  \omega\, E_2$, where $E_1$ and $E_2$ are regular or extended regular expressions (normalized if appropriate) and $\omega$ is a set operator such as $\cap$, $\setminus$, and  $\bigtriangleup$. Expressions of the form $! E$, where $!$ stands for the complement, are also considered.
For simplicity, extended expressions are not considered in this paper.\\

\begin{figure}[t]
\caption{Set of strings $\Lang(E)$ denoted by a normalized expression}
\label{fig.lang.expressions}

$$
\begin{array}{|ccll|}\hline&&&\\[-0.5em]
\Lang(0)& = & \{\} & \\
\Lang(1) & = & \{1\}  & \\[0.5em]
\Lang(x) & = & \{x\}  & \\[0.5em]
\:\:
\Lang\:{(\E_1\, +\, \dots\, +\, \E_n)} & = & 
  \Lang(\E_1)\, \cup\, \dots\, \cup\, \Lang(\E_n) & (n \ge 2)\:\:\\[0.5em]
\Lang{(\E_1\: \cdot\, \E_2)} & = &

  \Lang(E_1)\:\cdot\:\Lang(\E_2)
                              & \\[0.5em]
  \Lang{\:(\E^*)} & = & (\Lang(\E))^* & \\[0.5em]
  
  \hline
\end{array}
$$

\end{figure}

 The set of normalized expressions can be seen as a subset of the set of all regular expressions by identifying every union $\E_1\, +\, \dots\, +\, \E_n$ to the plain regular expression $\E_1\, +\, (\E_2\, + \dots\, +\, \E_n)$ (using right associativity). Thus, the language defined by a normalized expression $E$ is equal to $\Lang (E')$, where $E'$ is the regular expression to which $E$ is identified. Explicit rules defining  $\Lang (E)$ are given in Figure \ref{fig.lang.expressions}.
When we write normalized expressions on paper, we write them as the plain expressions to which they are identified and we can drop parentheses as  is usually done for plain regular expressions. As a typical example, the expression $$\reg{ b(a + b(1 + a + b*b))((a + b)a*)*}$$ has to be parsed as $E_1\,  (E_2\,  E_3)$ where $E_1 = \reg{b}$, $E_2 = \reg{a + b(1 + a + b*b)}$, and $E_3 = \reg{((a + b)a*)*}$.

\subsubsection{Operations on normalized expressions}
\label{normalized.expressions.operations}

We introduce three operations on normalized expressions. They can be used, at a high level, for describing algorithms working on normalized expressions. Since $0$, $1$, and the letters already are normalized, normalizing plain expressions can be  done recursively, using these operations, by induction on the structure of plain expressions. 

The operations \funion, \fconcat, and \fstar\/ take one or two normalized expressions as argument(s), and return a uniquely defined normalized expression denoting the union, the concatenation, or the iteration of the language(s) denoted by their argument(s). For the sake of beauty, we usually write these operations as the infix operators \union\/ (\funion) and \concat\/ (\fconcat), and the postfix operator $\star$ (\fstar).

\begin{definition}{Operations on normalized regular expressions}
\label{normalizing.operations}

\ \\
Let $E$, $E_1$, and $E_2$ be normalized expressions.

\begin{itemize}

\item The operation \funion\ ($\union$)
\begin{enumerate}
\item $\funion(0, E) = \funion(E, 0) = E$
\item Assume that $E_1$ and $E_2$ are different from $0$.

For $i = 1, 2$, let $S_i\, =\, \{E_{i\,1},\, \dots,\, E_{i\,{n_i}}\}$ where $E_i\, = E_{i\,1}\,+\, \dots\, +\, E_{i\,{n_i}}$, if $E_i$ is a union, and let $S_i\, =\,\{E_i\}$, otherwise. Let $F_1,\,\dots,\,F_m$ be the strictly ordered sequence of normalized expressions such that $S_1 \cup S_2\ =\ \{F_1,\,\dots ,\,F_m\}$. Then, by definition, $\funion(E_1, E_2)$ is the normalized expression $F_1\,+\,\dots\,+\,F_m$, if $m \ge 2$. Note that we can have $m = 1$. In that case, the result simply is $F_1$, which is not a union by definition of normalized expressions.
\end{enumerate}

\item The operation \fconcat\ ($\concat$)

\begin{enumerate}
\item $\fconcat(0, E) = \fconcat(E, 0) = 0$
\item $\fconcat(1, E) = \fconcat(E, 1) = E$
\item Assume that $E_1$ and $E_2$ are different from $0$ and $1$.

If  $E_1$ is not a concatenation, then 
$\fconcat(E_1,\,E_2)$ is the concatenation $E_1\,.\,E_2$.
Otherwise, $E_1$ can be written as $F_1\,.\,F_2$, where $F_1$ is not a concatenation. In that case, $\fconcat(E_1,\,E_2)\,=\,F_1\,.\,G$ where $G\,=\, \fconcat(F_2, E_2)$.
\end{enumerate}

\item The operation \fstar\ ($\star$)

\begin{enumerate}
\item $\fstar(0) = \fstar(1) = 1$
\item If $E$ is an iteration, $\fstar(E)\, = \, E.$
\item If $E$ is not an iteration and is different from $0$ and $1$, $\fstar(E)\, = \, E^{*}.$
\end{enumerate}
\end{itemize}
\end{definition}

 We conclude this section by two theorems (proven in Appendix \ref{normalized.expressions.complements}) stating the main mathematical properties of normalized expressions and their operations. It is important to understand that the symbol $=$, used between two meta-expressions denoting  normalized expressions, denotes strict syntactic equality of these normalized expressions. 

\begin{theorem}{Properties of the operations \union\/, \concat, and $\star$}
\label{theorem1}
\begin{enumerate}
\item The operation \funion\/ is associative, commutative, and idempotent. (So, we can freely write $E_1\,\union\,E_2\,\dots\,\union\,E_n$ without paying attention to the position of the expressions $E_i$ in the whole expression. Moreover, it does not matter if we repeat the same subexpression several times.)

\item The operation $\fconcat$ is associative. So, for any normalized expressions $E_1$, $E_2$, $E_3$, we can write:
$$E_1\,\concat\,E_2\,\concat\,E_3 \ = \ 
(E_1\,\concat\,E_2)\,\concat\,E_3 \ = \ E_1\,\concat\,(E_2\,\concat\,E_3).$$

\item Let $E$, $E_1$, and $E_2$ be normalized expressions. The following equalities hold:
$$
\begin{array}{ccccccc}
\Lang\:{(\E_1\union \E_2)}\: = \:
  \Lang\:{\E_1}\, \cup\, \Lang\:{\E_2} &&&

\Lang{(\E_1\: \concat\, \E_2)} \:= \:

  (\Lang{\:\E_1})\:.\:(\Lang{\:\E_2}) &&&

  \Lang{\:(E\star)} \:= \:(\Lang(\E))^* 
  
\end{array}
$$
\end{enumerate}
\end{theorem}

\begin{definition}{A congruence relation on regular expressions}
\label{congruence.expressions}

\ \\
We define the relation $\cong$ as the least congruence on the set of plain regular expressions that contains the following equivalences, for any regular expressions $E$, $E_1$, $E_2$, and $E_3$:
$$\begin{array}{ccccccc}
 E + E \: \cong\: 0 + E\: \cong\: E\: \cong\: E + 0

 &&&  0\cdot E\: \cong\: 0 \: \cong\:E\cdot 0  
&&&   1\cdot E\: \cong\: E \: \cong\:E\cdot 1 \\[1em]

 E_1 + (E_2 + E_3) \: \cong\: (E_1 + E_2) + E_3 
&&&  E_1 + E_2  \: \cong\: E_2 + E_1 &&&\\[1em]

 E_1 \cdot  (E_2 \cdot  E_3) \: \cong\: (E_1 \cdot  E_2) \cdot  E_3 
&&&  0^* \: \cong\: 1^* \: \cong\: 1 
&&&  (E^*)^* \: \cong\: E^*

\end{array}$$
\end{definition}

\begin{theorem}{Plain regular expressions that are equivalent with respect to the relation $\: \cong\:$ are normalized to the same expression.}
\label{theorem2}

\end{theorem}

\subsection{Implementation of normalized expressions and their operations}
\label{normalized.expressions.implementation}

The mathematical properties of the operations \union, \concat, and $\star$ ensure that we can compute a deterministic finite automaton from any normalized expression by computing its syntactic derivatives. An algorithm for doing so is described and proven correct in \cite{BLC5}. This algorithm can be seen as a variant, and an improvement, of the methods proposed in \cite{Antimirov,Brzozowski,Conway}. The main improvement is the fact that syntactic equality provides a sufficient condition for the equivalence of expressions that ensures that the set of syntactic derivatives is finite. As a matter of fact, we can go further by providing an implementation of normalized expressions in which syntactic equality can be checked in $O(1)$ time, by associating a unique integer identifier to any expression represented in the system. This identifier is attributed on demand and may change from one run of the system to the next.
This issue is the subject of this section. Notice also that, in this section, I write expression instead of normalized expression, for simplicity.

\subsubsection{Main implementation choices}
\label{implementation.choices}

As mentioned above, I choose to identify each expression represented in the system by a unique integer. The number of available identifiers is decided at the start of the system. We denote it by the letter $M$. The available identifiers are then: $0$, $1$, \dots, $M - 1$.

It is also worth mentioning in passing that the system is implemented in Java, as this programming language offers a very flexible notion of array, which is used extensively. In contrast, we make no use of more advanced features of Java, so that the system could easily be reimplemented in most other imperative programming languages, such as C. Since we make relatively little use of objects, except of course arrays, the system also makes relatively little use of the Java garbage collector. In the following, we denote an array of $n$ elements by $t[n]$, where $t$ is its name. Its elements are denoted by $t[i]$, where $i$ denotes an integer such that $0\leq i < n$. We  use the notation $\{v_0, \dots, v_{n - 1}\}$ for describing an array of $n$ values.

\subsubsection{Low-level data structures}
\label{low.level.data.structures}

Before explaining how expressions are represented, we need to describe some low-level data structures that are extensively used  in the system. The most useful of them is called \MultiList. A single object of type \MultiList\ is specified by two positive integers $n$ and $m$, and it implements $n$ \emph{disjoint} lists $\liste(0)$, $\liste(1)$, \dots, $\liste(n - 1)$ of integers greater or equal to $0$ and less than $m$. Most of the time $m = M$. When an object
$\MultiList(n, m)$ is created, all its lists are empty. It is possible to add a number into a list, if it does not belong to any list beforehand. It is also possible to remove a number from a list to which it belongs beforehand. These operations are executed in constant time.
For convenience, it is allowed to make an attempt to add a number to a list even if it already belongs to it or to another list of the same $\MultiList$. In that case, no list is modified. Similarly, an attempt to remove a value not belonging to any list does not change anything.
We can also traverse a list to get all its elements in  $O(\ell)$ time, where $\ell$ is the number of elements in the list. Adding and removing elements can be done during the traversal of one or possibly several lists. A newly added element becomes the first of the list. Removing an element does not change the ordering of the others. It is also possible to check in $O(1)$ time whether a number belongs to some unspecified list.

An object $\MultiList(n, m)$ is implemented thanks to three arrays of integers $\first[n]$, $\pred[m]$, and $\next[m]$. The values in these arrays are determined by the following rules: $\first[i]$ is the first element in  $\liste(i)$, if the list is not empty; it is equal to $-1$, if the list is empty. If an integer $i$ belongs to some list, $\pred[i]$ is the number preceding $i$ in the list, if $i$ is not the first element in the list, and $\pred[i] = - 1$, otherwise. Similarly, $\next[i]$ is the number following $i$ in the list, if $i$ is not the last element in the list, and $\next[i] = - 1$, otherwise. If an integer $i$ such that $0 \leq i < m$ does not belong to any list, the equality $\pred[i]= 0=\next[i]$ holds. On the basis of these rules, it is easy to implement the operations specified above within the announced complexity constraints.

 The implementation of expressions also uses two other kinds of objects. An object $\OneList(m)$ is equivalent to an object $\MultiList(1, m)$ except that the operations need one argument less, since the list number implicitly is $0$. 
On the other hand, we need an object called $\TwoLists(m)$, which is used as a pool of identifiers of expressions (with $m = M$). It is implemented with an object $\MultiList(2, m)$. The list $\liste(0)$ contains the identifiers currently in use, while $\liste(1)$ contains the identifiers that are free to be used. We can choose an identifier to use, either explicitly, if it is not in use, or implicitly as the first free identifier. We can also return a no longer needed  identifier to the free list. This is needed to implement a hand-crafted garbage collector for the system (see Appendix~\ref{garbage.collection}).

\subsubsection{Internal representation of expressions}
\label{internal.representation}

Let us assume a finite set of expressions that we call ``expressions currently represented in the system''. It is required that all their subexpressions also belong to this set. Moreover, we assume that each expression $E$ has received an identifier, i.e.\ an integer  $\iE$ such that $0\leq \iE < M$, different for different expressions.
Such a set is implemented thanks to the following objects. 

{\setlength{\leftmargini}{0em}

\begin{description}

\item[${\tabNexpr[M][]}$] is an array of integer arrays. Let $\iE$ be the identifier of an expression $E$, currently represented in the system. Then, ${\tabNexpr[\iE]}$ is an array of integers containing the identifiers of the direct subexpressions of $E$.

\item[${\type[M]}$] is an array of small integers representing the types of the expressions. Types are the following constant values: \ZERO, \ONE, \LETTER, \UNION, \CONCAT, \STAR.

\item[${\tabCode[M]}$] is an array of integers. With the same conventions as above, the integer ${\tabCode[\iE]}$ is the hash code of the expression $E$. Hash codes are computed as integer functions of the identifiers of the direct subexpressions of expressions, and of the integer representing their type (except that, for letters, the identifier of the letter is used; see below).

\item[${\iExprList(M)}$] is a $\TwoLists(M)$ object, determining the set of identifiers of the expressions currently represented in the system, and, complementarily, the set of free identifiers.
The atomic expressions $0$, $1$, $a$, \dots, $z$ are given the identifiers $0$, $1$, $2$, \dots, $27$, respectively. Those values are added to \iExprList\ before creating any other expression.

\item[${\hashTable(n, M)}$] is a $\MultiList(n, M)$ object used as a hash table to determine if an expression currently is represented in the system. The value of $n$ is chosen big enough to ensure that checking if an identifier is contained in the hash table can be done in  $O(1)$ time, on the average.

\end{description}}

\subsubsection{Implementation of the operations on expressions}
\label{implementation.operations}

We have already said how atomic expressions are represented. Their identifiers are the same in all executions of the system.
Now, let us explain how the operations \union, \concat, and $\star$ are implemented. The algorithms are low-level implementations of the abstract algorithms of Definition \ref{normalizing.operations}. The fundamental change is the fact that the arguments and result are identifiers of expressions instead of expressions. 
It is also necessary to check if the result of an operation already exists or if it has to be created, which is meaningless in the abstract case.

 The first step of each operation consists of determining the type of the resulting expression and computing an array containing the identifiers of its direct subexpressions. This is done as in Definition \ref{normalizing.operations} but pairs of expressions and strictly ordered sequences of expressions are represented by arrays of integers. The total order on expressions, left unspecified in Definition \ref{definition.expressions}, is thus materialized by the usual order on integers. This implementation choice is the most efficient possible.%
\footnote{As a consequence of this choice, the same expression can be represented differently in different runs of the system, depending on what expressions have been created beforehand.}
Having  the type of the result and the array containing the identifiers of its subexpressions, its hash code $h$ is computed, and used to traverse the list $\liste(i)$ of the hash table, where $i = \mod(h, n)$ for some appropriate function $\mod$ ensuring that $0 \leq i < n$. For all identifiers $\iE$ in $\liste(i)$, the value 
${\type[\iE]}$ and possibly the array ${\tabNexpr[\iE]}$ are
respectively compared to the type and to the array freshly computed. In case of equality, the result already exists so that the identifier $\iE$ is returned by the operation. If no comparison succeeds, a new identifier $\iE$ is chosen from \iExprList, and the elements ${\tabNexpr[\iE]}$, ${\type[\iE]}$, and ${\tabCode[\iE]}$ are initialized with the newly computed array, type, and hash code.
Also, the identifier $\iE$ is added to $\liste(i)$ and, finally, returned as the result of the operation.

\subsubsection{Complexity of normalization}
\label{complexity.normalization}

In this paper, normalized expressions are proposed as an efficient representation of plain regular expressions  for solving problems such as computing DFAs (\cite{BLC5}) and simplifying expressions (\cite{BLC4}). They constitute the first layer of a more expressive data structure, described in the next section. In any case, as a first step, it is interesting to provide arguments about its efficiency as a stand-alone data structure. Therefore, we first discuss the time and space complexity of executing a single operation \union, \concat, or $\iter$. Next, we consider the overall complexity of algorithms based on these operations. Finally, statistics on concrete experiments are presented. 

 Let us call the \emph{terms} of a normalized expression $E$ the expressions $E_1$, \dots, $E_n$ such that $E$ is syntactically equal to $E_1\, + E_2\, + \, \dots \, + \, E_n$ $(n \ge 0)$ (see Definition \ref{definition.expressions}). Similarly, let us call the \emph{factors} of a normalized expression $E$, the expressions $E_1$, \dots, $E_n$ such that $E$ is syntactically equal to $E_1\, \cdot (E_2\, \cdot (  \, \dots \, E_n))$ $(n \ge 0)$. (The case $n =0$ holds when $E = 1$, and the case $n = 1$ takes place when $E$ is not a concatenation.)
With this terminology, the time and space complexity of computing $E_1\union\E_2$ is  $O(n_1 + n_2)$, where $n_1$ and $n_2$ are the numbers of terms in $E_1$ and $E_2$. Similarly, the time and space complexity of computing $E_1\concat\E_2$ is $O(n_1)$, where $n_1$ is the number of factors in $E_1$. Also, the complexity of computing $E\star$ is $O(1)$. In all cases, the complexity is due to building new arrays ${\tabNexpr[\iE]}$, computing hash codes, and traversing lists in the hash table. The results hold on the average provided that the lists contain $O(1)$ elements.

 Now, let us discuss the complexity of algorithms working on normalized expressions and making intensive use of the operations \union, \concat, and $\star$. In fact, it is not possible to draw any general conclusion since the kind of expressions used may differ from an application to another (see, for instance, \cite{BLC5,BLC4}). But, as a first approach, we can consider the simple problem of normalizing a plain regular expression chosen at random. For such expressions, the probability that a subexpression is  normalized to an expression containing many terms or many factors is low. Therefore, normalizing a plain expression has an average complexity close to that of constructing a copy of the plain expression, i.e.\  it is close to linear in the size of the expression. However, there are a few worst cases where the complexity is quadratic. An example is the expression $(C_1\, +\, (C_2\, + \,\dots +\, (C_{n-1}\, + \,C_n)\,\dots\,))$, where the $C_i$ are small expressions that are not unions. For such examples, it is better to first build an array $\{\iC^{\ '}_1, \, \iC^{\ '}_2, \,\dots \,, \iC^{\ '}_n\}$ of identifiers by normalizing $C_1$,  $C_2$, \dots,  $C_n$, independently. This array can be sorted in $O(n \log n)$ time, which allows us to normalize the whole expression at worst in $O(s \log s)$ time, where $s$ is the size  of the expression. A similar optimization can be done for concatenation with a large number of factors, but, in this case, it is sufficient to apply the operator $\concat$ from right to left, to ensure a linear complexity.

 The discussion above suggests that it is useful to implement two $n$-ary versions of the operations $\funion$ and $\fconcat$. The space and time complexity of computing $\funion(\iE_1,\,\dots\,, \iE_n)$ is $O((m_1 + \dots + m_n)\log n)$, where $m_i$ is the number of terms of $E_i$ $(1 \leq i \leq n)$. The complexity of $\fconcat(\iE_1,\,\dots\,, \iE_n)$ is linear in the total number of factors of $E_1$, \dots, $E_n$.

\begin{table}[t]
\caption{Complexity of normalization}\label{computing.normalization.table}
\begin{center}
$\setlength{\arraycolsep}{0.35em}
\renewcommand{\arraystretch}{1.2}
\begin{array}{|r||r|r|r||r||r|r|r|r|r|}\hline
%& & &  & &  &  & &  & & &  &  \\[-0.5em]
\textit{size} & t_\textnormal{r}\ & t_T\  & t_N\  & |\,\textit{sub}\,| & \card \iE \hspace{1em}  & \card \textit{empty} \hspace{1em}& \card \textit{nempty}\hspace{1em} &\card \textit{iter}\hspace{1em}&\card \textit{dejaVu}\hspace{1em}\\[0.5em]\hline\hline
8 & 0.17 & 0.40 & 0.44 & 1.84 & 2,022 & 1,994 & 30,542 & 30,542 & 30,542 \\ 
16 & 0.20 & 0.85 & 0.89 & 2.01 & 20,234 & 20,203 & 50,577 & 50,577 & 50,574 \\ 
32 & 0.30 & 1.61 & 1.91 & 2.10 & 61,333 & 60,848 & 89,799 & 90,015 & 89,342 \\ 
64 & 0.45 & 2.85 & 3.77 & 2.14 & 137,937 & 135,991 & 175,819 & 176,881 & 173,901 \\ 
128 & 0.79 & 5.91 & 7.52 & 2.17 & 286,821 & 281,059 & 352,482 & 378,982 & 346,748 \\ 
256 & 1.36 & 11.2 & 15.7 & 2.18 & 578,190 & 561,321 & 716,449 & 722,859 & 699,608 \\ 
512 & 2.62 & 21.9 & 33.2 & 2.19 & 1,147{\scriptsize\times{10}^3} & 1,092{\scriptsize\times{10}^3} & 1,476{\scriptsize\times{10}^3} & 1,517{\scriptsize\times{10}^3} & 1,420{\scriptsize\times{10}^3} \\ 
1K & 5.24 & 45.4 & 70.4 & 2.19 & 2,260{\scriptsize\times{10}^3} & 2,067{\scriptsize\times{10}^3} & 3,078{\scriptsize\times{10}^3} & 3,464{\scriptsize\times{10}^3} & 2,885{\scriptsize\times{10}^3} \\ 
2K & 10.2 & 88.4 & 152 & 2.20 & 4,435{\scriptsize\times{10}^3} & 3,751{\scriptsize\times{10}^3} & 6,548{\scriptsize\times{10}^3} & 7,810{\scriptsize\times{10}^3} & 5,864{\scriptsize\times{10}^3} \\ 
4K & 21.5 & 169 & 349 & 2.20 & 8,709{\scriptsize\times{10}^3} & 6,379{\scriptsize\times{10}^3} & 14,224{\scriptsize\times{10}^3} & 17,582{\scriptsize\times{10}^3} & 11,894{\scriptsize\times{10}^3} \\ 
8K & 51 & 365 & 783 & 2.20 & 17,086{\scriptsize\times{10}^3} & 11,602{\scriptsize\times{10}^3} & 29,620{\scriptsize\times{10}^3} & 39,757{\scriptsize\times{10}^3} & 24,136{\scriptsize\times{10}^3} \\ 
\hline
\end{array}$
\end{center}
\end{table}

\begin{table}[t]
\caption{Complexity of normalization (opt)}\label{computing.normalization.table.opt}
\begin{center}
$\setlength{\arraycolsep}{0.35em}
\renewcommand{\arraystretch}{1.2}
\begin{array}{|r||r|r|r||r||r|r|r|r|r|}\hline
%& & &  & &  &  & &  & & &  &  \\[-0.5em]
\textit{size} & t_\textnormal{r}\ & t_{T}\  & t_{N}\  & |\,\textit{sub}\,| & \card \iE \hspace{1em}  & \card \textit{empty} \hspace{1em}& \card \textit{nempty}\hspace{1em} &\card \textit{iter}\hspace{1em}&\card \textit{dejaVu}\hspace{1em}\\[0.5em]\hline\hline
8 & 0.16 & 0.40 & 0.61 & 1.84 & 2,022 & 1,994 & 27,209 & 27,209 & 27,209 \\ 
16 & 0.21 & 0.82 & 1.33 & 1.99 & 18,494 & 18,460 & 40,361 & 40,361 & 40,355 \\ 
32 & 0.30 & 1.60 & 2.54 & 2.06 & 50,697 & 50,427 & 69,580 & 70,471 & 69,338 \\ 
64 & 0.44 & 2.99 & 4.97 & 2.09 & 108,839 & 107,648 & 135,461 & 135,825 & 134,298 \\ 
128 & 0.74 & 5.72 & 10 & 2.10 & 220,628 & 217,027 & 270,320 & 272,447 & 266,747 \\ 
256 & 1.29 & 11.7 & 20.5 & 2.11 & 440,462 & 429,868 & 547,967 & 556,454 & 537,401 \\ 
512 & 2.59 & 22 & 41.7 & 2.11 & 867,830 & 836,295 & 1,121{\scriptsize\times{10}^3} & 1,197{\scriptsize\times{10}^3} & 1,090{\scriptsize\times{10}^3} \\ 
1K & 4.90 & 42.7 & 86.5 & 2.12 & 1,705{\scriptsize\times{10}^3} & 1,595{\scriptsize\times{10}^3} & 2,323{\scriptsize\times{10}^3} & 2,443{\scriptsize\times{10}^3} & 2,214{\scriptsize\times{10}^3} \\ 
2K & 10.1 & 84.7 & 176 & 2.12 & 3,338{\scriptsize\times{10}^3} & 2,946{\scriptsize\times{10}^3} & 4,891{\scriptsize\times{10}^3} & 5,479{\scriptsize\times{10}^3} & 4,500{\scriptsize\times{10}^3} \\ 
4K & 21.1 & 166 & 382 & 2.13 & 6,549{\scriptsize\times{10}^3} & 5,181{\scriptsize\times{10}^3} & 10,499{\scriptsize\times{10}^3} & 12,307{\scriptsize\times{10}^3} & 9,130{\scriptsize\times{10}^3} \\ 
8K & 88.9 & 358 & 982 & 2.13 & 12,835{\scriptsize\times{10}^3} & 8,261{\scriptsize\times{10}^3} & 23,098{\scriptsize\times{10}^3} & 31,419{\scriptsize\times{10}^3} & 18,523{\scriptsize\times{10}^3} \\ 
\hline
\end{array}$
\end{center}
\end{table}

Experiments carried out to support the previous considerations are presented in Tables \ref{computing.normalization.table} and~\ref{computing.normalization.table.opt}.
 Large files of $10,000$ plain expressions of sizes $8$, $16$, \dots, $8K (= 8,192)$ using at most two letters have been normalized. (Additional information about the test environment is given in Section \ref{experiments}.) Expressions in the files are written in infix notation on a single line. They are read one after the other and parsed to give a plain regular expression represented in the form of a binary tree. Then, the tree is traversed recursively to normalize the expression. All normalized expressions generated by this process are kept in memory. The following values are collected in the tables: the size of expressions in the file (\textit{size}), the average times needed to read an expression from the file $(t_\textnormal{r})$, parse the line and build the binary tree $(t_T)$, and normalize the expression $(t_N)$. Times are given in micro-seconds. The column $|\,\textit{sub}\,|$ gives the average number of direct subexpressions of an expression, i.e.\ the number of elements of the arrays 
${\tabNexpr[\iE]}$. The column $\card \iE$ contains the total number of identifiers in use at the end of the normalization process. The last four columns provide information about the hash table usage: the columns $\card \textit{empty}$ and  $\card \textit{nempty}$ respectively give the number of times that an empty or non empty list was searched for an existing identifier. The column $\card \textit{iter}$ gives the  number of iterations in the lists of the hash table. Finally, the column $\card \textit{dejaVu}$ provides the number of times that a newly normalized expression was found syntactically equal to an existing one, so that no new identifier was needed. Table \ref{computing.normalization.table} corresponds to the case where normalization is done by a simple recursion on the structure of plain expressions, while Table \ref{computing.normalization.table.opt} uses the optimization, explained in Section \ref{complexity.normalization}, allowing us to avoid a quadratic complexity in worst cases.

 An examination of the results in the tables confirms that normalization is an efficient and useful process. It also shows that the assumptions stated to justify the average complexity of the method actually hold for expressions chosen at random. For instance, the time needed for normalization is less than two times the time needed to build a plain regular expression, except for large expressions. The times needed for normalization grow almost linearly with the sizes of expressions. They are even better without the optimization.\footnote{This fact suggests that the best method could be dynamically chosen depending on the number of terms in unions. It would also be more efficient to use an insertion sort for computing the union of a short list of expressions.}
This is greatly explained by the values in the column $|\,\textit{sub}\,|$, which show that the number of terms in expressions is small, on the average.
In fact, a more interesting benefit of the optimization is that fewer normalized expressions (mainly unions) are constructed, saving memory space (compare the columns $\card \iE\,$). Finally, the figures in the last four columns fully support the claim that the time complexity of using the hash table is $O(1)$ on the average: The identifiers are almost perfectly distributed in the lists and most iterations in the lists are due to the normalization of a plain expression into an existing normalized expression.

\section{Second layer: equivalence classes and equations}
\label{second.layer}

Here, we go a step further toward our goal of building a ``world'' where regular languages have a best representation, both simple and unique. To reach this goal, regular languages are given an integer identifier corresponding to both a short normalized expression and the initial state of a DFA for the language.
Original techniques used to make it possible are described here but, to ensure that different regular languages are given different identifiers, the implemented system also uses well-known algorithms for minimizing DFAs \cite{Parsing,Hopcroft,Moore}. In this section, we  focus solely on the new techniques. The new and old methods are used together in Section~\ref{experiments},  devoted to experiments, showing that the overall goal is achievable.

A key idea of the approach described below is to maintain a large set of normalized expressions that are systematically simplified and linked to a DFA. This set constitutes a kind of database of regular languages from which useful information can be extracted to efficiently simplify  other expressions and compute DFAs for them. In the following, we call this database the \emph{background}. In Section~\ref{background.highlevel}, the background is described at a high level, with the main operations required to extend it and maintain its consistency.
Section~\ref{implementation.background} deals with its implementation, which is efficient and constitutes the most original and challenging contribution of this document. 

\subsection{High-level description of the background}
\label{background.highlevel}

\subsubsection{Content of the background: expressions, equivalence classes, and equations}
\label{background.content}

We define the background as a ``mutable'' abstract object containing at each moment a finite set of normalized expressions grouped into equivalence classes. The background also contains equations relating expressions in it.

Expressions belonging to the same equivalence class must denote the same regular language. The converse does not hold, in general, although it can be achieved, as shown later in Section~\ref{simplifying.expressions}.
Morever, each equivalence class contains a unique best expression, called the \emph{representative} of the class. If $E$ is an expression in the background, we denote the representative of its equivalence class by $\rep(E)$. The representatives are chosen in such a way that $\size(\rep(E)) \leq \size(E)$.
The size of an expression can be defined recursively in an obvious way but several variations can be imagined. That is why no strict definition is given here; it will be for the experiments, in Section~\ref{experiments}.

\emph{Equations} are constructs of the form $E\:=\:o + \dots + x\cdot E_x + \dots$, where $o\in \{0, 1\}$, the $x$ are distinct letters, and $E$ and the $E_x$ are expressions in the background. We say that $E$ is the \emph{left part} of the equation, while $o + \dots + x\cdot E_x + \dots$ is its \emph{right part}.
Notice that, technically, a right part is \emph{not} a normalized expression but another kind of formal object built with  $0$ or $1$, letters, and normalized expressions.
For every equation in the background, it is required that

\begin{enumerate}
\item $E=\rep(E)$ and $E_x=\rep(E_x)$, for all $x$.

\item $\Lang(E)\: =\: \Lang(o)\: \cup \dots \cup \: \{x\}\,.\,\Lang(E_x)\:\cup \dots\:$.
\end{enumerate}
We call these conditions the \emph{invariant} of the background. See Section~\ref{background.example} for examples.
 We can see that, intuitively, each equation relates an expression to its direct derivatives but the expressions $E_x$ are not necessarily the exact syntactic derivatives of $E$, with respect to $x$, as defined in \cite{BLC5}. Let $E$ be an expression belonging to the background. We say that $E$ has an equation in the background if $\rep(E)$ is the left part of an equation in the background.
However, it is not necessarily the case that an expression in the background has an equation. Thus, in general, the number of equations in the background is smaller than the number of expressions. However, this is not always true because some equations may overlap:
 We say that two different equations in the background \emph{overlap} if either their left parts or their right parts are equal (i.e.\  syntactically identical). A background containing overlapping equations can be refined by merging some equivalence classes. Otherwise, we say that the background is \emph{reduced}. This desirable property should be enforced as soon as possible but it can be locally violated in some operations, as explained in the next subsection.

Sets of equations in the background can be used to represent deterministic finite automata, as suggested in \cite{Brzozowski,BLC5,Conway}. We say that a set of equations is \emph{complete} if every expression $E_x$ used in the right part of an equation is the left part of an equation in this set.
A complete set of equations determines a DFA for all expressions that are the left part of an equation of this set: Left parts are the states and right parts define transition functions to next states and indicate if the corresponding left parts are accepting states or not. If $E$ is the left part of an equation belonging to a complete set of equations, we say that the smallest complete set of equations containing this equation is \emph{the} DFA of $E$ in the background. If ${E}^\prime$ is another expression such that $E = \rep(E')$, we say that the DFA of $E$ is \emph{a} DFA for $E'$.

\subsubsection{Operations on the background}
\label{background.operations}

The background can be extended with new expressions and new equations. On the other hand, it can be made more informative by merging equivalence classes and refining equations accordingly.

New expressions can be added to the background by applying the operations \union, \concat, and $\star$ to expressions in the background. A newly added expression defines a new equivalence class containing this expression only. Similarly, new equations can be added to background by specifying a left part $E$ and a right part $o + \dots + x \cdot E_x + \dots$. They must respect the invariant specified in Section~\ref{background.content}. The first condition of the invariant can be easily enforced, but the second condition is under the responsibility of the user of the system. It can be ensured by using the algorithm to compute derivatives described in \cite{BLC5}. A reduced background may lose this property after the addition of a new equation.

Two equivalence classes can be unified to a single one by using the operation \unify. Let $E_1$ and $E_2$ be two expressions belonging to two different equivalence classes of the background. The operation $\unify(E_1, E_2)$ unites the two classes into one, as follows: Assuming that $\rep(E_1)$ is better than $\rep(E_2)$,%
\footnote{I intentionally leave undefined the predicate ``is better than'' here because many definitions make sense. See the discussion at the end of Section~\ref{related.work} and also Appendix \ref{most.frequent.simplified}. A desirable definition could be ``is easier to understand'' but it is not formalizable. }
 \  $\rep(E_1)$ becomes the new representative of the new equivalence class. Moreover, $\rep(E_2)$ is replaced by $\rep(E_1)$ in every equation using it. The old equations are removed from the background. It is the user's responsability to make sure that the precondition $\Lang(E_1) = \Lang(E_2)$ holds, since the invariant of the background can be violated otherwise. 

Unifying equivalence classes and adding equations may result in a non reduced background. Then, a reduced background can be obtained by applying the operation $\merge$ to it: If the background is not reduced, the operation selects two overlapping equations $E_i = o_i + \dots + x\cdot E_{i\: x} + \dots$ $(i = 1,2)$. If $E_1 \neq E_2$, it unifies $E_1$ and $E_2$; otherwise, it selects two expressions $E_{1\: x}$ and $E_{2\: x}$ that are not equal, and it unifies them. Afterwards, it iterates the same process until a reduced background is obtained.
The execution of the operation $\merge$ terminates since the number of equivalence classes decreases by one at each iteration. Moreover, the invariant of the background is maintained, since it is maintained after each execution of $\unify$. In addition, the final set of equations is made of all equations $\rep(E)\:=\:o + \dots + x \cdot \rep(E_x) + \dots$ such that an equation $E\:=\:o + \dots + x \cdot E_x + \dots$ was in the initial background. Consequently, each complete set of equations is replaced by a new complete set of equations, representing a possibly smaller DFA for the same expressions.
It can also be proven that the final set of equivalence classes produced by the operation $\merge$ does not depend on what particular pairs of overlapping equations are chosen at each iteration of the algorithm (see Section~\ref{confluency.reduce}). This confluence property is nice but not really essential for most applications of the work presented in this paper.

\subsubsection{Example}
\label{background.example}

I give an example to show how the background evolves when it is extended with new expressions and equations, and subsequently refined by using the operations $\unify$ and $\reduce$. 

Let us start with a minimal background, only containing the atomic expressions
$0$, $1$, $a$, \dots, $z$ spread into $27$ equivalence classes. Let us add to the background the expression $E = (1 + a)(a b^*)^*$. Now let us compute the derivatives of $E$ and of its subexpressions using the algorithm of \cite{BLC5}.%
\footnote{To focus on the important things, we do not develop the computation of the derivatives of $1$, $a$, $b$, $1 + a$, and $b^*$.} The principle of this algorithm boils down to unfolding expressions to the left until they start with a letter. Expressions starting with the same letter are then grouped to form the derivatives. For the expression $F = (a b^*)^*$, we get:
$$
\begin{array}{rclc}
F & = & 1 + a b^* (a b^*)^* &\\ 
  & = & 1 + a b^* F & \\
  & = & 1 + a \cdot G & (1)\\
  &&&\\
G & = &   b^* F & \\
  & = & F + b b^* F & \\
  & = & 1 + a \cdot G + b\cdot G & (2)\\
\end{array}
$$
The equations $(1)$ and $(2)$ may thus be added to the background. They constitute a complete set of equations, i.e.\ a DFA, and even an MDFA, for $F$.

\paragraph{Note} The above calculation of  equations for $F$ and $G$ makes use of the equality symbol ($=$) in an intuitive and ``loose'' way. It certainly does not denote syntactic equality as in Definition \ref{definition.expressions} but we can see that the left and right parts of equations denote the same regular language, which intuitively justifies the conclusion.
The algorithm in \cite{BLC5} obtains the same result, by working on normalized expressions using the operations $\union$, $\concat$,  $\iter$, and a set of derivation rules for normalized expressions.\\

Let us continue by computing derivatives and equations for the expression $E$:
$$
\begin{array}{rclc}
E & = & (1 + a) F &\\ 
  & = & F + a F & \\
  & = & 1 + a b ^* F + a F & \\
  & = & 1 + a \cdot (F + b^* F) & \\
  & = & 1 + a \cdot H & (3)\\
  &&&\\
H & = & F + b^* F  & \\
  & = & 1 + a b^* F + b b^* F & \\
  & = & 1 + a \cdot G + b\cdot G & (4)\\
\end{array}
$$
At this point, we have a DFA for $E$, made of the three equations $(3)$, $(4)$, and $(2)$. However, the equations $(2)$ and $(4)$ overlap (as defined in Section \ref{background.content}). Since $\size(G) < \size(H)$, we replace $H$ by $G$ in all equations. Therefore,  equation $4$ is removed from the background, and equation $3$ is replaced by
$$
\begin{array}{rclc}
E & = &  1 + a \cdot G & (3')\\
\end{array}
$$
Now the equations $(1)$ and $(3')$ overlap. Thus, the expressions $E$ and $F$ are unified, which means that $E$ is simplified to $F$. Equation $(3')$ is removed from the background and we are left with equations $(1)$ and $(2)$, constituting an MDFA for $E$ and $F$. We also see that the expressions $E$, $F$, $G$, and $H$ are distributed into two equivalence classes $\{E, F \}$ and $\{G, H \}$, of which $F$ and $G$ are the representatives. Notice that exactly the same classes and equations would have been obtained if the derivatives of $E$ were computed before those of $F$. But computing the derivatives of only $E$ would have failed to simplify $E$ into $F$ because $F$ is not a derivative of $E$.

Finally, let us add the expression $U = (a + b)^*$ to the background and compute its derivatives. We get a single new equation:
$$
\begin{array}{rclc}
U & = &  1 + (a + b)\, U & \\
  & = &  1 + a \cdot U + b \cdot U & (5)\\
\end{array}
$$
Although $\Lang (U) \: = \: \Lang( G)$, the expressions $U$ and $E$ are kept in different equivalence classes because equations $(2)$ and $(5)$ do not overlap. However, we can apply a minimization algorithm \cite{Hopcroft,Moore} to the set of all equations to conclude that  $\Lang (U) \: = \: \Lang( G)$, and thus to safely execute $\unify(U, E)$, resulting into the final set of equations:
$$
\begin{array}{rclc}
F & = &  1 + a \cdot U & (1')\\
U & = &  1 + a \cdot U + b \cdot U & (5)\\
\end{array}
$$
and the equivalence classes $\{E, F \}$ and $\{G, H, U \}$, of which $F$ and $U$ are the representatives.

\subsection{Implementation of the background}
\label{implementation.background}

The high-level description of the background, as a tool to represent regular languages in a unified way, looks promising but the reader may doubt that it can be implemented  efficiently enough to be usable in practice. In the rest of this section, I describe an efficient implementation.
The main issues to be solved and the principles of the techniques used to do so are discussed in Section~\ref{background.implementation.principles}.  The used data structures are presented in Section~\ref{background.implementation.datastructures}.
Then, the main algorithms are precisely described
in Section~\ref{background.implementation.algorithms}. Finally, Section~\ref{background.example.continued} shows that the example of Section~\ref{background.example} is correctly
 carried out at the implementation level.
\subsubsection{Principles}
\label{background.implementation.principles}
In summary, we need to optimally implement the operations $\unify$ and $\reduce$. Two subproblems need a special attention: Choosing pairs of overlapping equations and replacing an expression identifier by another, everywhere in the set of all equations. In our implementation,  the first subproblem is solved in time $O(1)$, and the second one in time $O(\nEq)$ where $\nEq$ is the initial number of equations containing the replaced identifier, in the set of equations.\footnote{We use the fact that the number of letters in the alphabet is bounded.} 

Equations can be represented as pairs made of an identifier $\iE$ for the left part, and an array $\tIE$ of identifiers for the right part. However, 
we want to represent them uniquely, as done for expressions, to safely reason on sets of equations. Thus, we decide to assign an identifier $\iEq$ to every equation represented in the background. Moreover, we also choose to assign an identifier $\iR$ to every right part of an equation, because this is very convenient to detect that two equations overlap. All in all, we represent an equation by a pair $\langle \iE, \iR \rangle$ of identifiers, to which we assign a ``top-level'' identifier $\iEq$. Now, it can be explained how detecting and choosing a pair of overlapping equations is done. For every identifier $\iE$ and every identifier $\iR$, we maintain a list of all identifiers $\iEq$ of equations, i.e.\ pairs $\langle \iE, \iR \rangle$, using it. When the background is reduced, all those lists contain at most one element. We also provide a list of the identifiers $\iE$ corresponding to expressions $E$ that are the left part of at least two equations, and a similar list for the identifiers $\iR$ of right parts. Using all this, we can detect in $O(1)$ time that at least two equations overlap and retrieve their identifiers  as the first two elements of a list.

Now, let us explain how to solve the problem of replacing an identifier $\iE_2$ by another identifier $\iE_1$ in all equations. We need to quickly find all identifiers $\iEq$ of an equation of which $\iE_2$
is the left part or having a right part containing $\iE_2$. For the left part, we already have a list of all equations using $\iE_2$. To find equations with right parts containing $\iE_2$, we provide, for every letter $x$ and every identifier $\iE$, a list of all identifiers $\iEq$ of equations with a right part containing an expression $E_x$ of which $\iE$ is the identifier. Using those lists, we can retrieve all equations using $\iE_2$ in a number of steps equal to the number of these equations. Each time  a pair $\langle \iE, \iR \rangle$ is retrieved, it is removed from the background and its identifier is removed from all the lists containing it. It is replaced by another pair $\langle \iE', \iR' \rangle$ using $\iE_1$ instead of $\iE_2$,%
\footnote{We set $\iE' := \iE_1$, if $\iE = \iE_2$, and  $\iE': = \iE$, otherwise. Moreover, $\iR'$ is the identifier of a right part obtained by replacing $E_2$ by $E_1$ in the right part identified by $\iR$. Thus, $\iR' = \iR$, if this right part does not use $E_2$.}
 with a new identifier, unless this other pair already exists in the background. All necessary operations can be done in $O(1)$ time if we consider that the number of letters is bounded.

Finally, we also need a way to represent equivalence classes of expressions with an efficient method to merge them. Unsurprisingly,
we use the well-known Union-Find method \cite{Knuth}, which only requires a single array of $M$ integers.

\subsubsection{Data structures}
\label{background.implementation.datastructures}

{\setlength{\leftmargini}{0em}

\begin{itemize}

\item[] $\tabIE[M]$, $\tabIR[M]$, and $\tabTIE[M][]$ are arrays representing the equations in the background. With the notation of Section~\ref{background.implementation.principles} the following equalities must hold:
\begin{center}$\tabIE[\iEq] = \iE$,\:\: $\tabIR[\iEq] = \iR$,\:\: and\:\: $\tabTIE[\iR] = \tabIE$,
\end{center}
where $\iE$ is the identifier of the left part of an equation, $\tabIE$ is an array of integers representing its right part, and $\iR$ is the identifier of $\tabIE$. The array $\tabIE$ represents the right part of the equation as follows: $\tabIE[0] = o$,%
\footnote{We identify the symbols $0$ and $1$ with their identifiers, i.e.\ the integers $0$ and $1$.}
 $\tabIE[1] = \iE_a$, $\tabIE[2] = \iE_b$, \dots, where $\iE_x$ stands for the identifier of $E_x$.

\item[] $\hashEq(n, M)$, and $\hashTIE(n, M)$ are $\MultiList(n, M)$ objects serving as hash tables for the identifiers of equations and right parts of equations. The hash codes are computed similarly as for expressions, based on the identifiers of expressions in the equations. The value of $n$ is chosen as for expressions (see Section~\ref{internal.representation}).

\item[] $\nextIR(M)$ and $\nextIEq(M)$ are $\TwoLists(M)$ objects determining the sets of identifiers of right parts and equations currently represented in the system. They are used similarly to ${\iExprList(M)}$ for expressions (again, see Section~\ref{internal.representation}).

\item[] $\listIEQIE(M, M)$ and $\listIEQIR(M, M)$ are $\MultiList(M, M)$ objects containing the lists of identifiers of equations corresponding to each identifier $\iE$ of a left part, and to each identifier $\iR$ of a right part.

\item[] $\listIEQIEx[\nl](M, M)$ is an array of $\nl$ objects $\MultiList(M, M)$ providing, for every letter $x$ and  every expression identifier $\iE$, the list of identifiers $\iEq$ of equations such that $\iE$ is the identifier of $E_x$ in the equation. The integer $\nl$ is the number of different letters allowed in expressions and it is assumed that those are the first $\nl$ letters in the alphabet. 
\item[] $\listIEtwoIEq(M)$ and $\listIRtwoIEq(M)$ are $\OneList(M)$ objects respectively providing the list of all identifiers $\iE$ of expressions that are the left part of at least two equations and the list of all identifiers $\iR$ of right parts used by at least two equations.
\item[] $\tree[M]$ is an array of integers representing the equivalence classes of the background. Identifiers corresponding to a single equivalence class are grouped to form a tree of which the representative of the class is the root. If $\iE$ is such an identifier, the condition $\tree[\iE]<0$ must hold. For another identifier, it is required that $\tree[\iE]=\iE'$ where $\iE'$ is above $\iE$ in the tree. When merging two classes, we choose the smallest one as the new representative. This is not the optimal way to maintain a minimal depth of the tree. In practice however, this method is acceptable and the depth seldom is greater than $1$ since we redirect nodes to the representative every time it is computed.
\item[] $\sizett[M]$ is an array of long integers such that $\sizett[\iE] = \size(E)$, where $\iE$ is the identifier of $E$. It is wise to use long integers because syntactic derivatives of expressions can be very large.

\end{itemize}}

\subsubsection{Algorithms}
\label{background.implementation.algorithms}

We are now in a position to provide a description of the algorithms implementing the operations $\unify$ and $\reduce$ as well as some subproblems helping their implementation. Ordinary language is prefered to pseudocode but, to show that it is sufficently precise, we provide an example of Java code in Appendix \ref{java.code}.

{\setlength{\leftmargini}{0em}

\begin{itemize}

\item[] $\findIEq(\iE, \iR)$ and $\findIR(\tabIE)$ respectively check if a pair $\langle \iE, \iR \rangle$ representing an equation or an array $\tabIE$ representing the right part of an equation, are represented in the background. Their identifier is returned when the check is positive. The integer $-1$ is returned, otherwise.

A hash code $h$ is first computed for $\langle \iE, \iR \rangle$ or $\tabIE$. Then, it is reduced to $i = \mod(h, n)$.
Afterwards, the list $\liste(i)$ of the object $\hashEq(n, M)$ or $\hashTIE(n, M)$ is traversed, depending on the case. Each iteration delivers an integer $\id$ that is used to check if $\langle \iE, \iR \rangle$ or $\tabIE$ already exists in the background by checking if $\tabIE[\id] = \iE$ and $\tabIR[\id] = \iR$, or if $\tabTIE[\id] = \tabIE$. As soon as the check succeeds, the identifier $\id$ is returned. The integer $-1$ is returned if the check never succeeds.

\item[] $\addEq(\iE, \tabIE)$ adds an equation to the background, unless it already belongs to it. The integer $\iE$ is the identifier of the left part of the equation, and $\tabIE$ is an array representing its right part, as explained in Section~\ref{background.implementation.datastructures}. It is assumed that all expressions used in the equation are representatives.

\begin{enumerate}
\item We call $\findIR(\tabIE)$ to get $\iR$. If $\iR \neq - 1$, we call $\findIEq(\iE, \iR)$ to get $\iEq$; then, if $\iEq \neq - 1$, we stop since the equation already exists in the background.

\item If $\iR = - 1$, we choose a new identifier in $\nextIR$ as the actual value of $\iR$. We set $\tabTIE[\iR] := \tabIE ;$ and we add $\iR$ to the list $\liste(i)$ of $\hashTIE$, where $i = \mod(h, n)$ and $h$ is the hash code of $\tabIE$.

\item We call $\findIEq(\iE, \iR)$ to get $\iEq$. If $\iEq = - 1$, we choose a new identifier in $\nextIEq$ as the actual value of $\iEq$, we set $\tabIE[\iEq] := \iE$ and $\tabIR[\iEq] := \iR$, and we add $\iEq$ to the list $\liste(i')$ of $\hashEq$, where $i' = \mod(h', n)$ and $h'$ is the hash code of $\langle \iE, \iR \rangle$.

\item We add $\iEq$ to the list $\liste(\iE)$ of $\listIEQIE$ and to the list $\liste(\iR)$ of $\listIEQIR$.

\item If the list $\liste(\iE)$ of $\listIEQIE$ contains two elements or more, we add $\iE$ to the list $\listIEtwoIEq$. Similarly, we add $\iR$ to  $\listIRtwoIEq$ if the list $\liste(\iR)$ of $\listIEQIR$ contains at least two elements.

\item We add $\iEq$ to the list $\liste(\tabIE[i + 1])$ of every $\MultiList$ $\listIEQIEx[i]$ where $0 \leq i < \nl$.

\end{enumerate}

%remove EQ
\item[] $\removeEq(\iEq)$ removes the equation identified by $\iEq$ from the background.

\begin{enumerate}
\item We retrieve the pair $\langle \iE, \iR \rangle$ and the array $\tabIE$, representing the equation by setting $\iE := \tabIE[\iEq]$, $\iR := \tabIR[\iEq]$, and $\tabIE := \tabTIE[\iR]$.

\item We remove $\iEq$ from $\hashEq$, $\listIEQIE$, and $\listIEQIR$ (details are left to the reader). 

\item We remove $\iEq$ from the list $\liste(\tabIE[i + 1])$ of every $\MultiList$ $\listIEQIEx[i]$ where $0 \leq i < \nl$.

\item We return $\iEq$ to the list of free identifiers in $\nextIEq$.

\item If the list $\liste(\iR)$ of $\listIEQIR$ is empty, we remove $\iR$ from $\hashTIE$ and we return $\iR$ to the list of free identifiers in $\nextIR$.

\item If the list $\liste(\iE)$ of $\listIEQIE$ no longer contains two elements or more, we remove $\iE$ from the list $\listIEtwoIEq$. Similarly, we remove $\iR$ from  $\listIRtwoIEq$ if the list $\liste(\iR)$ of $\listIEQIR$ does not contain more than one element anymore.

\end{enumerate}

\label{page.substitute}
\item[] $\substitute(\iE_1, \iE_2)$ replaces all occurrences of $\iE_2$ by $\iE_1$ in all equations of the background. To be precise, we should say that, after executing the operation, the data structures implementing the 
background correcty implement the ``abstract'' background obtained by replacing all occurrences of the expression $E_2$ by $E_1$ in the initial ``abstract'' background. It is assumed that, initially, $E_1$ and $E_2$ are representatives of their equivalence classes.

\begin{enumerate}
\item We traverse the list $\liste(\iE_2)$ of $\listIEQIE$ and, for every identifier $\iEq$ in the list, we do the following:

\begin{enumerate}
\item We set $\iR := \tabIR[\iEq]$ and $\tabIE := \tabTIE[\iR]$. 
\item We execute $\removeEq(\iEq)$.
\item We create a new array $\tabIE^\textit{\scriptsize new}$ in which we copy the corresponding elements of $\tabIE$ except the elements equal to $\iE_2$, which are replaced by $\iE_1$. 
\item We execute $\addEq(\iE_1, \tabIE^\textit{\scriptsize new})$.
\end{enumerate}

\item For all $i$ such that $0\leq i < \nl$, we traverse the list $\liste(\iE_2)$ of $\listIEQIEx[i]$ and, for every identifier $\iEq$ in the list, we do the same operations as above with the change that $\iE_1$ is replaced by $\iE$, where 
$\iE = \tabIE[\iEq]$, in the call to $\addEq$.
\end{enumerate}

\textbf{Remark} The attentive reader will have noticed that, in step 1, $E_2$ necessarily is the left part of all equations retrieved by the loop, so that  $\addEq$ must be called with $iE_1$ as first argument. To the contrary, in step 2, $E_2$ cannot be the left part of any retrieved equation because all such equations have already been retrieved and modified in step 1. Therefore, it is correct to call $\addEq$ with $\iE$. It can possibly be the case that $\iE = \iE_1$. Moreover, no equation modified in step 1  can be retrieved again in step 2.

\item[] $\reduce$ implements the abstract operation $\reduce$ specified in Section~\ref{background.operations} (see also Appendix~\ref{confluency.reduce}).

While at least one of the two lists $\listIEtwoIEq$ and $\listIRtwoIEq$ is not empty, we do the following:

\begin{enumerate}
\item We select two expression identifiers $\iE_1$ and $\iE_2$ as follows: 

\begin{enumerate}
\item If $\listIEtwoIEq$ is not empty, let $\iE$ be its first element, and let $\iEq_1$ and $\iEq_2$ be the first two elements of the list $\liste(\iE)$ of $\listIEQIE$; we compare the corresponding elements of the arrays $\tabTIE[\iEq_1]$ and $\tabTIE[\iEq_2]$ until two different elements are found; we choose them as the values of $\iE_1$ and $\iE_2$.

\item Otherwise, the list $\listIRtwoIEq$ must be non empty: Let $\iR$ be its first element, and let $\iEq_1$ and $\iEq_2$ be the first two elements of the list $\liste(\iR)$ of $\listIEQIR$; we set $\iE_1 := \tabIE[\iEq_1]$ and $\iE_2 := \tabIE[\iEq_2]$.

\end{enumerate}
\item We assign to $\iF_1$ and $\iF_2$ two different values such that $\sizett[\iF_1] \leq \sizett[\iF_2]$, chosen among $\iE_1$ and $\iE_2$.

\item We execute $\substitute(\iF_1, \iF_2)$.

\item We set $\tree[\iF_2] := \iF_1$.
\end{enumerate}

\item[] $\repit(\iE)$ implements the operation $\rep(E)$.

\begin{enumerate}
\item We set $\iB := \iE$.
\item While $\tree[\iB] \geq 0$, we set $\iB := \tree[\iB]$.
\item While $\tree[\iE] \neq \iB$, 
we set $\iF := \tree[\iE];\:\: \tree[\iE] := \iB;\:\: \iE := \iF$.
\item We return $\iB$.
\end{enumerate}

\item[] $\unify(\iE_1, \iE_2)$ implements the abstract operation $\unify(E_1, E_2)$ specified in Section~\ref{background.operations}.

\begin{enumerate}
\item We set $\iE_1 := \repit(\iE_1)$ and $\iE_2 := \repit(\iE_2)$. If $\iE_1 = \iE_2$, we stop here.
\item We assign to $\iF_1$ and $\iF_2$ two different values such that $\sizett[\iF_1] \leq \sizett[\iF_2]$, chosen among $\iE_1$ and $\iE_2$.

\item We execute $\substitute(\iF_1, \iF_2)$.

\item We set $\tree[\iF_2] := \iF_1$.
\item We execute $\reduce$.
\end{enumerate}
\end{itemize}}

\paragraph{Note}

The correctness of the algorithms presented above can be checked by symbolic execution, based on the high-level description of the background in Section~\ref{background.highlevel} and its representation by low-level data structures  made precise in \ref{background.implementation.datastructures}. It would be somewhat tedious to write down the details of this check explicitly, however, but, as an exercise, the reader could, for instance, check that no explicit update of the objects $\listIEQIE$, $\listIEQIR$, $\listIRtwoIEq$, and $\listIEtwoIEq$ is needed in the description of the algorithm $\reduce$ because all necessary changes are performed by the algorithm $\substitute$, through the use of $\addEq$ and $\removeEq$. It could also be reassuring to look at the example below.

\subsubsection{Example (continued)}
\label{background.example.continued}

I complement the example of Section \ref{background.example} by illustrating that everything
works fine at the low level. I focus on the execution of $\reduce$ and $\unify$, leaving aside the computation of derivatives, which falls within the scope of \cite{BLC5}. Executing the algorithms step by step would be tedious and anything but enlightening for most people. So, I only raise the tricky points and show that they are correctly solved.

In fact, every high-level situation presented in Section \ref{background.example} can be readily translated to a low-level description without requiring any explicit hand simulation of the algorithms of Section \ref{background.implementation.algorithms}, because the translation is completely determined by the representation rules given in Section \ref{background.implementation.principles}. Thus, I \emph{a priori} provide the low-level translations of three key high-level situations and I discuss separately the pitfalls to be avoided in moving from one to the other. 

The three situations that we consider are 1) the content of the background after the computation of the equation ($4$), 2) the content of the background after the execution of $\reduce$, and 3) the content of the background after executing $\unify(U, G)$. Their translations are depicted in Figures \ref{situation1}, \ref{situation2}, and \ref{situation3}.
In these figures,
expressions, equations, and right parts of equations are replaced, i.e.\ represented, by identifiers that we consistently denote by $iE$, $iF$, \dots, for expressions, $iEq1$, $iEq2$, \dots, for equations, and $iR1$, $iR2$, \dots, for right parts. Identifiers are chosen in the free lists of the objects $\iExprList$, $\nextIEq$, $\nextIR$, respectively. The figures only depict the non empty lists. Moreover, the hash tables 
$\hashEq$ and $\hashTIE$ respectively contain all identifiers of equations and right parts that are currently in use.

\begin{figure}[t]
\caption{Low-level description of the first situation}
\label{situation1}
$$
\begin{array}{rclcrclcrcl}
\tabIE[iEq1] & = & iF &&
\tabIR[iEq1] & = & iR1 &&
\tabTIE[iR1] & = & \{1,\: iG,\: 0\}\\
\tabIE[iEq2] & = & iG &&
\tabIR[iEq2] & = & iR2 &&
\tabTIE[iR2] & = & \{1,\: iG,\: iG\} \\
\tabIE[iEq3] & = & iE &&
\tabIR[iEq3] & = & iR3 &&
\tabTIE[iR3] & = & \{1,\: iH,\: 0\}\\
\tabIE[iEq4] & = & iH &&
\tabIR[iEq4] & = & iR2 &&
\end{array}
$$
$$
\begin{array}{lclcl}
\liste(iE) & = & (iEq3) & \textrm{in} & \listIEQIE\\
 \liste(iF) & = & (iEq1) & \textrm{in} & \listIEQIE\\
 \liste(iG) & = & (iEq2) & \textrm{in} & \listIEQIE\\ 
 \liste(iH) & = & (iEq4) & \textrm{in} & \listIEQIE\\[0.6em] 
 
 \liste(iR1) & = & (iEq1) & \textrm{in} & \listIEQIR\\ 
 \liste(iR2) & = & (iEq4,\ iEq2) & \textrm{in} & \listIEQIR\\ 
  \liste(iR3) & = & (iEq3) & \textrm{in} & \listIEQIR\\[0.6em]
  
  \liste(iG) & = & (iEq4,\ iEq2,\  iEq1) & \textrm{in} & \listIEQIEx[0]\\
  \liste(iH) & = & (iEq3) & \textrm{in} & \listIEQIEx[0]\\[0.6em]
  
 \liste(iG) & = & (iEq4,\ iEq2) & \textrm{in} & \listIEQIEx[1]\\
\end{array}
$$

$$\listIRtwoIEq\:\:\: =\:\:\: (iR2)$$
\end{figure}

 Let us consider the passing from the first to the second situation, done by the operation $\reduce$. The fact that  equations $(2)$ and $(4)$ overlap is correctly identified by the fact that $\listIRtwoIEq\: =\: (iR2)$. Thus, $H$ must be replaced by $G$ in equation $(4)$, which must be removed from the background. But the resulting equation is identical to  equation $(2)$. There is a risk to create a copy of $(2)$ here, but the operation $\reduce$ correctly avoids doing so: A copy of the array $\{1,\: iG,\: iG\}$ is created and its hash code is computed to look in the table $\hashTIE$ whether an identifier exists for it. It must be noticed that $iR2$, still belongs to this hash table because, 
 after removing the equation~$(4)$, $\liste(iR2) = (iEq2)$ in $\listIEQIR$. Thus, the identifier $iR2$ is returned. Afterwards, an identifier for the new equation is searched in the hash table $\hashEq$, based on the pair of identifiers $\langle iG,\ iR2\rangle$. The identifier $iEq2$ is returned, showing that the ``new'' equation already exists. No duplicate is created. Now, after considering the case of  equation $(4)$, the operation $\reduce$ looks at the list $\liste(H) = (iEq3)$ in $\listIEQIEx[0]$ to detect that the right part of equation~$(3)$ contains an occurrence of $H$. A modified copy of $\tabTIE[iR3]$, equal to $\{1,\: iG,\: 0\}$, is created and its identifier $iR1$ is retrieved from $\hashTIE$. Then, it is checked that no identifier exists for $\langle iE, iR1\rangle$. So a new identifier $iEq3'$ is chosen in $\nextIEq$ and added everywhere it should be. 
At this point, $\listIRtwoIEq\: =\: (iR1)$; the identifiers $iEq1$ and $iEq3'$ are selected in the list $\liste(iR1)$ of $\listIEQIR$. So, $E$ is replaced by $F$ in the equation $(3')$. The identifier $iEq3'$ is removed from all lists to which it belongs and returned to the free list of $\nextIEq$. The  equation $(3')$, modified,  in fact is $(1)$. This is detected as before for $(2)$ and $(4)$. Consequently, the execution of $\reduce$ stops, giving the situation in Figure \ref{situation2}.

\begin{figure}[t]
\caption{Low-level description of the second situation}
\label{situation2}
$$
\begin{array}{rclcrclcrcl}
\tabIE[iEq1] & = & iF &&
\tabIR[iEq1] & = & iR1 &&
\tabTIE[iR1] & = & \{1,\: iG,\: 0\}\\
\tabIE[iEq2] & = & iG &&
\tabIR[iEq2] & = & iR2 &&
\tabTIE[iR2] & = & \{1,\: iG,\: iG\} \\
\end{array}
$$
$$
\begin{array}{lclcl}
 \liste(iF) & = & (iEq1) & \textrm{in} & \listIEQIE\\
 \liste(iG) & = & (iEq2) & \textrm{in} & \listIEQIE\\ 
 \\[0.6em] 
 
 \liste(iR1) & = & (iEq1) & \textrm{in} & \listIEQIR\\ 
 \liste(iR2) & = & (iEq2) & \textrm{in} & \listIEQIR\\ 
 \\[0.6em]
  
\liste(iG) & = & (iEq2,\  iEq1) & \textrm{in} & \listIEQIEx[0]\\[0.6em]
  
 \liste(iG) & = & (iEq2) & \textrm{in} & \listIEQIEx[1]\\
\end{array}
$$

$$\tree[iE]\:=\:iF \hspace{2em} \tree[iH]\:=\:iG$$
\end{figure}

\begin{figure}[!b]
\caption{Low-level description of the third situation}
\label{situation3}
$$
\begin{array}{rclcrclcrcl}
\tabIE[iEq1'] & = & iF &&
\tabIR[iEq1'] & = & iR1' &&
\tabTIE[iR1'] & = & \{1,\: iU,\: 0\}\\
\tabIE[iEq5] & = & iU &&
\tabIR[iEq5] & = & iR5 &&
\tabTIE[iR5] & = & \{1,\: iU,\: iU\} \\
\end{array}
$$
$$
\begin{array}{lclcl}
 \liste(iF) & = & (iEq1') & \textrm{in} & \listIEQIE\\
 \liste(iU) & = & (iEq5) & \textrm{in} & \listIEQIE\\ 
 \\[0.6em] 
 
 \liste(iR1') & = & (iEq1') & \textrm{in} & \listIEQIR\\ 
 \liste(iR5) & = & (iEq5) & \textrm{in} & \listIEQIR\\ 
 \\[0.6em]
  
\liste(iU) & = & (iEq1',\  iEq5) & \textrm{in} & \listIEQIEx[0]\\[0.6em]
  
 \liste(iU) & = & (iEq5) & \textrm{in} & \listIEQIEx[1]\\
\end{array}
$$

$$\tree[iE]\:=\:iF \hspace{2em} \tree[iH]\:=\:iG
\hspace{2em} \tree[iG]\:=\:iU
$$
\end{figure}

 Now, let us consider the passing from the second situation to the third. New identifiers $iU$, $iEq5$, and $iR5$ are chosen and a new array $\{1,\ iU,\ iU\}$ is created to represent the equation $(5)$. Then, the operation $\unify(iU, iG)$ is executed after detecting that $\Lang(U) = \Lang(G)$, by the minimization algorithm (from \cite{Moore} or \cite{Hopcroft}, not to be described here). Since $\sizerm[iU] < \sizerm[iG]$, the equations $1$ and $(2)$ are removed from the background and the equation $(1)$ is replaced by $(1')$ giving raise to the situation of Figure \ref{situation3}.

\section{Experimental evaluation}
\label{experiments}

Now, I report on experiments conducted to illustrate the power of the implementation framework described in this paper, as well as its computational efficiency. Remember that the first layer of the framework efficiently implements   normalized expressions by associating a unique identifier to each of them, which is enough to ensure that the set of their syntactic derivatives is finite (see \cite{BLC5}). This improves the methods proposed in \cite{Brzozowski,Conway} and can be seen as an optimal implementation of the method proposed in \cite{Antimirov}. The algorithm to compute the DFA corresponding to an expression is defined and proven correct in \cite{BLC5}. Therefore, it is not reexplained here. However, the implementation of normalized expressions is not explained in \cite{BLC5}, nor is the implementation of the background. Those issues are the subject of the present paper.
The second layer of the framework, i.e.\ the background, has also already been used in another paper \cite{BLC4} dedicated to the simplification of regular expressions, using different kinds of elaborate simplification rules. That paper also fails to explain how the background is implemented.  This shortcoming is once again remedied here.

To be consistent with the explanations above, we are not going to repeat the experiments previously conducted in \cite{BLC5,BLC4}. Instead, we focus on experiments that allow us to emphasize the intrinsic strength of the system, namely the fact that expressions and DFAs are intimately integrated thanks to their common unique identifier, which makes it possible to build a ``world'' (a background) where every represented regular \emph{language} has a unique identifier which is, at the same time, the identifier of an MDFA and  a small (often minimal) expression for this language.
In Section \ref{computing.dfas}, experiments are conducted to show that building DFAs from expressions very much benefits from simplifying expressions beforehand.
In Section \ref{simplifying.expressions}, we consider a straightforward  simplification algorithm exploiting the fact that, given a background containing sufficiently many randomly generated expressions, most expressions represented in the background are equivalent to another, small and often even minimal expression, also in the background. The integrated structure of the background makes it possible to detect all such equivalences.  
We show that this method also is quite efficient, on the average, provided that a few  simplification rules are applied beforehand.
In Section \ref{distribution.minimal.size}, we analyze the structure of the background created by applying the algorithm of Section \ref{simplifying.expressions} to a relatively large set of randomly chosen expressions.
This study further explains why the algorithm celebrated here is accurate and fast, on the average. 
Icing on the cake, it gives us precise information on the statistical distribution of expressions with respect to their simplified if not minimal size.

Remember that the system is programmed in Java. The tests are run on an old MacBook Pro (early 2015) with an Intel Core i5 dual-core running at 2.7 GHz with 8GB of memory. Java version 1.8.0\_131 is used. We also fix the number of identifiers available for normalized expressions to $5,000,000$ in all tests. Times are wall-clock times, as they are measured using the method \textnormal{System.nanoTime()}.

The expressions used in these experiments contain at most two different letters. This  restrictive choice makes the experimental results more striking and interesting to analyze. In particular, no bounds are imposed on the size of expressions (mostly derivatives) created by the algorithms, in spite of the fact that the problems are PSPACE-complete \cite{Meyer}. Other experimental results, for expressions using up to eight letters, are given in Appendix \ref{appendix.experimental.results}.

\subsection{Computing a DFA from an expression}
\label{computing.dfas}
\renewcommand{\arraystretch}{1.3}

We address the problem of building a DFA or an MDFA from a given expression. Remember that DFAs are represented by complete sets of equations of the form $E = o + \dots + x\cdot E_x + \dots$, where the $x$ are letters. In this case, $x \in\{a, b\}$. Roughly speaking, the $E_x$ ``are'' the derivatives of $E$ with respect to the letter $x$. Actually, two variants are considered according to the layer of the framework that is used. The fundamental algorithm only requires the first layer and it computes the $E_x$ as the \emph{syntactic derivatives} of $E$, as defined in \cite{BLC5}. These syntactic derivatives are uniquely defined normalized expressions. This algorithm is called $E$ below. A second algorithm, called $B$, needs to use the second layer of the framework, i.e.\ the background: Every equation computed by the fundamental algorithm is rewritten as $\rep(E) = o + \dots + x\cdot \rep(E_x) + \dots$ to be compatible with the existing equations of the background. The possibly modified
equation is added to the background, which is then reduced. The set of equations obtained by this second algorithm generally is smaller than the set of derivatives computed by algorithm $E$, but it
is not guaranteed to be minimal. Thus, in order to compute minimal sets of equations (MDFAs), we consider a third algorithm, called $M$, which computes the equivalence classes of the expressions, used in the equations, denoting the same regular language
(see \cite{Parsing}, pages $124-128$, for the description of such an algorithm).
Expressions in the same equivalence class are unified using the $\unify$ operation of the background, resulting in a minimal set of equations.

 To assess the efficiency of these three algorithms, we have created files of $10,000$ randomly generated expressions of equal sizes $8$, $16$, $\dots$, $8192$ ($8K$). The algorithms have been applied to every file, as follows: The expressions are read and normalized one by one; their DFA is then computed (their MDFA, for algorithm $M$). Importantly, the previously read expressions are kept in the system and their computed DFAs as well. Thus, the results of previous computations can be reused, if applicable. Statistics about these experiments are depicted in Tables \ref{computing.DFAs.sizes} and \ref{computing.DFAs.times}. 
\begin{table}[t]%\vspace{-2cm}
\caption{Computing DFAs from normalized expressions (sizes)}\label{computing.DFAs.sizes}
\begin{center}
$\setlength{\arraycolsep}{0.3em}
\renewcommand{\arraystretch}{1.2}
\begin{array}{|r||r|r|r|r||r|r||r|r|r||r|r||r|r|}\hline
%& & &  & &  &  & &  & & &  &  \\[-0.5em]
\textit{size} & \card D_E & \card D_B & \card D_O &\card C_O & \card \textit{Eq}_B & \card \textit{Eq}_M &
\textit{Max}_E & \textit{Max}_B & \textit{Max}_M &
 \textit{\footnotesize ma}_E  & \textit{\footnotesize ma}_B &  \textit{\footnotesize vu}_E &  \textit{\footnotesize vu}_B\\[0.5em]\hline\hline
8 & 1 & 1 & 0 & 2 & 3 & 3 & 7 & 7 & 7 & 63\% & 100\% & 83\% & 86\%\\
16 & 5 & 4 & 2 & 2 & 4 & 4 & 18 & 17 & 17 & 40\% & 94\% & 9\% & 17\%\\
32 & 8 & 8 & 6 & 1 & 6 & 6 & 34 & 32 & 30 & 15\% & 71\% & 4 & 9\\
64 & 15 & 15 & 14 & 1 & 10 & 8 & 74 & 66 & 66 & 2\% & 34\% & 0 & 0\\
128 & 32 & 32 & 31 & 1 & 20 & 9 & 273 & 195 & 188 & 5 & 6\% & 0 & 0\\
256 & 85 & 85 & 84 & 1 & 50 & 11 & 1,875 & 1,458 & 1,336 & 0 & 13 & 0 & 0\\
512 & 309 & 307 & 307 & 1 & 182 & 11 & 115e^2 & 6,527 & 2,939 & 0 & 0 & 0 & 0\\
1K & 1,773 & 1,767 & 1,767 & 1 & 1,055 & 11 & 910e^2 & 645e^2 & 2,608 & 0 & 0 & 0 & 0\\\hline
2K & 206e^2 & 206e^2 & 206e^2 & 1 & 126e^2 & 16 & 228e^4 & 147e^4 & 7,207 & 0 & 0 & 0 & 0\\
4K & 480e^3 & 572e^3 & 572e^3 & 0 & 297e^3 & 8 & 432e^4 & 252e^4 & 86 & 0 & 0 & 0 & 0\\
\hline
\end{array}$
\end{center}
\end{table}
Table \ref{computing.DFAs.sizes} provides information about the size of the computed DFAs and the reusability of previous results. The column \textit{size} 
gives the size of the input expressions before normalization. 
The columns $\card D_E$ and $\card D_B$ give the average  number of iterations executed by  algorithms $E$ and $B$ (each iteration computes a new derivative). The column $\card D_O$ gives the same information for a variant of algorithm $B$ that detects the fact that a computed derivative already has an equation in the background, i.e.\  the fact that its representative is the left part of an equation. The column $\card C_O$ is the average number of times that this check saves an iteration in the algorithm.
The columns $\card \textit{Eq}_B$ and  $\card \textit{Eq}_M$ provide the average number of equations computed by algorithms $B$ and $M$. 
(For algorithm $E$, this number is equal to  
$\card D_E$.)
The values in
$\textit{Max}_E$,  $\textit{Max}_B$, and $\textit{Max}_M$ are the maxima of these numbers. The columns
 $\textit{ ma}_E$ and $\textit{ ma}_B$ are the number of times that algorithms $E$ and $B$ produce a minimal set of equations (or derivatives, for $E$), on the average, in percent or in actual value, when appropriate. Finally, 
$\textit{ vu}_E$ and  $\textit{ vu}_B$ are the number of times that a normalized expression has been seen previously. But the corresponding input expressions may be different. In such a case, no computation is performed for the newly read expression, except, of course, the normalization. All values are rounded, and a value such as $432e^4$  is an abbreviation for $432\times{10}^4$.

 Some comments are worth making. We see that the number of iterations made by the algorithms grows very fast (exponentially), and the number of equations computed by  algorithm $B$ as well, but this is not the case after minimization by  algorithm $M$, for which this number stabilizes at $11$.%
\footnote{Since the algorithms need too much space and take too much time when \textit{size} goes beyond $1K$, we have limited the number of input expressions to $1000$ for $\textit{size}=2K$, and to $100$ for $\textit{size}=4K$. No attempt has been made to apply the algorithms to expressions such that $\textit{size}=8K$.}
This suggests that many input expressions are equivalent, and can possibly be simplified to much smaller expressions, as will be shown in Section \ref{simplifying.expressions} . We can  also note that  algorithm $B$ often provides
minimal sets of equations when the size of expressions is not too large.

\begin{table}[t]
\caption{Computing DFAs from normalized expressions (times)}\label{computing.DFAs.times}
\begin{center}\small
$
\begin{array}{|r||r|r|r|r||r|r|r|r||r|r|r||r|}\hline
\textit{size} & t^\textit{\tiny dfa}_E & t^\textit{\tiny dfa}_B & t^\textit{\tiny dfa}_O & t^\textit{\tiny dfa}_M & t^\textit{\tiny der}_E & t^\textit{\tiny der}_B & t^\textit{\tiny der}_O & t^\textit{\tiny der}_M & t^\textit{\tiny red}_B & t^\textit{\tiny red}_O & t^\textit{\tiny red}_M &  t^\textit{\tiny min}_M \\[0.5em]\hline\hline
8 & 11\,\mu & 5.0\,\mu & 3.7\,\mu & 22\,\mu & 1.2\,\mu & 1.4\,\mu & 858\,\nu & 882\,\nu & 1.4\,\mu & 844\,\nu & 1.1\,\mu & 1.2\,\mu\\
16 & 19\,\mu & 16\,\mu & 13\,\mu & 13\,\mu & 9.0\,\mu & 5.5\,\mu & 4.9\,\mu & 5.3\,\mu & 3.8\,\mu & 2.9\,\mu & 3.4\,\mu & 4.3\,\mu\\
32 & 28\,\mu & 51\,\mu & 31\,\mu & 46\,\mu & 17\,\mu & 17\,\mu & 15\,\mu & 14\,\mu & 7.6\,\mu & 6.9\,\mu & 8.3\,\mu & 8.3\,\mu\\
64 & 66\,\mu & 111\,\mu & 87\,\mu & 102\,\mu & 54\,\mu & 64\,\mu & 40\,\mu & 56\,\mu & 14\,\mu & 14\,\mu & 21\,\mu & 16\,\mu\\
128 & 174\,\mu & 193\,\mu & 190\,\mu & 158\,\mu & 160\,\mu & 146\,\mu & 142\,\mu & 116\,\mu & 22\,\mu & 23\,\mu & 35\,\mu & 26\,\mu\\
256 & 597\,\mu & 640\,\mu & 622\,\mu & 576\,\mu & 580\,\mu & 505\,\mu & 505\,\mu & 488\,\mu & 49\,\mu & 52\,\mu & 93\,\mu & 65\,\mu\\
512 & 3.4\,m & 3.8\,m & 3.6\,m & 3.6\,m & 3.3\,m & 3.4\,m & 3.2\,m & 3.2\,m & 192\,\mu & 193\,\mu & 415\,\mu & 296\,\mu\\
1K & 38\,m & 40\,m & 39\,m & 39\,m & 37\,m & 37\,m & 36\,m & 36\,m & 1.6\,m & 1.5\,m & 4.1\,m & 2.7\,m\\\hline
2K & 1.0\,s & 1.3\,s & 1.2\,s & 1.3\,s & 998\,m & 1.2\,s & 1.1\,s & 1.2\,s & 54\,m & 53\,m & 141\,m & 90\,m\\
4K & 91\,s & 126\,s & 116\,s & 109\,s & 91\,s & 118\,s & 109\,s & 104\,s & 3.9\,s & 3.6\,s & 6.9\,s & 4.2\,s\\
\hline
\end{array}$
\end{center}
\end{table}

 Table \ref{computing.DFAs.times} provides information about the running time of the same algorithms as above. Depending on the case, times are specified in nanoseconds ($\nu$), microseconds ($\mu$), milliseconds ($m$), or seconds ($s$). The columns $t^\textit{\tiny dfa}_E$,  $t^\textit{\tiny dfa}_B$, $t^\textit{\tiny dfa}_O $, and 
$t^\textit{\tiny dfa}_M$ provide the average running times of algorithms $E$, $B$, $O$, and $M$ (without the time spent in minimization of the DFA).
The columns $t^\textit{\tiny der}_E $, $t^\textit{\tiny der}_B $, $t^\textit{\tiny der}_O $, and $t^\textit{\tiny der}_M$ are the times spent in computing syntactic derivatives.
The columns $t^\textit{\tiny red}_B $, $t^\textit{\tiny red}_O $, and $t^\textit{\tiny red}_M $ 
are the times spent in  algorithms  $\reduce$ and $\unify$.
Finally, the column $t^\textit{\tiny min}_M$ gives the time spent in  algorithm $M$ to minimize the DFA computed by algorithm $O$. This includes the time spent to compute the sets of equivalent expressions and the time spent to unify them in the background. It must be noted that the algorithm $\reduce$ is used by  algorithms $B$ and $M$ to reduce the background after adding an equation to it.
In addition, algorithm $M$ uses $\unify$ to put equivalent expressions into the same equivalence class and compute a best representative for them.

 We can see that all times grow exponentially but this is only due to the computation of the syntactic derivatives: Their number grows exponentially but also their size. The times spent in reducing or minimizing sets of equations remain ten times smaller.

Although the exponential behavior observed in Tables \ref {computing.DFAs.sizes} and \ref{computing.DFAs.times} is unavoidable in the worst case, we show in the next experiment that this is not true, on the average, if the randomly chosen expressions are  properly simplified beforehand. We first apply a straightforward simplification algorithm, called \emph{lifting}, inspired by the lifting method from \cite{Kahrs} and similar to the still simpler algorithm from \cite{Rotondo}. This algorithm is linear in the size of the expression. It is best to use it before normalization of the expressions, i.e.\ outside the background, to save identifiers needed for normalized expressions. As will be shown below, the simplification operated by lifting is sufficient to greatly improve the efficiency of the algorithms analyzed previously. More information about the lifting simplification is given in Appendix \ref{information.lifting}.

 Now, we analyze the performance of the previous algorithms modified as follows: After reading an expression, and before normalizing it, we simplify it by lifting. The corresponding results are presented in Tables \ref{computing.DFAs.sizes.lifted} and \ref{computing.DFAs.times.lifted}. It must be stressed that, this time, the algorithms can be, and are, applied to sets of $10,000$ expressions even for sizes equal to $2K$, $4K$, and $8K$.

\begin{table}[t]
\caption{Computing DFAs from lifted expressions (sizes)}\label{computing.DFAs.sizes.lifted}
\begin{center}\small
$\setlength{\arraycolsep}{0.3em}
\renewcommand{\arraystretch}{1.2}
\begin{array}{|r||r|r|r|r||r|r||r|r|r||r|r||r|r|}\hline
%& & &  & &  &  & &  & & &  &  \\[-0.5em]
\textit{size} & \card D_E & \card D_B & \card D_O &\card C_O & \card \textit{Eq}_B & \card \textit{Eq}_M &
\textit{Max}_E & \textit{Max}_B & \textit{Max}_M &
 \textit{\footnotesize ma}_E  & \textit{\footnotesize ma}_B &  \textit{\footnotesize vu}_E &  \textit{\footnotesize vu}_B\\[0.5em]\hline\hline
8 & 0 & 0 & 0 & 2 & 3 & 3 & 7 & 7 & 7 & 92\% & 100\% & 90\% & 90\%\\
16 & 4 & 3 & 1 & 2 & 4 & 4 & 18 & 17 & 17 & 73\% & 98\% & 37\% & 40\%\\
32 & 6 & 6 & 4 & 2 & 6 & 6 & 34 & 32 & 30 & 48\% & 87\% & 25\% & 25\%\\
64 & 10 & 10 & 9 & 2 & 9 & 8 & 74 & 66 & 66 & 36\% & 68\% & 30\% & 30\%\\
128 & 18 & 18 & 17 & 1 & 13 & 9 & 270 & 195 & 188 & 36\% & 54\% & 32\% & 32\%\\
256 & 32 & 32 & 31 & 1 & 23 & 11 & 1,874 & 1,391 & 1,336 & 36\% & 50\% & 33\% & 33\%\\
512 & 57 & 57 & 56 & 1 & 38 & 11 & 114e^2 & 6,527 & 2,939 & 37\% & 50\% & 34\% & 34\%\\
1K & 111 & 111 & 109 & 1 & 72 & 11 & 250e^2 & 180e^2 & 2,608 & 37\% & 50\% & 34\% & 34\%\\\hline
2K & 197 & 197 & 196 & 1 & 131 & 11 & 362e^3 & 268e^3 & 7,207 & 37\% & 50\% & 34\% & 34\%\\
4K & 206 & 205 & 205 & 1 & 133 & 11 & 123e^3 & 853e^2 & 1,767 & 38\% & 51\% & 35\% & 35\%\\
8K & 238 & 238 & 237 & 1 & 159 & 10 & 303e^3 & 212e^3 & 2,126 & 38\% & 51\% & 35\% & 35\%\\
\hline
\end{array}$
\end{center}
\end{table}

 Looking at Table \ref{computing.DFAs.sizes.lifted}, we observe that the values in the columns $\card D_E$, $\card D_B$, and $\card D_O$ grow less than linearly. This is because the average size of lifted expressions tends to a limit (see \cite{Rotondo}). This is also the case for DFAs. But the later phenomenon is independent of lifting: Lifting just makes it cheaper to actually compute the DFAs. The last four columns show that, after lifting, about $50\%$ of the DFAs computed by algorithm $B$ are minimal ($38\%$ for $E$), and that much more lifted expressions are recognized as ``d\'eja vu''. 

\begin{table}[t]
\caption{Computing DFAs from lifted expressions (times)}\label{computing.DFAs.times.lifted}
\begin{center}\small
$
\begin{array}{|r||r|r|r|r||r|r|r|r||r|r|r||r|}\hline
\textit{size} & t^\textit{\tiny dfa}_E & t^\textit{\tiny dfa}_B & t^\textit{\tiny dfa}_O & t^\textit{\tiny dfa}_M & t^\textit{\tiny der}_E & t^\textit{\tiny der}_B & t^\textit{\tiny der}_O & t^\textit{\tiny der}_M & t^\textit{\tiny red}_B & t^\textit{\tiny red}_O & t^\textit{\tiny red}_M &  t^\textit{\tiny min}_M \\[0.5em]\hline\hline
8 & 10\,\mu & 24\,\mu & 4.3\,\mu & 22\,\mu & 784\,\nu & 839\,\nu & 1.1\,\mu & 974\,\nu & 872\,\nu & 734\,\nu & 724\,\nu & 1.7\,\mu\\
16 & 9.2\,\mu & 13\,\mu & 13\,\mu & 25\,\mu & 4.4\,\mu & 4.5\,\mu & 3.9\,\mu & 4.2\,\mu & 2.6\,\mu & 2.9\,\mu & 3.2\,\mu & 5.5\,\mu\\
32 & 20\,\mu & 42\,\mu & 30\,\mu & 31\,\mu & 9.7\,\mu & 8.5\,\mu & 13\,\mu & 8.7\,\mu & 5.4\,\mu & 6.1\,\mu & 5.6\,\mu & 7.9\,\mu\\
64 & 36\,\mu & 61\,\mu & 61\,\mu & 60\,\mu & 25\,\mu & 38\,\mu & 21\,\mu & 19\,\mu & 9.0\,\mu & 9.5\,\mu & 12\,\mu & 13\,\mu\\
128 & 69\,\mu & 95\,\mu & 123\,\mu & 88\,\mu & 56\,\mu & 47\,\mu & 69\,\mu & 50\,\mu & 14\,\mu & 16\,\mu & 25\,\mu & 23\,\mu\\
256 & 132\,\mu & 182\,\mu & 187\,\mu & 188\,\mu & 124\,\mu & 131\,\mu & 121\,\mu & 125\,\mu & 22\,\mu & 22\,\mu & 39\,\mu & 33\,\mu\\
512 & 292\,\mu & 380\,\mu & 359\,\mu & 344\,\mu & 276\,\mu & 284\,\mu & 274\,\mu & 252\,\mu & 37\,\mu & 38\,\mu & 81\,\mu & 68\,\mu\\
1K & 849\,\mu & 969\,\mu & 967\,\mu & 914\,\mu & 828\,\mu & 809\,\mu & 812\,\mu & 745\,\mu & 84\,\mu & 72\,\mu & 186\,\mu & 152\,\mu\\\hline
2K & 2.1\,m & 2.5\,m & 2.5\,m & 2.4\,m & 2.1\,m & 2.1\,m & 2.2\,m & 2.1\,m & 146\,\mu & 146\,\mu & 513\,\mu & 401\,\mu\\
4K & 2.5\,m & 3.0\,m & 2.8\,m & 2.8\,m & 2.5\,m & 2.6\,m & 2.5\,m & 2.5\,m & 173\,\mu & 151\,\mu & 461\,\mu & 349\,\mu\\
8K & 3.6\,m & 4.1\,m & 3.8\,m & 4.0\,m & 3.6\,m & 3.6\,m & 3.4\,m & 3.5\,m & 229\,\mu & 176\,\mu & 688\,\mu & 527\,\mu\\
\hline
\end{array}$
\end{center}
\end{table}

 Table \ref{computing.DFAs.times.lifted} confirms the conclusions drawn from Table \ref{computing.DFAs.sizes.lifted}: the average time needed to compute DFAs and MDFAs from expressions of virtually any size is quite small. As an example, for $\textit{size} = 4K$, all times are reduced by more than $4$ orders of magnitude. For huge expressions, most of the time is spent in lifting, since the algorithm is linear while the average size of lifted expressions remains approximately the same ($70$). However, we must not forget that those are average values for perfectly randomly chosen expressions. Worst cases still remain and become harder and harder when the size grows, as suggested by the columns $\textit{Max}_E$, and $\textit{Max}_B$ of Table \ref{computing.DFAs.sizes.lifted}. Note that the  column $\textit{Max}_M$ seems to suggest that even the worst case for MDFAs size tends towards a limit.

\subsection{The simplifying power of the background}
\label{simplifying.expressions}

In this section, the power of the framework for simplifying expressions is both logically explained and experimentally illustrated. Importantly, the studied algorithms do not implement explicit simplification rules as in \cite{BLC4,Kahrs,Stoughton}. Such rules have actually been implemented in the system \cite{BLC4} but the goal here is to show that,  in a way, simplification of expressions is ``provided for free'' by the background being properly used. The key idea is to maintain a state of the background where all expressions denoting the same regular language have the same representative. Such a representative can be viewed as \emph{the} simplification of all expressions in its equivalence class. The power of the method is due to the observed fact that, inside a large set of randomly chosen expressions, including all their subexpressions, many expressions have an equivalent short one, which often is even minimal. This is further discussed in Section \ref{distribution.minimal.size}.

\subsubsection{The fundamental algorithm \PU}
\label{fundamental.algorithm}
 Let us assume a large set of randomly chosen expressions. In our experiments, these are sets of $10,000$ expressions of fixed sizes: $8$, $16$, \dots, $8K$ (the same as in Section \ref{computing.dfas}). 
We add the expressions to the background one by one, while normalizing them. In fact, it is much more efficient to first simplify them by lifting, as explained in Section \ref{computing.dfas}.
Then we apply the following steps to every subexpression of the newly added expression:
\begin{enumerate}
\item We simplify it by \emph{propagation}, that is: We (possibly) create a new equivalent expression by replacing its direct subexpressions by their representatives, using operators \union, \concat, and $\star$.
\item We build the MDFA of the simplified subexpression, as explained in Section \ref{computing.dfas}.
\item We unify the newly created MDFA with the global MDFA consisting of the union of the MDFAs of all expressions to which the three steps have already been applied. This can be done reasonably efficiently by associating a hashcode to the MDFA of every processed subexpression. This hashcode only depends on the ``shape'' of the MDFA, not on the actual values of its derivatives, except for those equal to $0$, $1$, or  a letter, since their identifiers are fixed in the system. For every hashcode $h$, a list of identifiers of the corresponding expressions is maintained. So we go through the list corresponding to the hashcode of the new subexpression and we check the equivalence of the new subexpression with the expressions in the list. If an equivalent expression is found, the two expressions are unified and a single one is kept in the list. Otherwise, the new subexpression is added to the list. The same treatment is applied to all expressions that are left parts of the equations constituting the MDFA, unless it is detected that these expressions already belong to the global MDFA.
\end{enumerate}

 Additional explanations are needed to understand how the method is implemented and why it is efficient.
Each subexpression encountered in the process need not  be considered more than once. This can speed up the process a lot. Moreover, we should not consider an expression before looking at all its subexpressions, because a lot of time can be saved in computing the MDFA of the expression if it is presimplified by propagation. To ensure both behaviors, we use a handmade data structure that maintains a dependency graph  between every encountered expression and its direct superexpressions. All expressions have a counter of their not yet processed subexpressions. When this counter is set to $0$, the expression is put into a list of all expressions ``ready to be processed''. The next expression to be processed is arbitrarily chosen in this list. The implementation of this data structure is similar to the data structure presented in \cite{Knuth}, pages $258-268$, for implementing a topological sort.

Table \ref{simplification.expressions.1} gives experimental results about the method just explained, applied to the same test data as in the previous subsection. The algorithm is called \textit{PU}, where $P$ recalls that propagation is used while $U$ recalls that unification of MDFAs is applied. Since normalization ($N$) and lifting ($L$) already can provide a substantial simplification, we also consider them as a basis for comparison. The three columns \textit{ssize (arith)} provide the average size of the simplified expressions, i.e.\ their arithmetic mean. We see that normalization reduces the size by $28\%$ on the average, independently of the input size. The behaviour of lifting is very different since the average size converges to a limit below $70$. Similarly, the average size for algorithm \textit{PU} tends to a value below $25.5$, which is much better. The two colums \textit{ssize (geo)} provide the geometric means, which are much smaller, indicating that many simplified expressions have a size well below the arithmetic mean.%
\footnote{This also allows the reader to compare the results with the figures reported in \cite{Kahrs}. For input expressions where $\textit{size} = 8$, the geometric mean is equal to $0$ because an expression is simplified to $1$, the size of which is equal to $0$.}%
The next three columns $t$ report on the average execution times. For normalization, they grow linearly, although  the theoretical time complexity of normalization is $O(s \log s)$, where $s$ is the input size. For lifting, the times are similar for small expressions but they become almost three times faster for the largest expressions. They grow less than linearly, while the theoretical complexity of lifting is linear.  For algorithm \PU, the times grow linearly at the beginning but seem to converge to approximately $4.5\,m$, for large expressions. This can be explained by the fact that \PU\/ is applied to lifted expressions, the average size of which tends to a constant ($70$). Note that we do not include the lifting time in the time for \PU. We can observe that the times for \PU\/ are not much bigger than the times for computing DFAs, reported in Table \ref{computing.DFAs.times.lifted}, in spite of the fact that algorithm \textit{PU} does much more work, and, in particular, computes MDFAs for all subexpressions of the input expressions. This can be explained, on the one hand, by the fact that a lot of the work made for previous expressions can be reused for new ones, and, on the other hand, by the fact that propagation often reduces very much the size of expressions before their DFA is computed. This claim is supported by the figures in the column $t^\textit{\tiny der}$. If we compare them with the figures in the corresponding columns in Table \ref{computing.DFAs.times.lifted}, we see that they are more than three times lower. Moreover, they constitute only a small part of the total time, contrary to what we see in Figure \ref{computing.DFAs.times.lifted}.
The last three columns of Table \ref{simplification.expressions.1} indicate the number of input expressions that are minimized by the algorithms, either in percent or in actual value for small numbers. 
The method to determine these numbers is described later on, in Section \ref{two.improvements}. We see that normalization can minimize only small expressions while lifting and \textit{PU} minimize up to $34\%$ and $60\%$ of the expressions, respectively. Once again, these proportions tend to a limit when the size of the input expressions grows.

\begin{table}[t]
\caption{Simplification of expressions (1)}\label{simplification.expressions.1}
\begin{center}
$\begin{array}{|r||r|r|r||r|r||r|r|r||r||r|r|r|}\hline
& \multicolumn{3}{c||}{\textit{ssize (arith)}}&
 \multicolumn{2}{c||}{\textit{ssize (geo)}}&
 \multicolumn{3}{c||}{t}& t^\textit{\tiny der}
 & \multicolumn{3}{c|}{\card\textit{min}}\\\hline
\textit{size} & N & L & \PU  
 & L & \PU  & {N} & {L}& \PU &
 \PU &
 N& L& \PU \\[0.5em]\hline\hline
8 & 71\% & 4.78 & 4.49 & 0.00 & 0.00 & 12\,\mu & 12\,\mu &7.9\,\mu & 727\,\nu & 43\% & 84\% & 99\% \\
16 & 70\% & 8.86 & 7.66 & 7.96 & 6.86 & 18\,\mu & 35\,\mu &34\,\mu & 4.3\,\mu & 14\% & 51\% & 88\% \\
32 & 71\% & 15.49 & 12.44 & 12.97 & 10.28 & 26\,\mu & 29\,\mu &98\,\mu & 11\,\mu & 9 & 27\% & 55\% \\
64 & 71\% & 24.62 & 18.00 & 17.43 & 12.64 & 65\,\mu & 48\,\mu &244\,\mu & 32\,\mu & 0 & 29\% & 50\% \\
128 & 71\% & 35.15 & 21.78 & 20.42 & 12.93 & 58\,\mu & 53\,\mu &589\,\mu & 62\,\mu & 0 & 31\% & 55\% \\
256 & 72\% & 46.66 & 24.12 & 22.97 & 12.99 & 125\,\mu & 99\,\mu &1.2\,m & 160\,\mu & 0 & 32\% & 57\% \\
512 & 72\% & 56.47 & 25.18 & 24.40 & 12.92 & 159\,\mu & 125\,\mu &2.0\,m & 315\,\mu & 0 & 33\% & 59\% \\
1K & 72\% & 63.66 & 25.53 & 25.25 & 12.84 & 287\,\mu & 154\,\mu &3.5\,m & 637\,\mu & 0 & 33\% & 59\% \\
2K & 72\% & 67.10 & 25.11 & 25.12 & 12.54 & 523\,\mu & 251\,\mu &4.1\,m & 925\,\mu & 0 & 33\% & 60\% \\
4K & 72\% & 69.62 & 25.31 & 25.23 & 12.47 & 1.0\,m & 387\,\mu &4.0\,m & 764\,\mu & 0 & 34\% & 60\% \\
8K & 72\% & 69.23 & 25.23 & 24.75 & 12.48 & 2.0\,m & 659\,\mu &4.4\,m & 1.1\,m & 0 & 34\% & 60\% \\
\hline
\end{array}$
\end{center}
\end{table}

\subsubsection{Three weaker variants of the fundamental algorithm}
\label{tree.variants}

 To better understand why algorithm \textit{PU} is effective, it is useful to compare it to three other algorithms, which implement only some of its functionalities. The algorithm \textit{PU} has been previously described by three steps 1, 2, and 3, executed for all subexpressions. Here, we consider algorithms $M$, which only applies the step 2, $P$, which applies steps 1 and 2, and $U$, which applies steps 2 and 3. Thus, all algorithms compute MDFAs for all subexpressions, but $P$ also concentrates on propagation while $U$ concentrates on unifying MDFAs. Experimental results are given in Table \ref{simplification.expressions.2}. As far as size reduction is concerned, we see that all three algorithms are much less effective than \textit{PU}. Moreover, they also are much less efficient, spending a lot of time in computing derivatives. However, algorithm $P$ is three times more efficient than the two others, for large expressions, confirming the role of propagation to speed up the computation of DFAs. Conversely, algorithm $U$ is much better than the two others to minimize expressions, confirming that the test data contain many minimal subexpressions that are equivalent to other large input expressions.

\begin{table}[t]
\caption{Simplification of expressions (2)}\label{simplification.expressions.2}
\setlength{\arraycolsep}{0.31em}
\begin{center}%\small
$\begin{array}{|r||r|r|r||r|r|r||r|r|r||r|r|r||r|r|r|}\hline

& \multicolumn{3}{c||}{\textit{ssize (arith)}}&
 \multicolumn{3}{c||}{\textit{ssize (geo)}}&
 \multicolumn{3}{c||}{t}& 
 \multicolumn{3}{c||}{t^\textit{\tiny der}}& 
 \multicolumn{3}{c|}{\card\textit{min}}\\\hline

\textit{size} & M & P & U  
 & M & P   & U  & {M} & {P}& U &
 M &  P &  U & 
 M&P&U \\[0.5em]\hline\hline
8 & 4.6 & 4.6 & 4.5 & 0.0 & 0.0 & 0.0 & 4.8\,\mu & 4.4\,\mu &6.2\,\mu & 750\,\nu & 556\,\nu & 704\,\nu & 97\% & 97\% & 99\% \\
16 & 8.2 & 8.2 & 7.7 & 7.4 & 7.3 & 6.9 & 14\,\mu & 25\,\mu &31\,\mu & 3.5\,\mu & 12\,\mu & 3.4\,\mu & 71\% & 73\% & 85\% \\
32 & 14.6 & 14.1 & 13.2 & 12.1 & 11.8 & 10.8 & 52\,\mu & 41\,\mu &76\,\mu & 22\,\mu & 9.1\,\mu & 12\,\mu & 33\% & 35\% & 49\% \\
64 & 23.3 & 22.0 & 20.6 & 16.4 & 15.6 & 13.9 & 89\,\mu & 103\,\mu &206\,\mu & 33\,\mu & 35\,\mu & 28\,\mu & 32\% & 33\% & 47\% \\
128 & 33.0 & 30.2 & 27.6 & 18.9 & 17.7 & 14.9 & 206\,\mu & 212\,\mu &563\,\mu & 99\,\mu & 102\,\mu & 84\,\mu & 36\% & 37\% & 52\% \\
256 & 43.6 & 38.1 & 34.9 & 21.0 & 19.2 & 15.8 & 500\,\mu & 462\,\mu &1.3\,m & 275\,\mu & 221\,\mu & 244\,\mu & 37\% & 38\% & 54\% \\
512 & 52.8 & 45.2 & 41.6 & 22.3 & 20.2 & 16.3 & 1.2\,m & 952\,\mu &2.8\,m & 832\,\mu & 591\,\mu & 882\,\mu & 38\% & 39\% & 55\% \\
1K & 59.3 & 49.3 & 45.4 & 23.0 & 20.7 & 16.5 & 3.7\,m & 2.3\,m &6.2\,m & 2.8\,m & 1.7\,m & 2.5\,m & 38\% & 39\% & 56\% \\
2K & 62.4 & 51.3 & 47.4 & 22.9 & 20.5 & 16.2 & 8.6\,m & 4.4\,m &12\,m & 6.5\,m & 3.2\,m & 6.5\,m & 38\% & 39\% & 57\% \\
4K & 64.8 & 52.4 & 48.9 & 22.9 & 20.4 & 16.1 & 11\,m & 6.0\,m &15\,m & 8.6\,m & 4.6\,m & 8.9\,m & 39\% & 39\% & 56\% \\
8K & 64.9 & 52.7 & 50.0 & 22.8 & 20.3 & 16.0 & 29\,m & 8.7\,m &35\,m & 22\,m & 6.8\,m & 24\,m & 38\% & 39\% & 57\% \\
\hline
\end{array}$
\end{center}
\end{table}

\subsubsection{Two improvements: computing minimal expressions and iterating}
\label{two.improvements}

 We now go a step further, following two possible ways of improving the simplifying power of algorithm \textit{PU}. First, we consider the problem of minimizing expressions, i.e.\ finding a shortest equivalent expression for a given expression. Let $L$ be a regular \emph{language}. We define the size of $L$, denoted by $\size(L)$, as the smallest integer $s$ such that there exists an expression $E$ such that $\size(E)=s$ and $\Lang(E)= L$. Then, we may say that an expression $E$ is \emph{minimal} if $\size(E)=\size(\Lang(E))$.
It is possible to compute a single minimal expression for all regular languages using a given number of letters $\nl$, up to a certain size. For $\nl = 2$, this is reasonably feasible, on the computer used for these experiments, up to $\size = 15$. Table \ref{number.minimal} provides the number of regular languages with a minimal expression of a given size $s$, for  $s \leq 15$ and $\nl = 2$. (Moreover, Table \ref{frequent.minimal}, in Appendix \ref{most.frequent.simplified}, displays a selection of such minimal expressions.)

\begin{table}[t]
\caption{Number of regular languages with a minimal expression of $\textit{size} \leq 15$ ($\nl = 2$)}\label{number.minimal}
\begin{center}
$\begin{array}{|r||r|r|r|r|r|r|r|r|r|r|r|r|}\hline
\textit{size} & \leq 4  & 5 & 6 & 7 & 8 & 9 & 10 & 11 & 12 & 13 & 14 & 15 \\[0.3em]\hline\hline
\card \textit{min} & 36 & 41 & 132 & 353 & 836 & 2.9K & 6.9K & 22K & 63K & 185K & 572K & 1.7M \\
\hline
\end{array}$
\end{center}
\end{table}

 To compute minimal expressions, we use two sets of lists of identifiers (i.e.\ two objects \MultiList). The first set is made of lists $\textit{list}_1(s)$, where $0 \leq s \leq 15$. They are to be filled with identifiers of minimal expressions of size $s$. The second set contains a larger set of lists $\textit{list}_2(h)$ containing minimal expressions having an MDFA with a hash code equal to $h$. 
The lists $\textit{list}_1(s)$ are filled bottom up for $s = 0, 1, \dots, 15$, by computing expressions $E_1 \union E_2$ and $E_1 \concat E_2$, where $E_i\in \textit{list}_1(s_i) (i= 1,2)$ with $s_1 + s_2 + 1 = s$, and expressions $E \star$, where $E\in \textit{list}_1(s- 1)$. To save computation time, some %``clever'' 
checks are applied to avoid creating several times the same expression. Then, the MDFA of the expression is computed together with its hash code $h$. Afterwards, the list  $\textit{list}_2(h)$ is traversed to check if an expression equivalent to the new one already exists. If it is so, the new expression is ignored, otherwise, it is added to both lists $\textit{list}_1(s)$ and $\textit{list}_2(h)$. In the end, all minimal expressions can be stored in an external file.

 It is now possible to ``improve'' algorithm \PU\/ by reading all minimal expressions from the external file and adding them to the background before doing anything else.%
\footnote{
To make the method efficient, we do not copy the minimal expressions into the file, but a binary image of their representation in the background. For the sake of brevity, we avoid giving further details.}
We use the letter $R$ to designate the improved algorithm. Experimental results about algorithm $R$ are shown in Table \ref{simplification.expressions.3}. 
We see that it is slightly more accurate than \textit{PS} (by approximately $7\%$), but less efficient by $50\%$. It also calculates minimal expressions in $5\%$ more cases. Note that, in algorithm $R$, minimality of expressions is deduced from their equivalence with an expression in the external file. For the other algorithms, we compared the size of the simplified expressions with the size of the corresponding expressions, as computed by algorithm $R$.

\begin{table}[t]
\caption{Simplification of expressions (3)}\label{simplification.expressions.3}
\setlength{\arraycolsep}{0.25em}
\begin{center}\small
$\begin{array}{|r||r|r|r||r|r|r||r|r|r||r|r|r||r|r|r|}\hline

& \multicolumn{3}{c||}{\textit{ssize (arith)}}&
 \multicolumn{3}{c||}{\textit{ssize (geo)}}&
 \multicolumn{3}{c||}{t}& 
 \multicolumn{3}{c||}{t^\textit{\tiny der}}& 
 \multicolumn{3}{c|}{\card\textit{min}}\\\hline

\textit{size} & R & \RI\: & \PUI\:  
 &R & \RI\:   & \PUI\: & {R} & \RI\: & \PUI\: &
 R & \RI\: &  \PUI\: &  
 R&\RI\:&\PUI\: \\[0.5em]\hline\hline
8 & 4.49 & 4.49 & 4.49 & 0.0 & 0.0 & 0.0 & 6.8\,\mu & 461\,\nu &702\,\nu & 806\,\nu & 0\,\nu & 0\,\nu & 100\% & 100\% & 100\% \\
16 & 7.44 & 7.44 & 7.59 & 6.7 & 6.7 & 6.8 & 59\,\mu & 917\,\nu &2.1\,\mu & 19\,\mu & 27\,\nu & 84\,\nu & 100\% & 100\% & 91\% \\
32 & 11.50 & 11.49 & 12.22 & 9.7 & 9.7 & 10.1 & 249\,\mu & 18\,\mu &20\,\mu & 56\,\mu & 236\,\nu & 533\,\nu & 73\% & 73\% & 59\% \\
64 & 16.70 & 16.68 & 17.67 & 12.0 & 11.9 & 12.5 & 623\,\mu & 5.4\,\mu &14\,\mu & 92\,\mu & 382\,\nu & 1.9\,\mu & 59\% & 59\% & 52\% \\
128 & 20.21 & 20.16 & 21.29 & 12.3 & 12.3 & 12.7 & 1.3\,m & 39\,\mu &54\,\mu & 136\,\mu & 1.3\,\mu & 14\,\mu & 62\% & 62\% & 57\% \\
256 & 22.37 & 22.23 & 23.41 & 12.4 & 12.4 & 12.8 & 2.4\,m & 117\,\mu &256\,\mu & 236\,\mu & 4.1\,\mu & 22\,\mu & 63\% & 63\% & 59\% \\
512 & 23.48 & 23.34 & 24.37 & 12.4 & 12.3 & 12.7 & 3.5\,m & 24\,\mu &242\,\mu & 402\,\mu & 5.4\,\mu & 135\,\mu & 64\% & 64\% & 60\% \\
1K & 23.91 & 23.71 & 24.72 & 12.3 & 12.3 & 12.6 & 5.3\,m & 273\,\mu &344\,\mu & 758\,\mu & 188\,\mu & 95\,\mu & 64\% & 64\% & 61\% \\
2K & 23.57 & 23.36 & 24.34 & 12.1 & 12.0 & 12.3 & 6.0\,m & 411\,\mu &669\,\mu & 1.1\,m & 186\,\mu & 211\,\mu & 65\% & 65\% & 61\% \\
4K & 23.63 & 23.38 & 24.38 & 12.0 & 11.9 & 12.2 & 6.1\,m & 285\,\mu &302\,\mu & 900\,\mu & 49\,\mu & 108\,\mu & 66\% & 66\% & 62\% \\
8K & 23.52 & 23.31 & 24.43 & 12.0 & 11.9 & 12.2 & 5.9\,m & 65\,\mu &482\,\mu & 935\,\mu & 22\,\mu & 293\,\mu & 65\% & 65\% & 62\% \\
\hline
\end{array}$
\end{center}
\end{table}

 A second improvement we can think of is to apply algorithms more than once to the input expressions, in order to take advantage of a richer state of the background. 
In fact, an expression considered at the beginning may be found equivalent to a smaller expression later, but this possible simplification may be missed if we output the simplified expression immediately. The same phenomenon can reduce propagation accuracy, and, in this case, is even more difficult to detect. To get better results, a simple solution consists of re-executing the algorithms with the state of the background obtained at the end of the first run. We designate such an algorithm by adding the letter $I$ after the name of the non iterating version. Experimental results are given for algorithms $\RI$ and $\PUI$, in Table \ref{simplification.expressions.3}. The times for these algorithms correspond to the reexecution phase only. They are much smaller than for the first execution. We observe that the simplified expressions are a bit smaller. Also, $\PUI$ detects $2\%$ more minimal expressions. Note that reexecution may be done in several ways. Here, we store the output of the initial execution in a file and we re-apply the algorithms to this file. Moreover, a few additional simplifications can be obtained by iterating the algorithms more than once, but this possibility is not evaluated here. Finally, it is possible to perform the reexecution without computing new derivatives, using only propagation. This method is faster but slightly less accurate.

\subsection{On the distribution of expressions with respect to their simplified size}
\label{distribution.minimal.size}

In these final experiments, we focus on two issues:
\begin{enumerate}
\item We further elaborate on the ``simplifying power'' of the algorithms presented in the previous subsection. We concentrate on the two most significant algorithms, \textit{PU} and \textit{R}. (Algorithms  \textit{PUI} and \textit{RI} are slightly more accurate but more difficult to analyze.)
\item  We use statistics on our experimental results to help  understand how regular expressions are distributed according to their minimum size. For expressions with a small minimum size, indisputable conclusions can be drawn. These expressions make up the majority.
\end{enumerate}
\begin{table}[t]
\caption{Distribution of expressions with respect to simplification (algorithm  $PU$)}\label{distribution.simpl.pu}
\begin{center}
\renewcommand{\arraystretch}{0.9}
$\begin{array}{|r||c|r|r|r|r|r|r|r|r|r|}\hline
\textit{size} & \textit{ssize} & _{0:2}& _{3:4}& _{5:8}& _{9:16}& _{17:32}& _{33:64}& _{65:128}& _{129:256}& _{257:512}\\[0.5em]\hline\hline
8 & \card \textit{expr} & 2,391 & 3,414 & 4,195 & 0 & 0 & 0 & 0 & 0 & 0 \\\hline  
 & \card \textit{diff} & 7 & 28 & 565 & 0 & 0 & 0 & 0 & 0 & 0 \\\hline\hline  
16 & \card \textit{expr} & 452 & 1,964 & 3,781 & 3,803 & 0 & 0 & 0 & 0 & 0 \\\hline  
 & \card \textit{diff} & 5 & 21 & 664 & 3,287 & 0 & 0 & 0 & 0 & 0 \\\hline\hline  
32 & \card \textit{expr} & 22 & 1,913 & 2,079 & 2,771 & 3,215 & 0 & 0 & 0 & 0 \\\hline  
 & \card \textit{diff} & 2 & 10 & 240 & 2,412 & 3,211 & 0 & 0 & 0 & 0 \\\hline\hline  
64 & \card \textit{expr} & 0 & 2,318 & 1,759 & 1,454 & 2,410 & 2,059 & 0 & 0 & 0 \\\hline  
 & \card \textit{diff} & 0 & 2 & 91 & 1,038 & 2,403 & 2,059 & 0 & 0 & 0 \\\hline\hline  
128 & \card \textit{expr} & 0 & 2,642 & 1,787 & 1,610 & 1,440 & 1,768 & 753 & 0 & 0 \\\hline  
 & \card \textit{diff} & 0 & 1 & 74 & 1,107 & 1,424 & 1,768 & 753 & 0 & 0 \\\hline\hline  
256 & \card \textit{expr} & 0 & 2,696 & 1,853 & 1,636 & 1,393 & 1,343 & 960 & 119 & 0 \\\hline  
 & \card \textit{diff} & 0 & 1 & 71 & 1,090 & 1,387 & 1,342 & 960 & 119 & 0 \\\hline\hline  
512 & \card \textit{expr} & 0 & 2,822 & 1,818 & 1,651 & 1,282 & 1,323 & 879 & 222 & 3 \\\hline  
 & \card \textit{diff} & 0 & 1 & 72 & 1,094 & 1,276 & 1,323 & 879 & 222 & 3 \\\hline\hline  
1K & \card \textit{expr} & 0 & 2,798 & 1,883 & 1,632 & 1,331 & 1,242 & 840 & 258 & 16 \\\hline  
 & \card \textit{diff} & 0 & 1 & 72 & 1,054 & 1,328 & 1,242 & 840 & 258 & 16 \\\hline\hline  
2K & \card \textit{expr} & 0 & 2,866 & 1,917 & 1,607 & 1,353 & 1,213 & 772 & 246 & 26 \\\hline  
 & \card \textit{diff} & 0 & 1 & 73 & 1,051 & 1,346 & 1,212 & 772 & 246 & 26 \\\hline\hline  
4K & \card \textit{expr} & 0 & 2,931 & 1,884 & 1,630 & 1,328 & 1,150 & 760 & 292 & 25 \\\hline  
 & \card \textit{diff} & 0 & 1 & 71 & 1,073 & 1,312 & 1,150 & 760 & 292 & 25 \\\hline\hline  
8K & \card \textit{expr} & 0 & 2,919 & 1,917 & 1,594 & 1,280 & 1,234 & 781 & 248 & 27 \\\hline  
 & \card \textit{diff} & 0 & 1 & 69 & 1,044 & 1,279 & 1,234 & 781 & 248 & 27 \\\hline\hline 
 \end{array}$
\end{center}
\end{table}

 In Tables \ref{distribution.simpl.pu} and \ref{distribution.simpl.r}, statistics are given about the number of input expressions that are simplified to a given size by algorithms \textit{PU} and \textit{R}.
The number contained in the cell of a column ${i\!:\!j}$ belonging to a row $\card \textit{expr}$ where $\textit{size} = s$, is the number of input expressions of size $s$ that are simplified  to an expression with a size between $i$ and $j$, inclusive. The corresponding number in  row $\card \textit{diff}$, is the number of corresponding simplified expressions that are different. A  logarithmic scale is used for the simplified sizes, which makes the statistics fairly easy to read.

\begin{table}[t]
\caption{Distribution of expressions with respect to simplification (algorithm  $R$)}\label{distribution.simpl.r}
\begin{center}
\renewcommand{\arraystretch}{1}
$\begin{array}{|r||c|r|r|r|r|r|r|r|r|r|}\hline 
\textit{size} & \textit{ssize} & _{0:2}& _{3:4}& _{5:8}& _{9:16}& _{17:32}& _{33:64}& _{65:128}& _{129:256}& _{257:512}\\[0.5em]\hline\hline
8 & \card \textit{expr} & 2,391 & 3,414 & 4,195 & 0 & 0 & 0 & 0 & 0 & 0 \\\hline  
 & \card \textit{diff} & 7 & 28 & 565 & 0 & 0 & 0 & 0 & 0 & 0 \\\hline\hline  
16 & \card \textit{expr} & 452 & 1,964 & 3,839 & 3,745 & 0 & 0 & 0 & 0 & 0 \\\hline  
 & \card \textit{diff} & 5 & 21 & 702 & 3,249 & 0 & 0 & 0 & 0 & 0 \\\hline\hline  
32 & \card \textit{expr} & 22 & 1,913 & 2,088 & 3,600 & 2,377 & 0 & 0 & 0 & 0 \\\hline  
 & \card \textit{diff} & 2 & 10 & 246 & 3,240 & 2,377 & 0 & 0 & 0 & 0 \\\hline\hline  
64 & \card \textit{expr} & 0 & 2,318 & 1,760 & 1,895 & 2,392 & 1,635 & 0 & 0 & 0 \\\hline  
 & \card \textit{diff} & 0 & 2 & 92 & 1,472 & 2,392 & 1,635 & 0 & 0 & 0 \\\hline\hline  
128 & \card \textit{expr} & 0 & 2,642 & 1,815 & 1,880 & 1,368 & 1,728 & 567 & 0 & 0 \\\hline  
 & \card \textit{diff} & 0 & 1 & 75 & 1,392 & 1,364 & 1,728 & 567 & 0 & 0 \\\hline\hline  
256 & \card \textit{expr} & 0 & 2,696 & 1,853 & 1,900 & 1,331 & 1,281 & 853 & 86 & 0 \\\hline  
 & \card \textit{diff} & 0 & 1 & 71 & 1,350 & 1,328 & 1,281 & 853 & 86 & 0 \\\hline\hline  
512 & \card \textit{expr} & 0 & 2,822 & 1,818 & 1,875 & 1,215 & 1,288 & 796 & 185 & 1 \\\hline  
 & \card \textit{diff} & 0 & 1 & 72 & 1,314 & 1,213 & 1,288 & 796 & 185 & 1 \\\hline\hline  
1K & \card \textit{expr} & 0 & 2,798 & 1,883 & 1,857 & 1,264 & 1,192 & 779 & 218 & 9 \\\hline  
 & \card \textit{diff} & 0 & 1 & 72 & 1,276 & 1,264 & 1,192 & 779 & 218 & 9 \\\hline\hline  
2K & \card \textit{expr} & 0 & 2,866 & 1,917 & 1,835 & 1,258 & 1,176 & 713 & 213 & 22 \\\hline  
 & \card \textit{diff} & 0 & 1 & 73 & 1,275 & 1,254 & 1,176 & 713 & 213 & 22 \\\hline\hline  
4K & \card \textit{expr} & 0 & 2,931 & 1,884 & 1,856 & 1,219 & 1,152 & 698 & 240 & 20 \\\hline  
 & \card \textit{diff} & 0 & 1 & 71 & 1,285 & 1,217 & 1,152 & 698 & 240 & 20 \\\hline\hline  
8K & \card \textit{expr} & 0 & 2,919 & 1,920 & 1,792 & 1,238 & 1,184 & 710 & 222 & 15 \\\hline  
 & \card \textit{diff} & 0 & 1 & 70 & 1,244 & 1,237 & 1,184 & 710 & 222 & 15 \\\hline\hline  
\end{array}$
\end{center}
\end{table}
 Several interesting observations can be made.
\begin{itemize}
\item In both Tables \ref{distribution.simpl.pu} and \ref{distribution.simpl.r}, the figures for input expressions of size $  \ge 128$ are very close, which means that the distribution of random expressions with respect to their simplified size is almost independent of their size, if this size is large enough.
\item The figures in Tables \ref{distribution.simpl.pu} and \ref{distribution.simpl.r} are very similar. It means that initializing the background with minimal expressions of size $ \le 15$ does not bring a big improvement.

\item For both algorithms \textit{PU} and \textit{R}, at least $65\%$  of the input expressions are simplified into an expression such that size $ \le 16$. Moreover, besides the fact that almost $50\%$ of the input expressions are simplified to a size $\le 8$, the expressions to which they are simplified are very few, only about $70$. Even more astonishingly, for an input size $ \ge 128$, all (more than $2,600$) expressions simplified to a size $ \le 4$ are simplified to the same expression (in fact: $\reg{(a + b)*}$). To the contrary, almost all input expressions simplified to  size $> 16$ are simplified to different expressions.

\end{itemize}

 Although the statistics presented above only are results about the simplifying power of algorithms \textit{PU} and \textit{R}, which are not able to minimize a regular expression in all cases, it can be argued that they give a rather good upper approximation of the number of expressions minimized to a given size. As a matter of fact, algorithm \textit{R} provides optimal results for expressions with  size $ \leq 16$, and its statistics (the figures in Table \ref{distribution.simpl.r}) are very close to the statistics for algorithm \textit{PU}. For expressions with a greater simplified size, nothing precise can be said but the results ``seem so good'' that we can be satisfied with them as an upper approximation. More information about the actual minimal expressions into which input expressions of size $8K$ are simplified is given in Table \ref{frequent.minimal} of Appendix \ref{most.frequent.simplified}.

%------------------------------------------------------------------------------------------------------------------------

Remember that algorithm \textit{PU} basically simplifies expressions in two ways: propagation and unification of MDFAs. The algorithm \textit{R} proceeds in the same way but, in addition, detects that certain expressions are equivalent to a minimal expression prerecorded in the background. Let us now concentrate on the relative importance of propagation and unification. Propagation often simplifies subexpressions but this simplification is primarily useful to speed up the computation of a DFA for the subexpression, which is then minimized and unified with the global MDFA of the background. Unification is extremely powerful because it ensures that the subexpression is simplified to the smallest equivalent expression in the background, and also because of two facts:
\begin{enumerate}
\item Since the input expressions are perfectly randomly chosen, many small expressions are registered in the background, as subexpressions of an input expression.
\item Generally speaking, most expressions are equivalent to a small expression.
\end{enumerate}

  In Tables \ref{distribution.pmin.pu} and \ref{distribution.pmin.r}, statistics are given about the number (noted $\card \textit{pmin}$) of input expressions that are simplified by unification to a subexpression of a lifted input expression. We call such subexpressions ``possibly minimal expressions'' since no shorter subexpression exists in the set of input expressions.

\begin{table}[t]
\caption{Distribution of possibly minimal expressions (algorithm  $PU$)}\label{distribution.pmin.pu}
\begin{center}
\renewcommand{\arraystretch}{1}
$\begin{array}{|r||c|r|r|r|r|r|r|r|r|r|}\hline
\textit{size} & \textit{ssize} & _{0:2}& _{3:4}& _{5:8}& _{9:16}& _{17:32}& _{33:64}& _{65:128}& _{129:256}& _{257:512}\\[0.5em]\hline\hline
8 & \card \textit{pmin} & 2,391 & 3,414 & 4,190 & 0 & 0 & 0 & 0 & 0 & 0 \\\hline  
16 & \card \textit{pmin} & 452 & 1,964 & 3,767 & 3,507 & 0 & 0 & 0 & 0 & 0 \\\hline  
32 & \card \textit{pmin} & 22 & 1,913 & 2,060 & 1,665 & 2,385 & 0 & 0 & 0 & 0 \\\hline  
64 & \card \textit{pmin} & 0 & 2,318 & 1,756 & 855 & 791 & 1,031 & 0 & 0 & 0 \\\hline  
128 & \card \textit{pmin} & 0 & 2,642 & 1,786 & 1,053 & 407 & 328 & 147 & 0 & 0 \\\hline  
256 & \card \textit{pmin} & 0 & 2,696 & 1,853 & 1,153 & 354 & 204 & 71 & 9 & 0 \\\hline  
512 & \card \textit{pmin} & 0 & 2,822 & 1,816 & 1,130 & 331 & 209 & 54 & 5 & 0 \\\hline  
1K & \card \textit{pmin} & 0 & 2,798 & 1,883 & 1,123 & 345 & 188 & 47 & 5 & 0 \\\hline  
2K & \card \textit{pmin} & 0 & 2,866 & 1,917 & 1,131 & 335 & 164 & 24 & 2 & 0 \\\hline  
4K & \card \textit{pmin} & 0 & 2,931 & 1,884 & 1,164 & 321 & 160 & 31 & 6 & 0 \\\hline  
8K & \card \textit{pmin} & 0 & 2,919 & 1,917 & 1,141 & 339 & 164 & 41 & 4 & 0 \\\hline  
 \end{array}$
\end{center}
\end{table}

  It can be seen in Table \ref{distribution.pmin.r} that almost all of them actually are minimal when their size is less than or equal to $8$. More than half of them  are minimal for a size between $9$ and $15$. The rest of those possibly minimal expressions are few and they are possibly unified to a minimal expression by algorithm \textit{R}. Other simplified expressions, which are significantly less numerous than the possibly minimal expressions, are simplified either by propagation or by unification to an expression simplified by propagation. Yet such simplified expressions can be minimal, sometimes. For instance, we can guess that many possibly minimal expressions with size $ > 16$ in Table \ref{distribution.pmin.r} actually are minimal. They are significantly fewer than in Table \ref{distribution.pmin.pu}, suggesting than the additional ones in Table \ref{distribution.pmin.pu} are minimized by algorithm \textit{R}. As a matter of fact, we can observe that the number of possibly minimal expressions in Table \ref{distribution.pmin.pu} ($6,525$ for ${\textit{size}= 8K}$) is almost equal to the number of minimal expressions in Table \ref{distribution.pmin.r} ($6,526$ for ${\textit{size}= 8K}$).

\begin{table}[t]
\caption{Distribution of possibly minimal expressions (algorithm  $R$)}\label{distribution.pmin.r}
\begin{center}
\renewcommand{\arraystretch}{1}
$\begin{array}{|r||c|r|r|r|r|r|r|r|r|r|}\hline 
\textit{size} & \textit{ssize} & _{0:2}& _{3:4}& _{5:8}& _{9:16}& _{17:32}& _{33:64}& _{65:128}& _{129:256}& _{257:512}\\[0.5em]\hline\hline
8 & \card \textit{pmin} & 2,391 & 3,414 & 4,186 & 0 & 0 & 0 & 0 & 0 & 0 \\\hline  
 & \card \textit{min} & 2,391 & 3,414 & 4,195 & 0 & 0 & 0 & 0 & 0 & 0 \\\hline\hline  
16 & \card \textit{pmin} & 452 & 1,964 & 3,778 & 2,397 & 0 & 0 & 0 & 0 & 0 \\\hline  
 & \card \textit{min} & 452 & 1,964 & 3,839 & 3,710 & 0 & 0 & 0 & 0 & 0 \\\hline\hline  
32 & \card \textit{pmin} & 22 & 1,913 & 2,068 & 1,052 & 1,144 & 0 & 0 & 0 & 0 \\\hline  
 & \card \textit{min} & 22 & 1,913 & 2,088 & 3,326 & 0 & 0 & 0 & 0 & 0 \\\hline\hline  
64 & \card \textit{pmin} & 0 & 2,318 & 1,756 & 719 & 436 & 511 & 0 & 0 & 0 \\\hline  
 & \card \textit{min} & 0 & 2,318 & 1,760 & 1,772 & 0 & 0 & 0 & 0 & 0 \\\hline\hline  
128 & \card \textit{pmin} & 0 & 2,642 & 1,814 & 894 & 235 & 161 & 48 & 0 & 0 \\\hline  
 & \card \textit{min} & 0 & 2,642 & 1,815 & 1,781 & 0 & 0 & 0 & 0 & 0 \\\hline\hline  
256 & \card \textit{pmin} & 0 & 2,696 & 1,853 & 1,040 & 202 & 91 & 23 & 4 & 0 \\\hline  
 & \card \textit{min} & 0 & 2,696 & 1,853 & 1,794 & 0 & 0 & 0 & 0 & 0 \\\hline\hline  
512 & \card \textit{pmin} & 0 & 2,822 & 1,816 & 1,050 & 214 & 117 & 15 & 0 & 0 \\\hline  
 & \card \textit{min} & 0 & 2,822 & 1,818 & 1,772 & 0 & 0 & 0 & 0 & 0 \\\hline\hline  
1K & \card \textit{pmin} & 0 & 2,798 & 1,883 & 1,034 & 212 & 86 & 15 & 1 & 0 \\\hline  
 & \card \textit{min} & 0 & 2,798 & 1,883 & 1,760 & 0 & 0 & 0 & 0 & 0 \\\hline\hline  
2K & \card \textit{pmin} & 0 & 2,866 & 1,917 & 1,055 & 187 & 94 & 9 & 0 & 0 \\\hline  
 & \card \textit{min} & 0 & 2,866 & 1,917 & 1,713 & 0 & 0 & 0 & 0 & 0 \\\hline\hline  
4K & \card \textit{pmin} & 0 & 2,931 & 1,884 & 1,084 & 202 & 85 & 15 & 0 & 0 \\\hline  
 & \card \textit{min} & 0 & 2,931 & 1,884 & 1,744 & 0 & 0 & 0 & 0 & 0 \\\hline\hline  
8K & \card \textit{pmin} & 0 & 2,919 & 1,920 & 1,065 & 199 & 95 & 14 & 0 & 0 \\\hline  
 & \card \textit{min} & 0 & 2,919 & 1,920 & 1,687 & 0 & 0 & 0 & 0 & 0 \\\hline\hline  
  \end{array}$
\end{center}
\end{table}

  To conclude, a few words can be said about possible ways to improve algorithms \textit{PU} and \textit{R}. First, we can notice that algorithm \textit{R} loses a lot of efficiency by prerecording all minimal expressions of size less than $16$ (see Table \ref{number.minimal}): Half of the identifiers available in the background are attributed to these expressions, while few of them are actually used to minimize a subexpression of an input expression. We could experimentally take note of the minimal expressions that are used and keep them in a file much smaller than the file currently used. Second, we can enhance both algorithms with explicit simplification rules similar to those of \cite{BLC4,Kahrs, Stoughton}. Propagation already is such a simplification rule, which is cheap and effective for randomly generated expressions. More elaborate simplification rules should be applied after propagation.

\section{Related work}
\label{related.work}

The work presented in this paper, and especially its implementation aspects, has many similarities to our previous work on what I called \emph{collections of structures} \cite{atindehouPhDThesis,BLC3,BLC2,BLC1}. A collection of structures is a data structure representing all terms of a congruence relation defined by a finite set of equations between ground terms \cite{Kozen77,Nelson80}. The terms are partitioned into  equivalence classes in such a way that a term $t$ can be minimized in time $O(\size(t))$ and that the equivalence of two terms $t_1$ and $t_2$ can be decided in time $O(\size(t_1) + \size(t_2))$. The method can be extended to theories defined by equations using variables, provided that the number of their equivalence classes is finite (and not too big), such as boolean formulas using a small number of letters (atomic propositional formulas). A complete study of collections of structures, including a theoretical complexity analysis, an experimental evaluation, and a comparison with related work can be found in \cite{atindehouPhDThesis}. I have attempted to extend the applicability of collections of structures to regular expressions, based on \cite{Kozen91,Salomaa66}, but the experimental results were disappointing. In this paper, collections of structures are abandoned in favor of the use of normalized expressions and the simpler Union-Find method to represent equivalence classes, which facilitates the integration of the two worlds of expressions and finite automata.\\

 As said before, the specific purpose of this paper is not to describe and discuss methods to simplify regular expressions. It is only to exactly explain how the framework works and to give evidence that it provides an integrated representation of regular languages that can be useful to solve various problems.
Nevertheless, other works mainly dedicated to the simplification of regular expressions present aspects comparable to specific points developed in this paper. We now discuss these issues.\\

 The work in \cite{Kahrs} is explicitly devoted to the problem of simplifying regular expressions. We compare it to our work from several viewpoints:
\begin{enumerate}
\item The implementation framework. In fact, this is the main issue to be addressed, as the implementation aspects are at the heart of this document.
\item The applicability of our framework to the implementation of the simplification methods presented in \cite{Kahrs}.
\item Some subtle differencies in the nature of the regular expressions that are considered and in the way they are randomly generated. This allows us to explain some differences in the experimental results.
\end{enumerate}

 As for the first point, let us note that the work in \cite{Kahrs} is implemented within the functional language Haskell, which makes it difficult and sometimes almost impossible to implement the low-level data structures extensively used in our work (see e.g.\ Section \ref{low.level.data.structures}). Instead, classical tree structures used to represent expressions are enhanced with so-called tags providing information on the denoted language and the simplification context. The fact that integers are used as identifiers in our system also speeds up some operations common to both systems, such as sorting terms in unions. In \cite{Kahrs}, a linear order on expressions is used to compare expressions, while we compare integers.
The fact that we use identifiers for expressions implies some overhead when new expressions are built but this overhead is low as explained in Section \ref{complexity.normalization} and has the benefit that the same expression is never built two times, which would often arise as shown in Tables  \ref{computing.normalization.table} and \ref{computing.normalization.table.opt}.
This can speed up simplification since the same expression need not be simplified two times. Derivatives of regular expressions are  also used in \cite{Kahrs} to check language equivalence and to compute a linear order on regular languages. This requires us to check the equivalence of expressions with respect to the relation $\cong$ of Definition \ref{congruence.expressions}. In \cite{Kahrs}, this is done by checking the syntactic equality of \emph{standardized} expressions, which correspond to normalized expressions in our terminology. In our work, however, this check amounts to checking the equality of two integers, which clearly is much more efficient.
It can also be observed that, in the framework described here, much information about work previously done is recorded in the background and can be reused to simplify and speed up further work. This does not seem to be the case in \cite{Kahrs} where expressions are simplified one by one. However, there is a proviso for ``catalogues'' mapping some expressions, or some languages represented by compacted automata inspired from \cite{Almeida}, to equivalent minimal expressions. The same effect can be achieved by the background as shown in Section \ref{simplifying.expressions} with algorithm \textit{R}. Moreover, comparable results can be obtained at a lower cost by simplifying expressions to subexpressions contained in the background, as shown by the analysis of algorithm \PU. Another difference is our use of a hashcode to search for equivalent expressions in a \MultiList\/ object. This is probably more efficient than using a linear order on regular languages as in \cite{Kahrs} if the hashing function is well chosen, i.e. sufficiently uniform. An advantage is that we do not have to keep a representation of the MDFAs of all minimal languages. An integer is enough.

Concerning the second point, it is quite clear that most of the simplification techniques presented in \cite{Kahrs} can be accommodated in our framework, including the tags giving information on the denoted languages and the simplification context: Such tags can be associated with identifiers using a few global arrays. The only problematic issue is their idea of renaming expressions to avoid representing several (or many) expressions equivalent modulo renaming inside the catalogues. It is not clear at this moment how this idea could be integrated in what is proposed here but it would probably entail a lot of additional algorithmic work and data structures, and a big loss in efficiency.

Regarding the third point, notice that the work in \cite{Kahrs} uses a different syntax of (plain) regular expressions, where expressions of the form $1 + E$ or $E + 1$ are replaced by so-called queries of the form $E ?$. This implies a very different set of randomly generated expressions. Moreover, they ``pre-standardize'' unions and concatenation when generating random expressions (S.\ Kahrs, personal communication, October 10, 2024). Such differences with the work presented here probably explain why some of their experimental results are very different from the measures in this paper. For instance, they report that $42.8\%$ of input expressions of size $2,050$ using two letters are simplified to $(a + b)^*$ by lifting, while we only get $26\%$ here.
In fact, those figures measure intrinsic properties of the chosen expressions, not the value of the simplification techniques (which are almost optimal for solving this problem).\\

 An elaborate method to simplify regular expressions is described  in \cite{Stoughton}, pages $92-120$. It is implemented in Standard ML and made available in \cite{StoughtonForlan}.
The approach is quite different from here and from \cite{Kahrs} because it is   \emph{mainly pedagogic, with no claims of either efficiency or completeness} (A. Stoughton, personal communication, October 14, 2022).
It uses a notion of regular expression complexity where the star-height, i.e.\ the maximum number of nested iterations in an expression, is the main parameter to minimize. For instance, the expression $1 + a(a + b)^*$ is considered simpler than $(a b^*)^*$, although its size is greater. A notion of weakly simplified expression, similar to standardized expressions in \cite{Kahrs}, is used as well as a set of $26$ simplification rules reducing the complexity of expressions. The rules use a partial decision algorithm to check the inclusion of regular languages. This algorithm proceeds by induction on the structure of expressions. Thus, the conceptual link with DFAs is not taken into account in that work. The simplification algorithm computes a sequence of weakly simplified expressions of strictly decreasing complexity. At each step, structural rules are applied to the current expression in an attempt to find one or several matching simplification rules. The application of structural rules potentially creates an exponential number of 
no longer weakly simplified variants of the current expression. Thus, the expression is weakly simplified again at the end of each step. 

 It would be possible to use the complexity measure and to adapt the simplification rules of 
\cite{Stoughton} to normalized expressions, inside an enhanced version of algorithm \PU. This would be more efficient. Moreover, it is the case that several of the simplification rules proposed in \cite{Stoughton} simplify an expression to one of its subexpression. Such simplifications are automatically and implicitly done by the algorithm \PU.

Finally, it is interesting to relate the results of Section \ref{distribution.minimal.size} to the theoretical work in \cite{Rotondo}. Using purely mathematical methods, they compute that the simplified lifting method (see Appendix \ref{information.lifting}) simplifies expressions to an average size of $78$, $496$, $2,\!518$, and $11,\!684$, respectively for $\nl = 2, 3, 4, 5$, when the input size of the expression tends to infinity. These figures can be compared to the average simplified sizes obtained by algorithms \Li\ and \PU\ for a size equal to $8K$. They respectively are equal to $69$, $439$, $1,\!508$ and $3,\!243$ for $\Li$ and to $25$, $334$, $1,\!400$, and $3,\!138$, for \PU. The values for \Li\/ and $\nl=2, 3$ are close to the figures in \cite{Rotondo}, indicating that $8K$ is sufficiently large to get good statistical results on expressions with $2$ or $3$ letters. Moreover, this shows that algorithm \PU\ simplifies such expressions much better than lifting, arguably giving interesting statistical information about minimization, in these cases. The fact that the values for $\nl= 4,5$ are well below $2,\!518$, and $11,\!684$ suggests that it would be necessary to consider input expressions much longer  than $8K$ to get interesting statistical information in these cases. This analysis is also confirmed by the fact that, in \cite{Rotondo}, the average proportion of expressions minimized to the universal one is
computed as at least equal to $31\%$, $13\%$,  $6.2\%$, and $2.8\%$, respectively for $\nl = 2, 3, 4, 5$, and that we get only slightly lower values for $\nl = 2, 3$ but much lower ones for $\nl= 4,5$ (see Table \ref{simplification.expressions.23458}).

\section{Conclusion and future work}
\label{conclusion}

I have presented an implementation framework for regular languages in which a large collection of regular languages can be represented as both regular expressions and DFAs, in an integrated way.
The main contribution consists of a meticulous description of  low-level data structures and algorithms designed to efficiently implement higher-level notions such as normalized expressions and equations between expressions and derivatives. 
In fact, this paper constitutes a prequel to my previous papers on simplifying regular expressions \cite{BLC4} and computing derivatives of regular expressions to build DFAs \cite{BLC5}. For lack of space, I was unable to describe the framework in a convincing way in those papers, which made their contribution difficult to assess. I guess that this is now possible.  

Another contribution lies in the experimental evaluation. It shows that the framework allows us to partition a large collection of regular languages in such a way that non equal regular languages belong to different equivalence classes; moreover, every represented language is given an identifier designating a small expression and an MDFA for the language, at the same time. Also, the collection of represented languages can be analyzed to give statistical information on the distribution of expressions with respect to their simplified size. These experiments are new with respect to those of \cite{BLC5,BLC4} as they assess the framework itself, independently of specific applications.

The work in this paper and in \cite{BLC5,BLC4} leaves open many research avenues such as relating the sizes of minimal expressions to the sizes of MDFAs, with the aim of optimally tuning the system, inventing new simplification rules for difficult expressions, and building a ``final'' well-documented version of the system. The experiments conducted in this paper suggest that the system could used to build a kind of simplification server that would improve itself over time by recording more and more ``interesting'' simplified expressions.

\subsection*{Artefact}
\addcontentsline{toc}{section}{Artefact}
The Java code of the algorithms presented in this paper can be found in the dropbox folder \href{https://www.dropbox.com/scl/fo/y7b03gynk7ynmfozp7gyf/AF2HFXF5PyCSaC_J7_j4lZY?rlkey=sf7bwx9vvqqamnsd94stm7b2v&st=32nofp4d&dl=0}
{\texttt{System Java Code}}, with some test data and explanations about running the test programs.

\subsection*{Author contribution and funding declaration}
\addcontentsline{toc}{section}{Author contribution and funding declaration}
I am the sole contributor of this paper. This research did not receive any specific grant from funding agencies in the public, commercial, or not-for-profit sectors.

\subsection*{Acknowledgements}

I would like to thank the people who have supported me during the many years I have devoted to this research and, in particular, Yves Deville, Pierre Flener, and Jos\'e Vander Meulen. Special thanks are due to Pierre Flener who made many suggestions to improve the final version of this paper.

I am also indebted to Stefan Kahrs and Alley Stoughton for answering my questions about their work and for the many leads they brought to my attention.

\bibliographystyle{plain}

\newpage
\addcontentsline{toc}{section}{References}
\bibliography{Biblio}

\appendix
\newpage
\section{Complements on normalized expressions}
\label{normalized.expressions.complements}

\subsection{Proof of Theorem \ref{theorem1}} 
\label{proof.theorem1}

\begin{enumerate}
\item The first part of the theorem is a consequence of the fact that the operation $\cup$ is associative, commutative, and idempotent. Let $E_1$, $E_2$, and $E_3$ be normalized expressions and let $S_1$, $S_2$, and $S_3$ be the corresponding sets of expressions according to Definition \ref{normalizing.operations}. The value of $E_1\union E_1$ is determined by $S_1 \cup S_1 $. Since $S_1 \cup S_1 = S_1$, it is equal to $E_1$. Similarly, $E_1\union E_2 = E_2\union E_1$ because $S_1\cup S_2 = S_2\cup S_1$, and
$E_1\union (E_2 \union E_3) = (E_1\union E_2) \union E_3$ because $S_1\cup (S_2 \cup S_3) = (S_1\cup S_2) \cup S_3$.

\item The second part of the theorem is immediate if one of the expressions $E_1$, $E_2$, and $E_3$ is equal to $0$ or $1$. Otherwise the proof 
can be done by induction on the syntactic structure of $E_1$:
\begin{enumerate}
\item Let us first suppose that $E_1$ is not a concatenation. Then,
$$\begin{array}{rcl}
(E_1\,\concat\,E_2)\,\concat\,E_3 & = & (E_1\,\cdot \,E_2)\,\concat\,E_3\\
& = & E_1\,\cdot \,(E_2\,\concat\,E_3) \\
& = & E_1\,\concat\,(E_2\,\concat\,E_3)
\end{array}$$
\noindent All equalities are immediate consequences of the definition of $\concat$.
\item Now, let us suppose that $E_1 = F_1\, \cdot  \,F_2$ where $F_1$ is not a concatenation. We have:
$$\begin{array}{rcl}
(E_1\,\concat\,E_2)\,\concat\,E_3 
& = & ((F_1\,\cdot \,F_2)\,\concat\,E_2)\,\concat\,E_3\\
& = & (F_1\,\cdot \,(F_2\,\concat\,E_2))\,\concat\,E_3\\
& = & (F_1\,\concat\,(F_2\,\concat\,E_2))\,\concat\,E_3\\
& = & F_1\,\concat\,((F_2\,\concat\,E_2)\,\concat\,E_3)\\
& = & F_1\,\concat\,(F_2\,\concat\,(E_2\,\concat\,E_3))\\
& = & F_1\,\cdot \,(F_2\,\concat\,(E_2\,\concat\,E_3))\\
& = & (F_1\,\cdot \,F_2)\,\concat\,(E_2\,\concat\,E_3)\\
& = & E_1\,\concat\,(E_2\,\concat\,E_3)\\
\end{array}$$
\noindent All equalities are immediate consequences of the definition except the fourth and the fifth ones, which use the induction on $E_1$ (first applied to $F_1$, and then to $F_2$).
\end{enumerate}

\item The proofs are direct consequences of the definitions.\\ Let use prove, for instance, that $\Lang(E_1 \union E_2) = \Lang( E_1)\: \cup\: \Lang( E_2)$. All terms (see Section \ref{complexity.normalization}) of $E_1 \union E_2$ are terms of $E_1$ or $E_2$, and conversely. Let $F_1$, \dots, $F_n$ be these terms $(n \ge 0)$. By definition of $\Lang$, in Figure \ref{fig.lang.expressions}, $\Lang(E_1 \union E_2) = \Lang( F_1)\: \cup\: \dots \:\cup\: \Lang (F_n) = \Lang (E_1)\: \cup \:\Lang (E_2)$.

\end{enumerate}

\subsection{Proof of Theorem \ref{theorem2}} 
\label{proof.theorem2}

Since the relation $\cong$ is a congruence, it is sufficient to show that all pairs of plain expressions respecting an equivalence of Definition \ref{congruence.expressions} are normalized to the same normalized expression. Let us denote the normalization of the plain regular expression $E$ by $\overline{E}$. All proofs are similar and direct consequences of the definitions and of Theorem \ref{theorem1}. As an example let us prove that 
$$\overline{E_1 + (E_2 + E_3)} = \overline{(E_1 + E_2) + E_3}.$$
Clearly:
$$
\begin{array}{rclcl}
\overline{E_1 + (E_2 + E_3)} & = & 
\overline{E_1} \union \overline{E_2 + E_3} && (\textrm{Definition of\ } \overline{E_1 + (E_2 + E_3)}\:)\\

& = &
\overline{E_1} \union (\overline{E_2} \union \overline{E_3})
&& (\textrm{Definition of\ } \overline{E_2 + E_3})\\

& = &
(\overline{E_1} \union \overline{E_2}) \union \overline{E_3}
&& (\textrm{Theorem \ref{theorem1} } )\\

& = &
(\overline{E_1 + E_2}) \union \overline{E_3}
&& (\textrm{Definition of\ } \overline{E_1 + E_2})\\

& = &
\overline{(E_1 + E_2) + E_3}
&& (\textrm{Definition of\ } \overline{(E_1 + E_2) + E_3}\:)\\

\end{array}
$$

\section{Complements on the background}

\subsection{Confluence of the operation \reduce}
\label{confluency.reduce}

The operation \reduce\/ modifies the background by merging some of its equivalence classes of expressions and reducing its set of equations by replacing every expression occurring in equations by its representative in the new equivalence classes. Our goal here is to prove that the new set of equivalence classes is the same for all possible executions of \reduce, 
independently of what pair of overlapping equations is selected at each step.
 Let $\Eq_0$ be the set of equations in the background before an execution of \reduce, and let $\Ecal$ be the set of expressions occurring in these equations. The execution of \reduce\/ computes an equivalence relation $\sim_\textnormal{q}$ on $\Ecal$ from which the new set of equivalence classes of the background can be deduced, and it is simpler to prove that $\sim_\textnormal{q}$ is the same in all possible executions than to reason on the complete set of equivalence classes. So we focus on the unicity of $\sim_\textnormal{q}$. The proof is organized in four steps, as follows:

\begin{enumerate}

\item We give an explicit characterization of the (non deterministic) sequence
$$\langle \sim_0,\, \Eq_0 \rangle \: \longrightarrow \:
\langle \sim_1,\, \Eq_1 \rangle \: \longrightarrow \:
\dots \: \longrightarrow \:
\langle \sim_\textnormal{q},\, \Eq_\textnormal{q} \rangle
$$
computed by the operation $\reduce$.
\item We define an equivalence relation $\sim$, in a declarative, i.e.\  non algorithmic, way.
\item We prove that\:\:\:
$\sim_0 \:\:\: \subseteq\:\:\: \sim_1\:\:\: \subseteq\:\:\: \dots\:\:\:  \subseteq\:\:\:  \sim_\textnormal{q}\:\:\: \subseteq\:\:\: \sim.
$
\item We observe that $\:\:\:\sim\:\:\: \subseteq\:\:\: \sim_\textnormal{q}.$
\end{enumerate}

\noindent So, we can conclude that all possible executions of \reduce\/ compute the same equivalence relation~$\sim$.

\paragraph{Step 1}

Let us start with some notation.
We abbreviate the construction $o + \, \dots\, + \,x\cdot E_x\,+\, \dots$ by $\rpart$. So equations can be written as $E = \rpart$, $E' = \rpart'$, and the like. We say that $E_x$ and $E'_x$ are corresponding expressions in $\rpart$ and $\rpart'$. If $\rho$ is a relation on $\Ecal$, we write $\rpart \:\rho\:\rpart'$ to mean that $E_x\:\rho\: E'_x$ for all corresponding expressions of $\rpart$ and $\rpart'$.

 Now, we proceed to characterize every pair $\langle \sim_i,\, \Eq_i \rangle$ $(0 \leq i \leq \textnormal{q})$, by recurrence on $i$. We remember that, for all $i$, a unique representative is chosen for every expression in its equivalence class. Moreover, the set $\Eq_i$ consists of all equations $E = \rpart$ only using representatives of expressions in $\sim_i$ such that  $E\sim_i E'$ and $\rpart \sim_i \rpart'$, for some equation $E' = \rpart'$ in $\Eq_0$.

\begin{enumerate}
\item For $i= 0$, the relation $\sim_0$ simply is the set of all pairs $\langle E, E \rangle$ such that $E\in \Ecal$. Moreover, for every equation $E = \rpart$ in $\Eq_0$, this equation itself is such that $E \sim_0 E$ and $\rpart \sim_0\rpart$, and it uses representatives only.

\item For any $i$ such that $i < \textnormal{q}$, let $F_1$ and $F_2$ be the expressions on which the chosen overlapping equations $E_1 = \rpart_1$ and $E_2 = \rpart_2$ disagree. Let us assume that the operation $\unify$ chooses to replace $F_2$ by $F_1$ in the equations of $\Eq_i$. Then:
 \begin{enumerate}
 \item The relation $\sim_{i + 1}$ is the smallest equivalence relation such that $\sim_i \subseteq \sim_{i + 1}$ and $\langle F_1, F_2 \rangle\: \in \:\sim_{i + 1}$. The equivalence classes of $F_1$ and $F_2$ are merged in $\sim_{i + 1}$ and $F_1$ becomes the representative of all expressions of the new class. For other classes, the representative remains the same.
\item Every equation $E = \rpart$ in $\Eq_i$ is modified into an equation $E^{''} = \rpart^{''}$ in $\Eq_{i+1}$, by replacing all occurrences of $F_2$ by $F_1$ in the first equation. Since $\Eq_0$ contains an equation $E' = \rpart'$ such that $E'\sim_i E$ and $\rpart' \sim_i \rpart$, the same equation is such that $E'\sim_{i+1} E^{''}$ and $\rpart' \sim_{i+1} \rpart^{''}$, by definition of $\sim_{i+1}$.
 
\end{enumerate}
\end{enumerate}

\paragraph{Step 2} Now, let us define the relation $\sim$.
We define it as the smallest equivalence relation on $\Ecal$ such that
$$E \sim E' \iff \rpart \sim \rpart'$$
for all pairs of equations $E = \rpart$ and $E' = \rpart'$ in 
$\Eq_0$.

\paragraph{Step 3} Next, let us prove that $\sim_0 \:\:\: \subseteq\:\:\: \sim_1\:\:\: \subseteq\:\:\: \dots\:\:\:  \subseteq\:\:\:  \sim_\textnormal{q}\:\:\: \subseteq\:\:\: \sim.
$
It is clear that $\sim_0\subseteq\sim$ since $\sim_0$ is the smallest equivalence relation on $\Ecal$. Now, let us prove that $\sim_i\subseteq\sim$ implies  $\sim_{i+1}\subseteq\sim$ for $i<\textnormal{q}$.
Using the same notation as above, the proof boils down to showing that 
$F_1\sim F_2$ since, then, $\sim_i \subseteq \sim$ and $\langle F_1, F_2 \rangle \in \sim_{i + 1}$ implies that $\sim_{i+1}\subseteq\sim$.
Two cases must be considered: Either $E_1 = E_2$, or $\rpart_1 = \rpart_2$.
In the first case, the set $\Eq_0$ contains two equations $E^{''}_1 = \rpart^{''}_1$ and $E^{''}_2 = \rpart^{''}_2$ such that $E^{''}_1 \sim_i E_1 = E_2 \sim_i E^{''}_2$. Thus $E^{''}_1 \sim E^{''}_2$, which implies that $\rpart^{''}_1\sim \rpart^{''}_2$, by definition of $\sim$. Finally, the latter equivalence implies that $F_1\sim F_2$, since $F_1$ and $F_2$ are equivalent modulo $\sim_i$ to two corresponding expressions in $\rpart^{''}_1$ and $\rpart^{''}_2$. The case $\rpart_1 = \rpart_2$ is similar since this equality successively implies  $\rpart^{''}_1\sim_i \rpart^{''}_2$, $\rpart^{''}_1\sim \rpart^{''}_2$, $E_1 \sim E_2$, and $F_1 \sim F_2$.

\paragraph{Step 4} Finally, to prove that $\sim\:\subseteq\:\sim_\textnormal{q}$, we can observe that the relation $\sim_\textnormal{q}$ satisfies the definition of $\sim$, except possibly minimality, i.e.\  that we have:
$$E_1 \sim_\textnormal{q} E_2 \iff \rpart_1 \sim_\textnormal{q} \rpart_2$$
for all pairs of equations $E_1 = \rpart_1$ and $E_2 = \rpart_2$ in 
$\Eq_0$. Consider two such equations and let $E'_1 = \rpart'_1$ and $E'_2 = \rpart'_2$ be their representatives in $\Eq_\textnormal{q}$. Let us assume first that  $E_1 \sim_\textnormal{q} E_2$. This implies that $E'_1 = E'_2$ since equivalent expressions have the same representative. But, since no distinct equations overlap in $\Eq_\textnormal{q}$, we must have $\rpart'_1 = \rpart'_2$, which implies $\rpart_1 \sim_\textnormal{q} \rpart_2$ because $\rpart_1 \sim_\textnormal{q} \rpart'_1$ and  $\rpart_2 \sim_\textnormal{q} \rpart'_2$. The proof in the other direction is similar.

\paragraph{Note 1}

Using the same notation as above, the set of equivalence classes of the background after an execution of the operation $\reduce$ can be constructed by removing, from the initial set, all classes containing an expression in $\Ecal$ and by adding to this set unions of these classes containing expressions equivalent with respect to $\sim$. In symbols:
$$\displaystyle
\Ccal = (\Ccal_0 \:\setminus\: \{C\in \Ccal_0\:|\:\:\Ecal\cap C\ \neq \{\}\})\:\cup\:
\Big\{\displaystyle\bigcup_{C'\cap\, C \neq \{\}} C\:\Big|\:\:C'\in\Ccal_{\sim}\:\:\&\:\:C\in\Ccal_0\Big\}$$
where $\Ccal_0$, $\Ccal$, and $\Ccal_\sim$ respectively are the initial and final set of equivalence classes of the background, and the set of equivalence classes of the relation $\sim$.

\paragraph{Note 2}

When two equations $E = \rpart$ and $E' = \rpart'$ overlap in the background, it is normally the case that $o = o'$, as a consequence of the invariant of the background stated in Section \ref{background.content}.  However, it is the user's responsibility to ensure that $\Lang( E) = \Lang ( E')$ when two expressions $E$ and $E'$ are unified. In case of a mistake, there is a high probability that the algorithm $\merge$ will find overlapping equations such that $o \neq o'$. This is even unavoidable if the MDFAs of $E$ and $E'$ are unified as in algorithm \PU. Therefore, it is useful to add a check that
$\tabTIE[\iEq_1][0] = \tabTIE[\iEq_2][0]$ at Step $1(a)$ of the algorithm $\reduce$ (see Section \ref{background.implementation.algorithms}). If the equality does not hold, it means that the user's program contains an error. This may be the case for erroneous simplification rules, for example.

\subsection{Garbage collection}
\label{garbage.collection}

When the object ${\iExprList(M)}$  no longer contains any free identifier (see Section \ref{internal.representation}), it is often possible to return certain identifiers to it, in order to continue the process underway. To do so, the background is equipped with an algorithm working as follows: It receives as inputs some data structures containing all identifiers needed to proceed. All identifiers reachable from these identifiers are marked. Afterwards, all non marked identifiers are returned to the list ${\iExprList(M)}$. Finally, the current process is resumed.

However, as we use a high-level programming language, it is neither really possible to automatically return to the exact position where the program was interrupted, nor to automatically determine the list of identifiers used by the current process. Thus, the following method is used: When no more free identifiers are available, an exception is thrown. The current process must catch the exception, call the garbage collector passing the information about the identifiers to be saved, and finally resume its work appropriately. In some cases, such as the computation of all derivatives of an expression, a big part of the interrupted process must often be resumed. Notice that it is the user's responsibility to decide which data in the current background are best to continue what is underway. For instance, in the experiments of Section \ref{experiments}, the algorithms attempt to keep all the identifiers used in the global MDFA: Representatives of equivalence classes are kept and other identifiers are freed unless they are needed by the current process. In some cases, it may make more sense to retain only some of the equations, or none at all. Sadly and obviously, it can always be the case that a given algorithm cannot be completed even by reinitializing all data structures from scratch.

\subsection{An example of Java code}
\label{java.code}

Figure \ref{java.substitute} depicts the Java code of the algorithm $\substitute$ described at page \pageref{page.substitute}. This Java method is part of the class \texttt{Background}, which contains the declarations of the arrays $\tabIE[M]$, $\tabIR[M]$, $\tabTIE[M][]$, and the objects $\listIEQIE(M, M)$ and $\listIEQIEx[\nl](M, M)$. It can be observed that the translation from ordinary language to Java is  straightforward.

\begin{figure}[!ht]
\caption{Java code of the method $\substitute$}
\label{java.substitute}

\begin{center}
\begin{tabular}{|c|}\hline
\begin{minipage}{14cm}
\begin{lstlisting}[language=java]	
void substitute(int iE1, int iE2)
{
  while (! list_IEQ_IE.isEmpty(iE2))
  {
    int iEq = list_IEQ_IE.choose(iE2) ;
    int iR = tabIR[iEq] ;
    int[] tabIE = tabTIE[iR] ;

    removeEq(iEq) ;

    int[] tabN = substitute(tabIE, iE1, iE2) ;

    addEq(iE1, tabN) ;  
  }

  int x = 1 ;
  while (x != list_IEQ_IE_x.length)
  {
    while (! list_IEQ_IE_x[x].isEmpty(iE2))
    {
      int iEq = list_IEQ_IE_x[x].choose(iE2) ;
      int iE = tabIE[iEq] ;
      int iR = tabIR[iEq] ;
      int[] tabIE = tabTIE[iR] ;
 
      removeEq(iEq) ;
  
      int[] tabN = substitute(tabIE, iE1, iE2) ;
  
      addEq(iE, tabN) ;  
    }
    x ++ ;
  }
}
\end{lstlisting}
\end{minipage}\\\hline
\end{tabular}
\end{center}
\end{figure}

\section{Additional experimental results}
\label{appendix.experimental.results}

I provide new  experimental results, complementary to those of Sections \ref{simplifying.expressions} and \ref{distribution.minimal.size}. We analyze the performance of algorithms \textit{N}, \textit{L}, and \PU\/ applied to expressions using up to $3$, $4$, $5$, and  $8$ letters. As previously the algorithms are applied to large files of $10,000$ randomly generated expressions of sizes ranging from $8$ to $8K$.
Since expressions become less and less simplifiable when their number $\nl$ of letters grows, an upper bound of $512$ has been imposed on the size of subexpressions for which an MDFA is computed after simplification by propagation. Otherwise, the time spent to compute the derivatives of a subexpression can soar, and the number of derivatives can even exceed the memory size. However, propagation is still applied to all subexpressions.

\subsection{Simplification of expressions}
\label{appendix.simplifying.expressions}

Table \ref{simplification.expressions.23458}  shows the arithmetic and geometric means of the sizes of expressions simplified by the algorithms \textit{N}, \textit{L}, and \PU, for various input sizes, and for allowed numbers of letters $\nl = 2, 3, 4, 5, 8$. It also depicts the numbers of expressions found equivalent to the universal one: $(a + b + \dots + \ell)^*$, where $\ell$ is the $\nl$-th letter in the alphabet. Lastly, it provides information about the 
mean execution times. Sizes of the simplified expressions are given in percent of the size of the input expressions if their ratio is greater than or equal to $\frac{2}{3}$; their actual values are given otherwise (rounded).

We can observe that the sizes of the simplified expressions quickly grow when the value of $\nl$ increases, which most probably indicates that the sizes of minimal expressions grow almost as quickly. As for the number of expressions found equivalent to the universal one, it decreases very quickly, and we can guess that the values found by algorithm \PU\/ are close to the actual average proportion of expressions that are equivalent to the universal expression.
Concerning execution times, we see that they grow linearly for algorithms \textit{N} and \textit{L} with respect to both $\size$ and $\nl$. For algorithm \PU, it is insightful to first look at the case $\nl = 3$: The execution times grow exponentially until $\size = 1K$ but less than linearly  afterwards. For $\size =8K$, they are greater by two orders of magnitude than the times for $\nl = 2$. This can be explained as follows: The algorithm \PU\/ applies to expressions first simplified by algorithm \textit{L}. The average size of these lifted expressions stabilizes below $70$ for $\nl = 2$; so, the execution times for \PU\/ also stabilize at $4.1\,m$. To the contrary, for $\nl = 3$, the average size of lifted expressions does not stabilize, although it would for larger sizes of input expressions. Therefore, the times for \PU\/ continue to grow but not so fast in the end due to the $512$ bound limiting the use of computing MDFAs. Nevertheless, since the average size of lifted expressions remains below $512$, the MDFAs of many subexpressions are computed. This can explain the fact that the sizes of simplified expressions still are much better for \PU\/ than for \textit{L}. Now, the results for $\nl = 4, 5, 8$ can be explained by the fact that many lifted expressions remain longer that $512$, on the average, which speeds up the execution times of \PU\/ but limits its simplification power, as can be observed in Table \ref{simplification.expressions.23458}. Increasing the $512$ bound could improve the simplification power of \PU\/ but at too high a cost in many cases.

\begin{table}[!h]
\caption{Simplification of expressions $(\nl = 2, 3, 4, 5, 8)$}\label{simplification.expressions.23458}
\setlength{\arraycolsep}{0.15em}
\renewcommand{\arraystretch}{1}
\begin{center}\small
$\begin{array}{|c|r||r|r|r||r|r|r||r|r|r||r|r|r||r|r|r|}\hline
&\nl\:\: & \multicolumn{3}{c||}{2}& \multicolumn{3}{c||}{3}& \multicolumn{3}{c||}{4}& \multicolumn{3}{c||}{5}& \multicolumn{3}{c|}{8}\\\hline
& \size\: & \textit{N}\:\:\: & \textit{L}\:\:\: & \textit{PU}\:\: & \textit{N}\:\: & \textit{L}\:\:\: & \textit{PU}\:\: & \textit{N}\:\:\: & \textit{L}\:\:\: & \textit{PU}\:\: & \textit{N}\:\:\: & \textit{L}\:\:\: & \textit{PU}\:\: & \textit{N}\:\:\: & \textit{L}\:\:\: & \textit{PU} \:\: \\ \hline \hline
\textrm{ } & 8 & 75\% & 5 & 4 & 75\% & 75\% & 5 & 88\% & 75\% & 75\% & 88\% & 75\% & 75\% & 88\% & 88\% & 88\% \\ \cline{2-17}
\textrm{a} & 16 & 69\% & 9 & 8 & 75\% & 69\% & 10 & 81\% & 75\% & 75\% & 88\% & 81\% & 75\% & 88\% & 88\% & 88\% \\ \cline{2-17}
\textrm{r} & 32 & 72\% & 15 & 12 & 78\% & 21 & 19 & 81\% & 75\% & 72\% & 84\% & 78\% & 78\% & 91\% & 88\% & 84\% \\ \cline{2-17}
\textrm{i} & 64 & 72\% & 25 & 18 & 78\% & 38 & 35 & 81\% & 72\% & 69\% & 84\% & 78\% & 77\% & 89\% & 86\% & 86\% \\ \cline{2-17}
\textrm{t} & 128 & 71\% & 35 & 22 & 78\% & 66 & 57 & 82\% & 68\% & 82 & 85\% & 77\% & 74\% & 90\% & 87\% & 86\% \\ \cline{2-17}
\textrm{h} & 256 & 71\% & 47 & 24 & 78\% & 110 & 90 & 82\% & 158 & 148 & 85\% & 74\% & 71\% & 89\% & 86\% & 86\% \\ \cline{2-17}
\textrm{ } & 512 & 72\% & 56 & 25 & 78\% & 170 & 133 & 82\% & 279 & 258 & 85\% & 69\% & 67\% & 89\% & 86\% & 85\% \\ \cline{2-17}
\textrm{m} & 1K & 72\% & 64 & 25 & 78\% & 246 & 186 & 82\% & 475 & 439 & 85\% & 655 & 634 & 90\% & 85\% & 84\% \\ \cline{2-17}
\textrm{e} & 2K & 72\% & 67 & 25 & 78\% & 312 & 238 & 82\% & 753 & 699 & 85\% & 1,165 & 1,127 & 90\% & 84\% & 83\% \\ \cline{2-17}
\textrm{a} & 4K & 72\% & 70 & 25 & 78\% & 380 & 289 & 82\% & 1,109 & 1,029 & 85\% & 1,987 & 1,922 & 90\% & 83\% & 82\% \\ \cline{2-17}
\textrm{n} & 8K & 72\% & 69 & 25 & 78\% & 439 & 334 & 82\% & 1,508 & 1,400 & 85\% & 3,243 & 3,138 & 90\% & 80\% & 80\% \\ \hline \hline 
\textrm{ } & 8 & 0 & 0 & 0 & 0 & 0 & 0 & 0 & 0 & 0 & 0 & 0 & 0 & 88\% & 88\% & 88\% \\ \cline{2-17}
\textrm{g} & 16 & 69\% & 8 & 7 & 75\% & 10 & 9 & 81\% & 75\% & 69\% & 81\% & 75\% & 75\% & 88\% & 88\% & 88\% \\ \cline{2-17}
\textrm{e} & 32 & 69\% & 13 & 10 & 78\% & 20 & 18 & 81\% & 72\% & 69\% & 84\% & 78\% & 75\% & 91\% & 88\% & 84\% \\ \cline{2-17}
\textrm{o} & 64 & 70\% & 17 & 13 & 78\% & 33 & 30 & 81\% & 69\% & 42 & 84\% & 77\% & 75\% & 89\% & 86\% & 86\% \\ \cline{2-17}
\textrm{m} & 128 & 71\% & 20 & 13 & 78\% & 50 & 42 & 82\% & 77 & 73 & 85\% & 73\% & 71\% & 89\% & 86\% & 85\% \\ \cline{2-17}
\textrm{ } & 256 & 71\% & 23 & 13 & 78\% & 72 & 56 & 82\% & 127 & 117 & 85\% & 68\% & 166 & 89\% & 86\% & 85\% \\ \cline{2-17}
\textrm{m} & 512 & 71\% & 24 & 13 & 78\% & 95 & 70 & 82\% & 200 & 180 & 85\% & 303 & 290 & 89\% & 85\% & 84\% \\ \cline{2-17}
\textrm{e} & 1K & 72\% & 25 & 13 & 78\% & 115 & 83 & 82\% & 300 & 271 & 85\% & 519 & 498 & 90\% & 83\% & 82\% \\ \cline{2-17}
\textrm{a} & 2K & 72\% & 25 & 13 & 78\% & 125 & 91 & 82\% & 415 & 375 & 85\% & 833 & 801 & 90\% & 81\% & 80\% \\ \cline{2-17}
\textrm{n} & 4K & 72\% & 25 & 12 & 78\% & 135 & 99 & 82\% & 521 & 475 & 85\% & 1,260 & 1,211 & 90\% & 78\% & 77\% \\ \cline{2-17}
\textrm{ } & 8K & 72\% & 25 & 12 & 78\% & 143 & 105 & 82\% & 613 & 558 & 85\% & 1,817 & 1,747 & 90\% & 74\% & 73\% \\ \hline \hline 
\textrm{ } & 8 & 171 & 719 & 721 & 41 & 98 & 98 & 4 & 4 & 4 & 0 & 0 & 0 & 0 & 0 & 0 \\ \cline{2-17}
\card  & 16 & 11 & 1,244 & 1,264 & 4 & 290 & 292 & 3 & 57 & 57 & 0 & 5 & 5 & 0 & 0 & 0 \\ \cline{2-17}
\textrm{u} & 32 & 0 & 1,814 & 1,871 & 0 & 560 & 560 & 0 & 197 & 198 & 0 & 55 & 55 & 0 & 0 & 0 \\ \cline{2-17}
\textrm{n} & 64 & 0 & 2,190 & 2,317 & 0 & 821 & 827 & 0 & 272 & 273 & 0 & 110 & 110 & 0 & 6 & 6 \\ \cline{2-17}
\textrm{i} & 128 & 0 & 2,435 & 2,642 & 0 & 1,009 & 1,024 & 0 & 388 & 389 & 0 & 139 & 139 & 0 & 6 & 6 \\ \cline{2-17}
\textrm{v} & 256 & 0 & 2,478 & 2,696 & 0 & 993 & 1,014 & 0 & 463 & 464 & 0 & 186 & 187 & 0 & 15 & 15 \\ \cline{2-17}
\textrm{e} & 512 & 0 & 2,561 & 2,822 & 0 & 1,008 & 1,024 & 0 & 476 & 480 & 0 & 197 & 197 & 0 & 19 & 19 \\ \cline{2-17}
\textrm{r} & 1K & 0 & 2,555 & 2,798 & 0 & 1,058 & 1,090 & 0 & 471 & 473 & 0 & 194 & 194 & 0 & 21 & 21 \\ \cline{2-17}
\textrm{s} & 2K & 0 & 2,612 & 2,866 & 0 & 1,126 & 1,155 & 0 & 466 & 469 & 0 & 219 & 219 & 0 & 17 & 20 \\ \cline{2-17}
\textrm{a} & 4K & 0 & 2,653 & 2,931 & 0 & 1,109 & 1,136 & 0 & 492 & 492 & 0 & 169 & 220 & 0 & 7 & 19 \\ \cline{2-17}
\textrm{l} & 8K & 0 & 2,666 & 2,919 & 0 & 1,138 & 1,156 & 0 & 464 & 464 & 0 & 104 & 203 & 0 & 4 & 19 \\ \hline \hline 
\textrm{ } & 8 & 15\,\mu & 12\,\mu & 8.1\,\mu & 22\,\mu & 25\,\mu & 17\,\mu & 13\,\mu & 14\,\mu & 16\,\mu & 11\,\mu & 13\,\mu & 19\,\mu & 9.3\,\mu & 34\,\mu & 46\,\mu \\ \cline{2-17}
\textrm{m} & 16 & 16\,\mu & 20\,\mu & 32\,\mu & 31\,\mu & 34\,\mu & 45\,\mu & 15\,\mu & 28\,\mu & 63\,\mu & 14\,\mu & 44\,\mu & 54\,\mu & 12\,\mu & 15\,\mu & 74\,\mu \\ \cline{2-17}
\textrm{e} & 32 & 25\,\mu & 46\,\mu & 103\,\mu & 20\,\mu & 35\,\mu & 121\,\mu & 39\,\mu & 29\,\mu & 139\,\mu & 44\,\mu & 30\,\mu & 116\,\mu & 44\,\mu & 26\,\mu & 169\,\mu \\ \cline{2-17}
\textrm{a} & 64 & 62\,\mu & 40\,\mu & 264\,\mu & 50\,\mu & 52\,\mu & 344\,\mu & 37\,\mu & 48\,\mu & 405\,\mu & 67\,\mu & 62\,\mu & 382\,\mu & 33\,\mu & 45\,\mu & 364\,\mu \\ \cline{2-17}
\textrm{n} & 128 & 83\,\mu & 56\,\mu & 623\,\mu & 64\,\mu & 85\,\mu & 1.2\,m & 102\,\mu & 96\,\mu & 1.0\,m & 96\,\mu & 110\,\mu & 1.1\,m & 58\,\mu & 103\,\mu & 852\,\mu \\ \cline{2-17}
\textrm{ } & 256 & 122\,\mu & 92\,\mu & 1.3\,m & 103\,\mu & 112\,\mu & 6.0\,m & 119\,\mu & 141\,\mu & 6.5\,m & 139\,\mu & 115\,\mu & 5.0\,m & 108\,\mu & 133\,\mu & 4.2\,m \\ \cline{2-17}
\textrm{t} & 512 & 230\,\mu & 138\,\mu & 2.0\,m & 184\,\mu & 170\,\mu & 23\,m & 195\,\mu & 209\,\mu & 42\,m & 234\,\mu & 254\,\mu & 27\,m & 182\,\mu & 254\,\mu & 14\,m \\ \cline{2-17}
\textrm{i} & 1K & 299\,\mu & 177\,\mu & 3.4\,m & 254\,\mu & 232\,\mu & 142\,m & 335\,\mu & 288\,\mu & 156\,m & 413\,\mu & 360\,\mu & 55\,m & 404\,\mu & 462\,\mu & 20\,m \\ \cline{2-17}
\textrm{m} & 2K & 581\,\mu & 235\,\mu & 4.1\,m & 715\,\mu & 320\,\mu & 225\,m & 720\,\mu & 483\,\mu & 154\,m & 674\,\mu & 617\,\mu & 81\,m & 657\,\mu & 853\,\mu & 33\,m \\ \cline{2-17}
\textrm{e} & 4K & 1.2\,m & 402\,\mu & 4.1\,m & 1.2\,m & 510\,\mu & 401\,m & 1.4\,m & 746\,\mu & 323\,m & 1.3\,m & 1.2\,m & 131\,m & 1.4\,m & 1.7\,m & 62\,m \\ \cline{2-17}
\textrm{ } & 8K & 2.4\,m & 734\,\mu & 4.1\,m & 2.3\,m & 832\,\mu & 535\,m & 2.9\,m & 1.2\,m & 289\,m & 2.5\,m & 1.9\,m & 206\,m & 2.6\,m & 2.7\,m & 123\,m \\ \hline 

\end{array}$
\end{center}
\end{table}

\begin{table}[!h]
\caption{Distribution of expressions with respect to simplification ($\nl = 3$)}\label{distribution.simpl.3}
\begin{center}\small
\setlength{\arraycolsep}{0.3em}
\renewcommand{\arraystretch}{0.95}
$
\begin{array}{|r||c|r|r|r|r|r|r|r|r|r|r|r|}\hline
\size & \textit{ssize}  & 6 & 12 & 24 & 48 & 96 & 192 & 384 & 768 & 1.5K & 3K & 6K \\ \hline \hline  
8 &  \card \textit{simp} & 7,482 & 2,518 & 0 & 0 & 0 & 0 & 0 & 0 & 0 & 0 & 0 \\ \hline 
 &  \card \textit{pmin} & 7,475 & 2,501 & 0 & 0 & 0 & 0 & 0 & 0 & 0 & 0 & 0 \\ \hline  \hline 
16 &  \card \textit{simp} & 1,862 & 5,525 & 2,613 & 0 & 0 & 0 & 0 & 0 & 0 & 0 & 0 \\ \hline 
 &  \card \textit{pmin} & 1,856 & 4,884 & 2,512 & 0 & 0 & 0 & 0 & 0 & 0 & 0 & 0 \\ \hline  \hline 
32 &  \card \textit{simp} & 691 & 1,134 & 6,035 & 2,140 & 0 & 0 & 0 & 0 & 0 & 0 & 0 \\ \hline 
 &  \card \textit{pmin} & 691 & 815 & 4,233 & 1,912 & 0 & 0 & 0 & 0 & 0 & 0 & 0 \\ \hline  \hline 
64 &  \card \textit{simp} & 828 & 649 & 851 & 6,352 & 1,320 & 0 & 0 & 0 & 0 & 0 & 0 \\ \hline 
 &  \card \textit{pmin} & 828 & 542 & 390 & 3,184 & 983 & 0 & 0 & 0 & 0 & 0 & 0 \\ \hline  \hline 
128 &  \card \textit{simp} & 1,024 & 752 & 723 & 1,121 & 5,869 & 511 & 0 & 0 & 0 & 0 & 0 \\ \hline 
 &  \card \textit{pmin} & 1,024 & 676 & 341 & 410 & 1,647 & 257 & 0 & 0 & 0 & 0 & 0 \\ \hline  \hline 
256 &  \card \textit{simp} & 1,014 & 796 & 739 & 968 & 1,608 & 4,779 & 96 & 0 & 0 & 0 & 0 \\ \hline 
 &  \card \textit{pmin} & 1,014 & 731 & 341 & 354 & 351 & 501 & 24 & 0 & 0 & 0 & 0 \\ \hline  \hline 
512 &  \card \textit{simp} & 1,024 & 791 & 721 & 945 & 1,295 & 1,979 & 3,240 & 5 & 0 & 0 & 0 \\ \hline 
 &  \card \textit{pmin} & 1,024 & 716 & 338 & 351 & 283 & 147 & 40 & 0 & 0 & 0 & 0 \\ \hline  \hline 
1K &  \card \textit{simp} & 1,090 & 793 & 671 & 833 & 1,130 & 1,557 & 2,177 & 1,749 & 0 & 0 & 0 \\ \hline 
 &  \card \textit{pmin} & 1,090 & 692 & 334 & 310 & 251 & 115 & 28 & 2 & 0 & 0 & 0 \\ \hline  \hline 
2K &  \card \textit{simp} & 1,155 & 735 & 683 & 809 & 1,145 & 1,372 & 1,714 & 1,720 & 667 & 0 & 0 \\ \hline 
 &  \card \textit{pmin} & 1,155 & 649 & 342 & 298 & 221 & 98 & 17 & 2 & 0 & 0 & 0 \\ \hline  \hline 
4K &  \card \textit{simp} & 1,136 & 765 & 686 & 795 & 1,072 & 1,307 & 1,525 & 1,549 & 1,065 & 100 & 0 \\ \hline 
 &  \card \textit{pmin} & 1,136 & 682 & 346 & 290 & 234 & 112 & 17 & 0 & 0 & 0 & 0 \\ \hline  \hline 
8K &  \card \textit{simp} & 1,156 & 749 & 661 & 836 & 1,010 & 1,242 & 1,431 & 1,468 & 1,138 & 307 & 2 \\ \hline 
 &  \card \textit{pmin} & 1,156 & 658 & 330 & 329 & 257 & 91 & 18 & 0 & 0 & 0 & 0 \\ \hline   
 \end{array}$
\end{center}
\end{table}

\subsection{Distribution of expressions with respect to their simplified size}
\label{appendix.distribution}

Although algorithm $\PU$ gives rather precise information about the distribution of expressions  with respect to their simplified size for $\nl = 2$, this is no longer the case for higher values of $\nl$. Notice that we are mainly interested by the distribution when the size of expressions tends to infinity because this distribution has an objective value and applies to input expressions of any size, except small ones. The main reason why precise information can be obtained for $\nl = 2$, but not beyond, is that algorithm $\PU$ applies to lifted expressions whose average size stabilizes at $70$
by $size = 1K$. Therefore, any large set of randomly chosen expressions of size at least $1K$ can be used to approximate the distribution. It is not so for higher values of $\nl$ since the average size of lifted expressions does not stabilize for sizes less than or equal to $8K$ (see Table \ref{simplification.expressions.23458}).
Moreover, we cannot go beyond, since we have  already imposed a bound of $512$ on the size of subexpressions for which an MDFA is computed, which is the second cause of inaccuracy of the modified algorithm \PU, analyzed here.

Anyway, the results provided by algorithm \PU, for $\nl = 3$, are presented in Table \ref{distribution.simpl.3}. We see that, compared to the case $\nl = 2$ (see Tables \ref{distribution.simpl.pu} and \ref{distribution.pmin.pu}), much fewer expressions are simplified into a ``possibly minimal'' expression, i.e.\ into a subexpression of the input set. This suggests that, to be more precise, we should consider (much) larger input sets.

\newpage
\subsection{Examples of most frequent simplified expressions}
\label{most.frequent.simplified}

Table \ref{frequent.minimal} depicts the set of  expressions into which the expressions of size $ 8K$ from our test data are the most frequently simplified (for $\nl = 2$). To keep the table reasonably small, only expressions selected at least $3$ times for simplification are collected. Since these expressions all have a size $\leq 15$, they are minimal since they are computed by algorithm $R$. Every expression is depicted with the number of input expressions simplified into it. For instance, $104$ expressions found in the input file are simplified (in fact, minimized) to \reg{(a + b)(a + b)*.} Statistically, this means that any expression of size $8K$ has a $1\%$ chance of being equivalent to  \reg{(a + b)(a + b)*,} and therefore of being simplified into it. In addition, we see that many expressions are ``similar''. If we define similar by the fact that two expressions are identical modulo a permutation of the letters and/or by reversing factors in concatenations, we see that all minimal expressions of size $5$ or $6$ are similar. Similar expressions have the same probability of being the simplification of an expression of any given input size. This similarity notion can be used to reduce the size of Table \ref{frequent.minimal}. For example, we could indicate that $443$ expressions, i.e.\ $4.43\%$ of the input expressions, are simplified into an expression similar to \reg{(ab*)*.}
We can also observe that some expressions in the table, although minimal, are not really illuminating as a description of the regular language they denote. For instance, the expression \reg{(a*ba*)*,} of size $8$, denotes the set of strings that are empty or contain at least one occurrence of the letter $b$. A clearer description is given by \reg{1 + (a + b)*b(a + b)*,} of size $12$, or by \reg{1 + a*b(a + b)*,} of size $10$. This illustrates the value of the star-height metric used by \cite{Stoughton}.

\begin{table}[!h]
\caption{Frequent simplified expressions ($\nl = 2$)}
\label{frequent.minimal}

\setlength{\arraycolsep}{0.15em}
\small
$$
\begin{array}{cccc}

\renewcommand{\arraystretch}{1.155}

\begin{array}{|r|c|}\hline
\multicolumn{2}{|c|}{ \size = 4 }\\ \hline
2,919 & \reg{ (a + b)*} \\\hline\hline

\multicolumn{2}{|c|}{ \size = 5 }\\ \hline
117 & \reg{ (b*a)*} \\
116 & \reg{ (a*b)*} \\
110 & \reg{ (ab*)*} \\
100 & \reg{ (ba*)*} \\
\hline\hline

\multicolumn{2}{|c|}{ \size = 6 }\\ \hline
192 & \reg{ b(a + b)*} \\
178 & \reg{ (a + b)*b} \\
168 & \reg{ a(a + b)*} \\
162 & \reg{ (a + b)*a} \\
\hline\hline

\multicolumn{2}{|c|}{ \size = 7 }\\ \hline
31 & \reg{ b + (ab*)*} \\
29 & \reg{ a + (ba*)*} \\
28 & \reg{ b + (b*a)*} \\
23 & \reg{ a + (a*b)*} \\
18 & \reg{ b(a*b)*} \\
10 & \reg{ a(b*a)*} \\
10 & \reg{ b(b*a)*} \\
9 & \reg{ (b*a)*a} \\
7 & \reg{ b(ab*)*} \\
6 & \reg{ b(ba*)*} \\
6 & \reg{ a(ba*)*} \\
6 & \reg{ (ba*)*a} \\
6 & \reg{ (a*b)*b} \\
5 & \reg{ (b*a)*b} \\
4 & \reg{ a(a*b)*} \\
4 & \reg{ (a*b)*a} \\
3 & \reg{ (ab*)*b} \\

\hline\hline
\multicolumn{2}{|c|}{ \size = 8 }\\ \hline
104 & \reg{ (a + b)(a + b)*} \\
33 & \reg{ (a*ba*)*} \\
32 & \reg{ (b*ab*)*} \\
27 & \reg{ b + a(a + b)*} \\
24 & \reg{ b + (a + b)*a} \\
24 & \reg{ a* + (ba*)*} \\
24 & \reg{ b* + (ab*)*} \\
22 & \reg{ (a + b)*aa} \\
21 & \reg{ b*(b*a)*} \\
\hline
\end{array}&

\renewcommand{\arraystretch}{1.144}
\begin{array}{|r|c|}\hline
\multicolumn{2}{|c|}{ \size = 8 }\\ \hline

20 & \reg{ a*(a*b)*} \\
18 & \reg{ a(a + b)*b} \\
17 & \reg{ a + (a + b)*b} \\
17 & \reg{ ab(a + b)*} \\
17 & \reg{ (a + b)*ab} \\
15 & \reg{ b(a + b)*a} \\
15 & \reg{ ba(a + b)*} \\
13 & \reg{ b(a + b)*b} \\
13 & \reg{ bb(a + b)*} \\
12 & \reg{ a + b(a + b)*} \\
12 & \reg{ (a + b)*bb} \\
10 & \reg{ 1 + a(b*a)*} \\
10 & \reg{ aa(a + b)*} \\
9 & \reg{ (a + b)*ba} \\
8 & \reg{ a(a + b)*a} \\
6 & \reg{ 1 + (b*a)*a} \\
6 & \reg{ 1 + b(a*b)*} \\
6 & \reg{ (a*b(1 + a))*} \\
5 & \reg{ 1 + (a*b)*b} \\
3 & \reg{ 1 + a(ab*)*} \\
3 & \reg{ (1 + b)(ab*)*} \\
3 & \reg{ (b*a)*(1 + b)} \\
3 & \reg{ ((1 + b)ab*)*} \\
3 & \reg{ ((1 + a)ba*)*} \\
3 & \reg{ (a + (a + b)b)*} \\
\hline\hline
\multicolumn{2}{|c|}{ \size = 9 }\\ \hline
52 & \reg{ a*b(a + b)*} \\
47 & \reg{ b*a(a + b)*} \\
10 & \reg{ 1 + ba(a + b)*} \\
8 & \reg{ 1 + ab(a + b)*} \\
7 & \reg{ 1 + a(a + b)*b} \\
6 & \reg{ 1 + (a + b)*ba} \\
6 & \reg{ 1 + bb(a + b)*} \\
6 & \reg{ 1 + b(a + b)*a} \\
5 & \reg{ 1 + (a + b)*ab} \\
5 & \reg{ 1 + a(a + b)*a} \\
5 & \reg{ 1 + b(a + b)*b} \\
5 & \reg{ (1 + b)a(a + b)*} \\
5 & \reg{ (a + b)*a(1 + b)} \\
4 & \reg{ 1 + (a + b)*aa} \\\hline

\end{array}
&
\renewcommand{\arraystretch}{1.137}
\begin{array}{|r|c|}\hline
\multicolumn{2}{|c|}{ \size = 9 }\\ \hline
4 & \reg{ b + a(b*a)*} \\
4 & \reg{ (1 + a)b(a + b)*} \\
4 & \reg{ (a + b)*b(1 + a)} \\
3 & \reg{ 1 + aa(a + b)*} \\
3 & \reg{ (b + (ab*)*)b} \\
3 & \reg{ ((a + bb)(1 + b))*} \\
3 & \reg{ (ba + ab*)*} \\
\hline\hline
\multicolumn{2}{|c|}{ \size = 10 }\\ \hline
16 & \reg{ a + (a*ba*)*} \\
12 & \reg{ (a + b)(a + b)*a} \\
12 & \reg{ (a + b)(a + b)*b} \\
11 & \reg{ b + (b*ab*)*} \\
11 & \reg{ b(a + b)(a + b)*} \\
11 & \reg{ a(a + b)(a + b)*} \\
6 & \reg{ b(a* + (ba*)*)} \\
4 & \reg{ a(b*ab*)*} \\
4 & \reg{ (ab*)*b*a} \\
4 & \reg{ (ab* + b*a)*} \\
3 & \reg{ 1 + a(b + (b*a)*)} \\
3 & \reg{ 1 + b + a(ab*)*} \\
3 & \reg{ b(b* + (ab*)*)} \\
3 & \reg{ ba*(a*b)*} \\
3 & \reg{ ba(a + b)*a} \\
3 & \reg{ bba(a + b)*} \\
3 & \reg{ bb(a + b)*b} \\
3 & \reg{ a(a*ba*)*} \\
3 & \reg{ a(a + b)*bb} \\
3 & \reg{ a(a + b)*ba} \\
3 & \reg{ a* + (a*b)*b} \\
3 & \reg{ b* + (b*a)*b} \\
3 & \reg{ (a + b)*(b + aa)} \\
3 & \reg{ ab* + (ba*)*} \\
3 & \reg{ (a + (a*b)*)(1 + a)} \\
3 & \reg{ (a + (a + b)(a + b))*} \\
\hline\hline
\multicolumn{2}{|c|}{ \size = 11 }\\ \hline
11 & \reg{ a*b(a + b)*b} \\
8 & \reg{ a + a*b(a + b)*} \\
8 & \reg{ (ba*)*(b*a)*} \\
8 & \reg{ (ab*)*(a*b)*} \\ 
7 & \reg{ bb*a(a + b)*} \\
\hline
\end{array}

&
\renewcommand{\arraystretch}{1.125}
\begin{array}{|r|c|}\hline
\multicolumn{2}{|c|}{ \size = 11 }\\ \hline
6 & \reg{ aa*b(a + b)*} \\
6 & \reg{ a*b(a + b)*a} \\
5 & \reg{ 1 + a(a + b)(a + b)*} \\
5 & \reg{ ((1 + a)(b + (a + b)a))*} \\
4 & \reg{ ba*b(a + b)*} \\
4 & \reg{ ab*a(a + b)*} \\
3 & \reg{ 1 + (a + b)*b(a + b)} \\
3 & \reg{ 1 + b(1 + (a*b)*b)} \\
3 & \reg{ b + b*a(a + b)*} \\
3 & \reg{ a(a + b)*a(1 + b)} \\
3 & \reg{ a*b*(b*a)*} \\
3 & \reg{ b*(b + (a + b)*a)} \\
3 & \reg{ (1 + a)b(a + b)*b} \\
3 & \reg{ (ab*)*b*a*} \\
3 & \reg{ ((1 + b)(a + (a + b)b))*} \\
3 & \reg{ (a + (a + b)*bb)*} \\
3 & \reg{ (b + (a + b)*aa)*} \\
\hline\hline
\multicolumn{2}{|c|}{ \size = 12 }\\ \hline
4 & \reg{ (a + b)(a + b)(a + b)*} \\
4 & \reg{ (a + b)*(a + (a + b)b)} \\
4 & \reg{ (aa + (1 + a)b(1 + a))*} \\
3 & \reg{ 1 + b*a(a + b)*b} \\
3 & \reg{ 1 + a*b(a + b)*a} \\
3 & \reg{ a + b(a + b)(a + b)*} \\
3 & \reg{ a* + ab + (ba*)*} \\
3 & \reg{ (a + b)*a(1 + b(1 + b))} \\
3 & \reg{ ((b + ab*a)(1 + a))*} \\
3 & \reg{ ((1 + b)(a + ba*b))*} \\
\hline\hline
\multicolumn{2}{|c|}{ \size = 13 }\\ \hline
6 & \reg{ (a + b)a*b(a + b)*} \\
5 & \reg{ (a + b)*ba(a + b)*} \\
4 & \reg{ 1 + (a + b)(a + b)(a + b)*} \\
3 & \reg{ (a + b)*bb(a + b)*} \\
\hline\hline
 
 \multicolumn{2}{|c|}{ \size = 14 }\\ \hline
3 & \reg{ 1 + (a + b)a*b(a + b)*} \\
3 & \reg{ b + ((a + ba*b)(1 + b))*} \\
3 & \reg{ a* + (ba*)*(a*b)*} \\
3 & \reg{ b* + (ab*)*(b*a)*} \\
\hline\hline

 \multicolumn{2}{|c|}{ \size = 15 }\\ \hline
3 & \reg{ ab + (ba*)*(b*a)*} \\
 \hline

\end{array}

\end{array}
$$
\end{table}

\newpage
\section{More information on lifting}
\label{information.lifting}

The lifting method was introduced in \cite{Kahrs}. A related but simplified method is described and theoretically studied in \cite{Rotondo}. The lifting algorithm, used in Section \ref{simplifying.expressions}, has  intermediate power and complexity. As said before, it is used as a pre-processing step in most of the algorithms studied in Section \ref{experiments}, i.e.\  it is applied ``outside the system'' to simplify an input expression before it is normalized and added to the set of expressions represented in the system. This ensures a more economical use of identifiers. Note that our lifting has a linear complexity, as does the algorithm of \cite{Rotondo}, while the algorithm of \cite{Kahrs} is quadratic because union ($+$) is an $n$-ary operator in their work.

The goal of this appendix is to describe the simplification rules that are used by our lifting algorithm. There are basically three rules.
Let $E$ be a (plain) regular expression. Let us denote the set of letters occurring in $E$ by ${\cal{A}}(E)$ and let us denote the set of letters $x$ such that $x\in\Lang(E)$ by ${\cal{A}}_1(E)$. Let  $x_1,\: x_2,\: \dots,\: x_\ell$ be letters. The rules are the following:

\begin{enumerate}
\item If ${\cal{A}}(E)\:=\:{\cal{A}}_1(E)$, then the expression $E^*$ can be simplified to $(x_1 + x_2 + \dots + x_\ell)^*$\\ where ${\cal{A}}_1(E) = \{x_1,\: x_2,\: \dots,\: x_\ell\}$.

\item If $1\in \Lang(E)$ and ${\cal{A}}(E)\subseteq \{x_1,\: x_2,\: \dots,\: x_\ell\}$ then $E\:\cdot\:(x_1 + x_2 + \dots + x_\ell)^*$ and
$(x_1 + x_2 + \dots + x_\ell)^*\:\cdot\:E$
both simplify to $(x_1 + x_2 + \dots + x_\ell)^*$.

\item If ${\cal{A}}(E)\subseteq \{x_1,\: x_2,\: \dots,\: x_\ell\}$ then $E\:+\:(x_1 + x_2 + \dots + x_\ell)^*$ and
$(x_1 + x_2 + \dots + x_\ell)^*\:+\:E$\\
both simplify to $(x_1 + x_2 + \dots + x_\ell)^*$.
\end{enumerate}

\noindent The rules can be applied bottom-up on subexpressions, since the sets ${\cal{A}}(E)$, ${\cal{A}}_1(E)$ and the fact that $1\in \Lang(E)$ can be computed bottom-up, as we go along. Since we use a limited number of letters, the sets can be represented by a single binary word and computed using logical operators on words. They are more general than the rules used in \cite{Rotondo} where a single ``universal'' expression $(x_1 + x_2 + \dots + x_\ell)^*$ using all letters possibly used in expressions is considered. 

 The fact that the rules in \cite{Rotondo} are effective for randomly generated expressions, especially for expressions using few letters, is a statistical property of regular expressions theoretically proved in \cite{Rotondo}. The greater generality of our rules does not bring much improvement in practice for such expressions.

\end{document}